\documentclass[aps, prd, amsmath, floatfix, superscriptaddress, preprintnumbers, preprint]{revtex4-1}

\usepackage{color}
\usepackage[colorlinks=true,linkcolor=blue,citecolor=blue,urlcolor=blue]{hyperref}
\usepackage[dvipsnames]{xcolor}
\usepackage{tabularx}
\usepackage{physics}
\usepackage{hhline}
\usepackage{rotating}
\usepackage{amsmath,amsfonts,amssymb,mathtools,cases,slashed,bm}
\usepackage{mathrsfs}
\usepackage[normalem]{ulem}
\usepackage{epsfig, graphicx}

\newcommand{\eref}[1]{Eq.~(\ref{#1})}

\newcommand{\diff}[1]{\mathrm{d}#1}
\newcommand{\xn}{x_{\rm N}}
\newcommand{\hel}{{^3 {\rm He}}}
\newcommand{\chired}{{\chi^2_{\rm red}}}
\newcommand{\Apa}{{$A_{\parallel}$}}
\newcommand{\Ape}{{$A_{\perp}$}}
\newcommand{\Atpe}{{$\widetilde{A}_{\perp}$}}
\newcommand{\Apajet}{{$A_{\parallel}^{\rm jet}$}}

\begin{document}

\title{Global QCD analysis of spin PDFs in the proton with high-$x$ \\ and lattice constraints}

\author{C.~Cocuzza}
\affiliation{\mbox{Department of Physics, William and Mary, Williamsburg, 
        Virginia 23185, USA}}
\author{N.~T.~Hunt-Smith}
\affiliation{CSSM and ARC Centre of Excellence for Dark Matter Particle Physics,
        Department of Physics, University of Adelaide, Adelaide 5005, Australia}
\author{W.~Melnitchouk}
\affiliation{Jefferson Lab, Newport News, Virginia 23606, USA \\
        \vspace*{0.2cm}
        {\bf JAM Collaboration \\ {\footnotesize \ (Spin PDF Analysis Group)}
        \vspace*{0.2cm} }}
\author{N. Sato}
\affiliation{Jefferson Lab, Newport News, Virginia 23606, USA \\
        \vspace*{0.2cm}
        {\bf JAM Collaboration \\ {\footnotesize \ (Spin PDF Analysis Group)}
        \vspace*{0.2cm} }}
\author{A.~W.~Thomas}
\affiliation{CSSM and ARC Centre of Excellence for Dark Matter Particle Physics,
        Department of Physics, University of Adelaide, Adelaide 5005, Australia}

\begin{abstract}
We perform a comprehensive global QCD analysis of spin-dependent parton distribution functions (PDFs), combining all available data on inclusive and semi-inclusive deep-inelastic scattering (DIS), as well as inclusive weak boson and jet production in polarized $pp$ collisions, simultaneously extracting spin-averaged PDFs and fragmentation functions.
Including recent Jefferson Lab DIS data at high $x$, together with subleading power corrections to the leading twist framework, allows us to verify the stability of the PDFs for $W^2 \geq 4$~GeV$^2$ and quantify the uncertainties on the spin structure functions more reliably.
We explore the use of new lattice QCD data on gluonic pseudo Ioffe-time distributions, which, together with jet production and high-$x$ DIS data, improve the constraints on the polarized gluon PDF.
The expanded kinematic reach afforded by the data into the high-$x$ region allows us to refine the bounds on higher twist contributions to the spin structure functions, and test the validity of the Bjorken sum rule.
\end{abstract}

\date{\today}
\preprint{JLAB-THY-25-4362, ADP-25-21/T1283}
\maketitle

\section{Introduction}
\label{s.intro}

The study of the spin structure of the proton has matured significantly over the past decade, with major progress witnessed on several fronts.
Ever more precise experimental data from modern accelerator facilities, combined with theoretical developments, including insights from lattice QCD simulations and improvements in analysis methodologies, have furthered our understanding of the decomposition of the proton’s spin into its quark and gluon (or parton) constituents over a large range of the momentum fractions $x$ carried by the partons.
Recently the need to simultaneously analyze parton distribution functions (PDFs) and fragmentation functions (FFs)~\cite{Ethier:2017zbq, Sato:2019yez, Moffat:2021dji} has been highlighted with the inclusion of semi-inclusive deep-inelastic scattering (DIS) data, which depend on both types of nonperturbative functions.
This also allows one to consistently determine unpolarized and polarized PDFs under the same set of conditions and approximations~\cite{Cocuzza:2022jye}.
Such analyses are naturally more complicated than conventional, single-purpose fits, requiring greater computing power and more sophisticated algorithms to provide  robust extractions of the various distributions while minimizing reliance on theoretical assumptions.

An illustrative example of this has been the attempt to extract the strange quark helicity distribution, $\Delta s$, utilizing semi-inclusive pion and kaon production data, which demonstrated that in the absence of assumptions about flavor SU(3) symmetry, the polarized strange content of the proton is consistent with zero~\cite{Sato:2019yez, Moffat:2021dji}.
Removing theoretical constraints associated with flavor SU(3) symmetry and assumptions about PDF positivity beyond leading order further revealed that the world data, including the important jet production data from polarized $pp$ collisions at RHIC~\cite{STAR:2014wox, STAR:2019yqm, STAR:2021mfd, STAR:2021mqa, PHENIX:2010aru}, were not sufficient to determine the sign of the gluon polarization, $\Delta g$~\cite{Zhou:2022wzm, Karpie:2023nyg}.
Only when combining the world data with constraints from lattice QCD simulations of pseudo-Ioffe time distributions sensitive to $\Delta g$, together with high precision large-$x$ data from Jefferson Lab, was it possible to eliminate the possibility of negative gluon polarization from the conventional positive solutions in a model independent manner~\cite{Hunt-Smith:2024khs}.
These studies affirm the need for a diverse set of observables, both experimental and, more recently, from lattice QCD, across a wide range of energies to provide detailed, model independent determinations of the parton helicity structure of the proton from low $x$ to high~$x$. \\

In previous global QCD analyses, the JAM collaboration has undertaken a systematic study of spin-dependent PDFs within a Bayesian Monte Carlo framework, using a holistic approach in which ultimately the polarized and unpolarized PDFs, along with fragmentation functions, and in some cases lattice observables, are fitted simultaneously~\cite{Jimenez-Delgado:2013boa, Sato:2016tuz, Ethier:2017zbq, Cocuzza:2022jye, Zhou:2022wzm, Karpie:2023nyg, Hunt-Smith:2024khs}.
Early JAM analyses focused on consistently describing the available DIS polarization asymmetry data, including subleading hadron and nuclear effects relevant particularly at high $x$ and low momentum transfers $Q^2$~\cite{Jimenez-Delgado:2013boa, Sato:2016tuz}.
Subsequent studies incorporated polarized SIDIS, single-inclusive $e^+ e^-$ annihilation (SIA), and polarized $W/Z$ production data to identify the flavor dependence of the polarized sea at small $x$~\cite{Ethier:2017zbq, Cocuzza:2022jye}, and polarized jet data along with lattice QCD matrix elements to constrain the polarization of the gluon~\cite{Zhou:2022wzm, Karpie:2023nyg, Hunt-Smith:2024khs}.

In this paper we combine the elements from these earlier studies and perform a new comprehensive global QCD analysis of helicity-dependent PDFs covering both the small-$x$ and large-$x$ regions, combining all available data from inclusive DIS on protons, deuterons, and $^3$He, semi-inclusive production of pion, kaons and other hadrons from protons and deuterons, as well as inclusive $W$ boson and jet production in singly and doubly polarized $pp$ collisions.
For the DIS data, we study the dependence of the fit results on the cut in the hadron final state invariant mass, $W$, to determine the extent to which high-$x$ data can provide stable results for the leading twist (LT) distributions.
This requires taking into account target mass effects and other power corrections to the LT formulation, which we determine phenomenologically from the data.
Along with the helicity-dependent PDFs, we also simultaneously fit the unpolarized PDFs and parton to hadron fragmentation functions, as in the earlier JAM analysis in Ref.~\cite{Cocuzza:2022jye}.
The resulting set of PDFs, which we refer to by the moniker \mbox{``JAMpol25''}, are constrained over the range $0.005 \lesssim x \lesssim 0.85$ for scales down to the charm quark mass, $\mu^2 = m_c^2$.

In the following, we begin by reviewing the formulas for observables used in this analysis in Sec.~\ref{s.observables}, including inclusive DIS and SIDIS, $W$-boson and jet production in polarized proton-proton collisions, as well as lattice QCD simulations of pseudo-Ioffe time distributions sensitive to the gluon polarization, $\Delta g$.
In Sec.~\ref{s.methodology} we outline our methodology, including a discussion of the parametrizations of the polarized PDFs and our treatment of subleading power corrections needed to account for higher twist effects at low $Q^2$.
In Sec.~\ref{s.globalQCDanalysis} we present the results of our global QCD analysis, performing a detailed comparison of data versus theory to assess the quality of the fit.
To establish the stability of the fitted results, we compare the $\chi^2$ values and $Z$-scores as a function of the cut on the invariant mass of the final state hadrons, from $W^2=3.5$~GeV$^2$ to 10~GeV$^2$.
The results for the helicity distributions are presented in Sec.~\ref{s.ppdfs}, while the residual higher twist contributions to the $g_1$ and $g_2$ structure functions are discussed in Sec.~\ref{s.HTnuc}.
Finally, in Sec.~\ref{s.conclusions} we summarize our results and discuss the improvements expected in the theoretical description and experimental data from future facilities.
For completeness, in Appendix~\ref{a.TMC-CF} we outline the derivation of the target mass corrections in collinear factorization.
In Appendix~\ref{a.data} we present the detailed data versus theory comparisons for all the polarized scattering observables used to constrain our global analysis. 

\section{Observables}
\label{s.observables}

Our analysis considers a set of spin-dependent observables that can be described theoretically using collinear factorization in the framework of fixed order perturbative QCD. 
For each observable we utilize short-distance partonic cross sections computed at next-to-leading order (NLO) accuracy in the strong coupling, $\alpha_s$, as described in the following subsections.
Recent studies~\cite{Borsa:2024mss, Cruz-Martinez:2025ahf} showed that effects from higher orders, such as next-to-next-to-leading order (NNLO), are surprisingly small.

\subsection{Polarized inclusive DIS}
\label{ss.pidis}

The classical observables measured in polarized lepton DIS from a target $T$, $\vec \ell\,\, \vec T \to \ell\, X$, where $T$ represents a proton $p$, deuteron $d$, or $^3$He nucleus, are the longitudinal and transverse double-spin asymmetries~\cite{Sato:2016tuz},
\begin{subequations}
\begin{eqnarray}
A_{\parallel} &=& \frac{\sigma^{\uparrow\Uparrow} - \sigma^{\downarrow\Uparrow}}{\sigma^{\downarrow\Uparrow} + \sigma^{\uparrow\Uparrow}}\, ,
\\
A_{\perp} &=& \frac{\sigma^{\uparrow\Rightarrow} - \sigma^{\downarrow\Rightarrow}} {\sigma^{\downarrow\Rightarrow} + \sigma^{\uparrow\Rightarrow}},
\end{eqnarray}
\end{subequations}
where the superscript $\uparrow$ ($\downarrow$) on the cross section $\sigma$ denotes the spin of the lepton along (opposite to) the beam direction, $\Uparrow$ denotes the spin of the target nucleon along the beam direction, and $\Rightarrow$ represents the spin of the target perpendicular to the beam direction. 
The cross sections are functions of the standard DIS Lorentz invariant kinematic variables, namely, Bjorken $x$, lepton inelasticity $y$, and the variable $\gamma^2$, defined as
\begin{equation}
x = \frac{Q^2}{2 P \cdot q}\, , ~~~~~~
\\
y = \frac{P \cdot q}{P \cdot k}\, , ~~~~~~
\\
\gamma^2 = \frac{4 M^2 x^2}{Q^2}\, ,
\label{e.kinematic}
\end{equation}
where $k$, $P$ and $q$ are the four-momenta of the incident lepton, target nucleon (of mass $M$), and exchanged virtual photon, respectively, with $Q^2 \equiv -q^2$.
In terms of these variables, the longitudinal and transverse spin asymmetries can be written as
\begin{subequations}
\begin{eqnarray}
A_{\parallel} &=& D\, \pqty{A_1 + \eta A_2}\, ,
\\
A_{\perp} &=& d\, \pqty{A_2 - \zeta A_1}\, ,
\label{e.AparAperp}
\end{eqnarray}
\end{subequations}
where the kinematic prefactors are given by
\begin{subequations}
\begin{eqnarray}
D &=& \frac{y\, \pqty{2-y} \pqty{2 + \gamma^2 y}} {2 (1+\gamma)^2 y^2 + \big[ 4 \pqty{1-y} - \gamma^2 y^2 \big] \pqty{1+R}}\, , 
\label{e.D}
\\
d &= & \frac{\sqrt{4(1-y) - \gamma^2 y^2}}{2-y} D\, ,
\\
\eta &=& \gamma \frac{4 \pqty{1-y} - \gamma^2 y^2}{\pqty{2-y} \pqty{2 + \gamma^2 y}}\, ,
\\
\zeta &=& \gamma \frac{2-y}{2 + \gamma^2 y}\, .
\end{eqnarray}
\end{subequations}
The virtual photoproduction asymmetries $A_1$ and $A_2$ can be written in terms of ratios of the spin-dependent $g_1$ and $g_2$ and spin-averaged $F_1$ structure functions,
\begin{subequations}
\begin{eqnarray}
A_1 &=& \frac{( g_1 - \gamma^2 g_2 )}{F_1}\, , \\
A_2 &=& \frac{\gamma ( g_1 + g_2 )}{F_1}\, .
\label{e.A}
\end{eqnarray}
\end{subequations}
At large $Q^2$ the $A_1$ asymmetry is dominated by the $g_1$ structure function, while the $A_2$ asymmetry depends on the sum $g_1+g_2$, which is sometimes referred to as the transverse spin structure function, $g_T \equiv g_1+g_2$.
Measurements of $A_2$ are therefore crucial for determining the $g_2$ structure function.
In Eq.~(\ref{e.D}) $R$ is the ratio of longitudinal to transverse photoproduction cross sections, and is related to the unpolarized $F_1$ and $F_2$ structure functions by
\begin{equation}
R = \frac{\rho^2 F_2 - 2x F_1}{2x F_1}\, ,
\label{e.R}
\end{equation}
where 
\begin{equation}
\rho^2 = 1 + \gamma^2\, . 
\label{e.rho}
\end{equation}
Each of the polarized and unpolarized structure functions is a function of two variables, which in this analysis we take to be the Bjorken variable $x$ and the four-momentum transfer squared $Q^2$.

Applying QCD collinear factorization, the spin-dependent $g_1$ structure function can be written, up to ${\cal O}(M^2/Q^2)$ power corrections, in terms of the LT spin-dependent quark ($\Delta q$) and gluon ($\Delta g$) PDFs,
\begin{eqnarray}
g_1^{\mbox{\tiny \rm LT}}(x,Q^2) 
&=& \frac12 \sum_q e_q^2 
\big[ \Delta C_q^{\mbox{\tiny \rm DIS}} \otimes \Delta q^+ 
   + 2\Delta C_g^{\mbox{\tiny \rm DIS}} \otimes \Delta g 
\big](x,Q^2)\, ,
\label{e.g1LT}
\end{eqnarray}
where $\Delta q^+ \equiv \Delta q + \Delta \bar{q}$, and the sum runs over all quark flavors $q$.
Here the symbol $\otimes$ denotes the convolution integral,
\begin{equation}
\big[ \Delta C_f^{\mbox{\tiny \rm DIS}} \otimes \Delta f \big](x,Q^2)
\equiv \int_x^1 \frac{\diff \xi}{\xi}\,
\Delta C_f^{\mbox{\tiny \rm DIS}}(\xi,Q^2)\, \Delta f \Big( \frac{x}{\xi},Q^2 \Big)\, ,
\label{e.convolution}
\end{equation}
and we have chosen the renormalization and factorization scales for the DIS process to be $\mu^2 = Q^2$.
The perturbatively calculable polarized DIS hard scattering coefficients, $\Delta C_f^{\mbox{\tiny \rm DIS}}$ ($f = q$, $g$), are expanded to NLO in the strong coupling $\alpha_s$,
\begin{eqnarray}
\Delta C_f^{\mbox{\tiny \rm DIS}} 
= \Delta C_f^{\mbox{\tiny \rm DIS(0)}} 
+ \frac{\alpha_s}{4 \pi} \Delta C_f^{\mbox{\tiny \rm DIS(1)}}
+ \mathcal{O}(\alpha_s^2)\, ,
\end{eqnarray}
with the coefficient functions taken from Ref.~\cite{Floratos:1981hs}.
Note that at leading order in $\alpha_s$ the hard scattering coefficient $\Delta C_f^{\mbox{\tiny \rm DIS(0)}}(\xi,Q^2) \propto \delta(1-\xi)$, and the Bjorken scaling variable $x$ in Eq.~(\ref{e.kinematic}) coincides with the parton momentum fraction, while at NLO and beyond these are different.
At leading twist, the $g_2$ structure function can be related to $g_1$ through the Wandzura-Wilczek relation~\cite{Wandzura:1977qf},
\begin{eqnarray}
g_2^{\mbox{\tiny \rm LT}}(x,Q^2)
= -g_1^{\mbox{\tiny \rm LT}}(x,Q^2) 
+ \int_x^1 \frac{\diff z}{z} g_1^{\mbox{\tiny \rm LT}}(z,Q^2)\, .
\label{e.WW}
\end{eqnarray}

For analysis of DIS data at relatively small values of $Q^2 \gtrsim m_c^2$ and invariant mass squared of the unobserved hadronic system $X$,
\begin{eqnarray}
W^2 = (P+q)^2 = M^2 + \frac{(1-x)}{x} Q^2\, ,
\label{e.W2}
\end{eqnarray}
power corrections of the type ${\cal O}(M^2/Q^2)$ may not be negligible for some lower energy datasets. Typically a cut in $W^2$ needs to be made to ensure that the theoretical framework is able to accommodate the data.
It is well known that one can extend the reach of a perturbative QCD description of DIS data by taking into account the kinematical target mass corrections (TMCs)~\cite{Jimenez-Delgado:2013boa, Sato:2016tuz, Schienbein:2007gr}, which can be applied to the polarized structure functions to incorporate $M^2/Q^2$ corrections.
These corrections scale with the Nachtmann variable $\xn$, where \cite{Nachtmann:1973mr, Greenberg:1971lpf}
\begin{eqnarray}
\xn = \frac{2x}{1+\rho}\, .
\end{eqnarray}
In the collinear factorization framework, the $g_1$ and $g_2$ structure functions, including TMCs, can be written as
\begin{subequations}
\begin{eqnarray}
g_1^{\mbox{\tiny \rm TMC}}(x,Q^2)
&=& \frac{1}{\rho^2} 
    g_1^{\mbox{\tiny \rm LT}}(\xn,Q^2)
 +  \frac{2(\rho-1)}{\rho^2} \int_{\xn}^1 \frac{\diff z}{z} 
    g_1^{\mbox{\tiny \rm LT}}(z,Q^2)\, ,
\label{e.CFtmc1}
\\
g_2^{\mbox{\tiny \rm TMC}}(x,Q^2) 
&=& -\frac{1}{\rho^2} 
    g_1^{\mbox{\tiny \rm LT}}(\xn,Q^2) 
+  \frac{2}{(1+\rho) \rho^2} \int_{\xn}^1 \frac{\diff z}{z} 
    g_1^{\mbox{\tiny \rm LT}}(z,Q^2)\, .
\label{e.CFtmc2}
\end{eqnarray}
\label{e.CFtmc}%
\end{subequations}
In the limit $M^2/Q^2 \to 0$, one has $\rho \to 1$ and $\xn \to x$, and the LT results in Eqs.~(\ref{e.g1LT}) and (\ref{e.WW}) are recovered,
    $g_{1,2}^{\mbox{\tiny \rm TMC}}(x,Q^2) \to 
     g_{1,2}^{\mbox{\tiny \rm LT}}(x,Q^2)$.
Note that the lower limit of the integral from the Wandzura-Wilczek relation, \eref{e.WW}, is now $\xn$ rather than $x$.
A derivation of Eqs.~(\ref{e.CFtmc}) is given in Appendix~\ref{a.TMC-CF}.

Alternative TMC prescriptions have also been used in the literature~\cite{Wandzura:1977qf, Matsuda:1979ad, Piccione:1997zh, Blumlein:1998nv, Accardi:2008pc}, the most common of which is that based on the operator product expansion, along the lines of the unpolarized TMC prescription of Georgi and Politzer~\cite{Georgi:1976ve}.
This prescription was used in the early JAM spin PDF analysis of inclusive DIS data at low $Q^2$~\cite{Sato:2016tuz}.
For consistency with the application of collinear factorization across other processes considered beyond inclusive DIS, such as those in the current analysis, we will utilize the TMC results for the collinear factorization given in Eqs.~(\ref{e.CFtmc}).

One immediate consequence of Eqs.~(\ref{e.CFtmc}), as in other TMC approaches, is that for nonzero $M^2/Q^2$, as $x \to 1$ one has $\xn < 1$, and hence a nonzero structure function at $x = 1$.
The fact that the TMC structure functions do not vanish as $x \to 1$ is known as the ``threshold problem" \cite{Georgi:1976ve, DeRujula:1976baf, DeRujula:1976ih} and has been discussed in the literature~\cite{Bitar:1978cj, Steffens:2006ds, Schienbein:2007gr, Steffens:2012jx}.
This problem is only relevant in the nucleon resonance region, well below values of $W^2$ where perturbative QCD is applicable, so in practice it will not be relevant for this analysis.

From phenomenological analyses it is also well established~\cite{Jimenez-Delgado:2013boa, Sato:2016tuz, Owens:2012bv, Accardi:2016qay, Melnitchouk:2005zr} that in the region where TMCs play a role, namely at low $Q^2$ and high $x$, other power corrections, such as those associated with higher twists, or multi-parton correlations, also need to be considered.
These are implemented in the standard manner by adding a phenomenological higher twist (HT) term.
For the $g_1$ structure function the HT contribution is power suppressed relative to the leading term, Eq.~(\ref{e.CFtmc1}), while for the $g_2$ structure function the HT correction to the LT term (\ref{e.CFtmc2}) is of the same order in $Q^2$~\cite{Jaffe:1990qh, Anselmino:1994gn},
\begin{eqnarray}
g_1^{\mbox{\tiny \rm HT}} 
= \frac{c^{\mbox{\tiny \rm HT}}_1}{Q^2}\, ,
\qquad
&&
g_2^{\mbox{\tiny \rm HT}} 
= c^{\mbox{\tiny \rm HT}}_2,
\label{e.g12HT}
\end{eqnarray}
with the shapes of the $x$-dependent functions 
    $c_{1,2}^{\mbox{\tiny \rm HT}} \equiv c_{1,2}^{\mbox{\tiny \rm HT}}(x)$
inferred from experimental data.
The total spin-dependent structure functions in our framework are given by the sum of the LT and HT contributions,
\begin{align}
g_{i} = g_{i}^{\mbox{\tiny \rm TMC}} + g_{i}^{\mbox{\tiny \rm HT}}\, ,
\qquad i=1,2.
\end{align}
In our numerical analysis both the parameter sets describing the LT and HT functions will be extracted from data.

For inclusive polarized DIS, our datasets include scattering from protons, as well as deuterium and $^3$He nuclei.
For nuclear targets, the structure functions are computed taking into account the wave functions of the nucleons in the polarized $A=2$ and 3 nuclei.
It was shown in earlier studies that this can be written in the form of a generalized convolution~\cite{Melnitchouk:1994tx, Kulagin:1994cj, Piller:1995mf, Kulagin:2007ph, Kulagin:2008fm, Jimenez-Delgado:2013boa},
\begin{eqnarray}
g_i^A(x,Q^2)
&=& \sum_N \big[ \Delta f_{ij}^{N/A} \otimes g_j^N \big](x,Q^2), \qquad
(i,j = 1,2), 
\label{e.giA}
\end{eqnarray}
where the spin-dependent smearing functions $\Delta f_{ij}^{N/A}$ are functions of the light-cone momentum fraction $y_N = p_N \cdot q / P_A \cdot q$ of the nucleus carried by the interacting nucleon, and the finite-$Q^2$ parameter $\gamma$.
The nuclear smearing effects are known to become important at large values of $x$~\cite{Kulagin:2007ph, Kulagin:2008fm, Jimenez-Delgado:2013boa}, but for $x \ll 1$ one may approximate the nuclear structure functions in terms of effective polarizations, 
$\int \dd{y_N}\, \Delta f_{ij}^{N/A}(y_N,\gamma) = \delta_{ij}\, {\cal P}_{N/A}\, \delta(1-y_N)$.
In this effective polarization approximation, the nuclear effects are independent of $x$.
In practice, however, since we include data in the high-$x$ region, our calculations are performed using the full convolution in Eq.~(\ref{e.giA}), with the polarized smearing functions normalized to the effective polarizations ${\cal P}_{N/A}$, as discussed in Refs.~\cite{Kulagin:2007ph, Kulagin:2008fm}.

Integrating the structure functions over $x$, we can define their lowest moments as
\begin{eqnarray}
\Gamma_i(Q^2) &=& \int_0^1 \dd{x} g_i(x,Q^2), \qquad i=1,2.
\end{eqnarray}
For the $g_1$ structure function of the proton, the lowest moment can be written, including ${\cal O}(\alpha_s)$ corrections, in terms of axial vector charges as
\begin{eqnarray}
\Gamma_1^{\mbox{\tiny \rm LT}}(Q^2)
&=& \bigg( 1 - \frac{\alpha_s(Q^2)}{\pi} \bigg) 
\bigg(
  \frac{1}{12}\, g_A + \frac{1}{36}\, a_8 + \frac{1}{9}\, \Delta\Sigma
\bigg),
\label{e.Gamma1LT}
\end{eqnarray}
where 
\begin{subequations}
\label{e.axialcharges}
\begin{eqnarray}
g_A &=& \int_0^1 \dd{x} [\Delta u^+ - \Delta d^+](x,Q^2), 
\label{e.axialcharges.gA}\\
a_8 &=& \int_0^1 \dd{x} [\Delta u^+ + \Delta d^+ - 2 \Delta s^+](x,Q^2), \\
\Delta\Sigma &=& \int_0^1 \dd{x} \sum_q \Delta q^+(x,Q^2),
\end{eqnarray}
\end{subequations}%
are the quark triplet, octet, and singlet axial charges, respectively.
Note that both $g_A$ and $a_8$ are scale invariant, while $\Delta\Sigma = \Delta\Sigma(Q^2)$ explicitly depends on the scale $Q^2$.
For the PDFs in the neutron, we use charge symmetry~\cite{Londergan:2009kj} to relate these to those in the proton, with the interchange $u \leftrightarrow d$.

For the lowest moment of the $g_2$ structure function, in the LT approximation $\Gamma_2^{\mbox{\tiny \rm LT}}$ satisfies the Burkhardt-Cottingham (BC) sum rule~\cite{Burkhardt:1970ti},
\begin{eqnarray}
\Gamma_2^{\mbox{\tiny \rm LT}}(Q^2) &=& 0.
\label{e.BCSR}
\end{eqnarray}
Phenomenologically, violation of the BC sum rule would suggest the presence of power corrections to $g_2$.
An additional combination of the spin structure functions which also directly reveals power corrections is the $d_2$ matrix element, defined in terms of the $x^2$-weighted moments of $g_1$ and $g_2$,
\begin{eqnarray}
d_2(Q^2) &=& \int_0^1 \dd{x} x^2\, \big[ 2 g_1(x,Q^2) + 3 g_2(x,Q^2) \big].
\label{e.d2}
\end{eqnarray}
From the Wandzura-Wilczek relation (\ref{e.WW}), one can immediately see that the LT part of this combination vanishes, $d_2^{\mbox{\tiny \rm LT}} = 0$, so that nonzero values can only arise from power corrections associated with target mass effects or higher twist contributions.
These contributions can be related to matrix elements describing the nucleon's color polarizability~\cite{Ji:1993sv, Stein:1994zk, Stein:1995si} or the transverse color force~\cite{Burkardt:2008ps} acting on quarks, and have been studied in earlier phenomenological analyses~\cite{Meziani:2004ne, Osipenko:2004xg, Deur:2004ti, Osipenko:2005nx}.

\subsection{Polarized semi-inclusive DIS}
\label{ss.psidis}

In polarized SIDIS, $\vec\ell\, \vec{T} \to \ell\, h\, X$, where $h$ is a hadron identified in coincidence with the scattered lepton, the usual observable  measured is the longitudinal double-spin asymmetry, $A_1^h$, which at large $Q^2$ is given by
\begin{eqnarray}
A_1^h(x,z,Q^2) = \frac{g_1^h(x,z,Q^2)}{F_1^h(x,z,Q^2)},
\label{e.A1h}
\end{eqnarray}
where $g_1^h$ and $F_1^h$ are the polarized and unpolarized semi-inclusive structure functions, respectively~\cite{Bacchetta:2006tn}.
In addition to the inclusive variables $x$ and $Q^2$, the SIDIS asymmetry also depends on the fraction of the virtual photon's energy carried by the hadron $h$, defined as
\begin{equation}
z = \frac{P \cdot p_h}{P \cdot q},
\end{equation}
where $p_h$ is the four-momentum of the produced hadron. 
In terms of these variables the invariant mass squared of the unobserved hadronic final state in the SIDIS process is given by
\begin{eqnarray}
W^2_{\mbox{\tiny \rm SIDIS}}
&=& (P + q - p_h)^2
\nonumber\\
&=& W^2 + m_h^2 - \frac{z}{x} Q^2 - 2 p_h \cdot q
\nonumber\\
&=& M^2 + m_h^2 + \frac{Q^2(1-x-z)}{x}
 + \frac{Q^4 z}{2 M^2 x^2} \big( \rho\, \rho_h - 1 \big),
\end{eqnarray}
where in analogy with Eq.~(\ref{e.rho}) we define $\rho_h$ by
\begin{equation}
\rho_h = \sqrt{1-\frac{4M^2 x^2\, m_{hT}^2}{z^2 Q^4}},
\end{equation}
with $m_{hT}$ the transverse mass of the produced hadron $h$,
\begin{equation}
m_{hT}^2 = m_h^2 + p_{hT}^2.
\end{equation}
For unidentified hadrons, we approximate that $m_h \approx m_{\pi}$.
Integrating over the transverse momentum of the produced hadron, $p_{hT}$, and expanding in powers of $M^2/Q^2$ and $m_h^2/Q^2$, the polarized SIDIS structure function $g_1^h$ can be written in terms of the helicity PDFs $\Delta q$, $\Delta g$ and single-hadron FFs $D_f^h$, 
\begin{eqnarray}
g_1^h(x,z,Q^2) 
&=& \frac12\sum_q e_q^2 
\big[ \Delta C_{qq}^{\mbox{\tiny \rm SIDIS}} \otimes \Delta q \otimes D_q^h 
    + \Delta C_{gq}^{\mbox{\tiny \rm SIDIS}} \otimes \Delta q \otimes D_g^h
\notag\\
& & \hspace*{1.2cm}
  +\, \Delta C_{qg}^{\mbox{\tiny \rm SIDIS}} \otimes \Delta g \otimes D_q^h 
\big](x,z,Q^2), 
\label{e.g1h}
\end{eqnarray}
where the sum $q$ runs over all quarks and antiquarks.
Generalizing the convolution in Eq.~(\ref{e.convolution}), we define the double convolution in Eq.~(\ref{e.g1h}) by
\begin{eqnarray}
\big[ \Delta C \otimes \Delta f \otimes D_{f'}^h \big](x,z)
\equiv \int_x^1 \frac{\diff \xi}{\xi} \int_z^1 
\frac{\diff \zeta}{\zeta}
  \Delta C(\xi, \zeta)\, 
  \Delta f\Big(\frac{x}{\xi}\Big)\, 
  D_{f'}^h\Big(\frac{z}{\zeta}\Big).
\end{eqnarray}
The perturbatively calculable polarized SIDIS hard scattering coefficients, $\Delta C_{ff'}^{\mbox{\tiny \rm SIDIS}}$ with $ff' = qq$, $gq$, $qg$, are expanded to NLO in the strong coupling constant,
\begin{eqnarray}
\Delta C_{ff'}^{\mbox{\tiny \rm SIDIS}} 
= \Delta C_{ff'}^{\mbox{\tiny \rm SIDIS(0)}}
+ \frac{\alpha_s}{4 \pi} 
  \Delta C_{ff'}^{\mbox{\tiny \rm SIDIS(1)}}
+ \mathcal{O}(\alpha_s^2),
\end{eqnarray}
with the hard coefficients taken from Ref.~\cite{Stratmann:2001pb}.
The renormalization and factorization scales are taken to be equivalent, $\mu_R = \mu_F = Q$, for the SIDIS process. 
For the neutron SIDIS structure function the same proton PDFs are used, except with the switch $u \leftrightarrow d$ that is derived from charge symmetry.
The unpolarized SIDIS structure function $F_1^h$ is also given by \eref{e.g1h} with the spin-dependent PDFs and hard coefficients replaced by their spin-averaged counterparts.

Since the SIDIS asymmetry and structure function in Eqs.~(\ref{e.A1h}) and (\ref{e.g1h}) are evaluated in the LT approximation, we exclude from our analysis of SIDIS data the regions of kinematics where higher twist and hadron mass corrections may contaminate the leading power description~\cite{Accardi:2009md, Guerrero:2015wha, Guerrero:2017yvf}.
We found that this can be achieved by applying cuts on the invariant mass of the unobserved hadrons in the SIDIS final state $W^2_{\mbox{\tiny \rm SIDIS}} > 20$~GeV$^2$ and energy fraction $z < 0.8$, while a cut on low-$z$ data, $z > 0.2$, avoids having to deal with final state hadron mass corrections.

For the nuclear corrections, one can use the same convolution framework as for inclusive DIS, assuming that there is no modification of the fragmentation function for nuclear targets compared with the proton.
However, since the SIDIS deuterium data typically do not extend to high values of $x$, $x \lesssim 0.3$, it is sufficient to approximate the nuclear corrections in terms of the effective polarizations discussed above in Sec.~\ref{ss.pidis}.
A dedicated analysis of SIDIS data in the low-$Q^2$ region, which includes these mass and higher twist corrections and nuclear smearing, will be given elsewhere~\cite{Adamiak2024}.

\subsection{$W$ production in polarized hadron collisions}
\label{ss.wproduction}

The longitudinal single spin asymmetry for $W$-boson production in proton-proton collisions is defined in terms of the helicity-dependent cross sections as
\begin{align}
A_L^{W}
= \frac{\diff \sigma^+ - \diff \sigma^-}{\diff \sigma^+ + \diff \sigma^-}
\equiv \frac{\diff \Delta \sigma}{\diff \sigma},
\label{e.ALW}
\end{align}
where $\diff \sigma^\pm$ is the $W$ production cross section for scattering a proton with helicity $\pm 1$ from an unpolarized proton.
The $W$ boson decays into a positron (or electron) and a neutrino (or antineutrino), which is unobserved, $\vec{p}\, p \to \ell X$.
In collinear factorization the cross section difference in the numerator of \eref{e.ALW} can be written in terms of the spin-dependent PDFs $\Delta f_a$ and spin-averaged PDFs $f_b$ in the polarized and unpolarized protons, respectively, and perturbative hard-scattering partonic cross sections $\diff \Delta \hat{\sigma}_{ab}$,
\begin{align}
\diff \Delta \sigma 
= &\sum_{a,b}\int \diff x_a \diff x_b\, 
  \Delta f_a(x_a,\mu_F)\, f_b(x_b,\mu_F)\, 
  \diff \Delta \hat{\sigma}_{ab}(x_a P_A, x_b P_B, p_\ell, \mu_R, \mu_F),
\label{e.deltasigma}
\end{align}
where $x_a$ and $x_b$ are the momentum fractions of the two partons from the colliding protons, $P_A$ and $P_B$ are the 4-momenta of the polarized and unpolarized protons, $p_\ell$ is the 4-momentum of the outgoing lepton, and the sum runs over quarks, antiquarks and the gluon.
In this analysis we choose the factorization and renormalization scales, $\mu_F$ and $\mu_R$, respectively, such that $\mu_R = \mu_F = p_{\ell, T} \equiv p_T$, the transverse momentum of the charged lepton.
In practice, however, the $W$-lepton asymmetry in \eref{e.ALW} has very little sensitivity to the scales~\cite{Ringer:2015oaa}.

The experimental measurements of the asymmetry are typically given in terms of $p_T$ and the lepton rapidity $\eta$, defined as 
$\eta \equiv \eta_\ell 
= \frac12\ln \big[ (p_\ell^0 + p_{\ell, z})/(p_\ell^0 - p_{\ell, z}) \big]$,
where $p_\ell^0$ and $p_{\ell, z}$ are the energy and longitudinal momentum of the lepton.
Introducing the usual Mandelstam variables
\begin{align}
&s \equiv (P_A+P_B)^2\, ,\qquad
 t \equiv (P_A - p_\ell)^2\, ,\qquad
 u \equiv (P_B - p_\ell)^2\, , 
\end{align}
we define the auxiliary variables
\begin{align}
&v \equiv 1 + \frac{t}{s}\, , \qquad
 w \equiv \frac{-u}{s+t}\, ,
\notag
\label{e.STU}
\end{align}
along with their partonic analogs,
\begin{align}
&\hat{s} \equiv (p_a + p_b)^2,\ \qquad
 \hat{t} \equiv (p_a - p_\ell)^2,\ \qquad
 \hat{u} \equiv (p_b - p_\ell)^2, 
\end{align}
and
\begin{align}
&\hat{v} \equiv 1 + \frac{\hat{t}}{\hat{s}}\, , \qquad
 \hat{w} \equiv \frac{-\hat{u}}{\hat{s}+\hat{t}}\, ,
\label{e.stu}
\end{align}
where the partonic momenta are given by
    $p_a = x_a P_a$ and $p_b = x_b P_b$.
In terms of the auxiliary hadronic and partonic variables, one can express the partonic momentum fractions $x_a$ and $x_b$ as
\begin{align}
x_a = \frac{vw}{\hat{v} \hat{w}}\, , \qquad
x_b = \frac{1-v}{1-\hat{v}}\, ,
\end{align}
and write the hadronic variables $v$ and $w$ in terms of the original $p_T$ and $\eta$,
\begin{align}
 v = 1 - \frac{p_T}{\sqrt{s}}e^{-\eta}\, , \qquad
vw = \frac{p_T}{\sqrt{s}}e^{\eta}.
\end{align}
With these relations between the hadronic and leptonic variables, one can now express the cross section in \eref{e.deltasigma} in terms of the lepton kinematics to first order in the strong coupling constant $\alpha_s$~\cite{Ringer:2015oaa},
\begin{align}
\frac{\diff^2 \Delta \sigma}{\diff p_T \diff \eta} 
= \frac{2}{p_T}
&\sum_{a,b}\int_{vw}^v \diff \hat{v} \int_{vw/\hat{v}}^1 \diff \hat{w}~
  x_a \Delta f_a(x_a,\mu_F)~x_b f_b(x_b,\mu_F) 
\notag\\
&\qquad 
\cross 
\bigg[
\frac{\diff \Delta \hat{\sigma}_{ab}^{(0)}(\hat{s},\hat{v})}{\diff \hat{v}}\delta(1-\hat{w})
+
\frac{\alpha_s(\mu_R)}{2\pi}
\frac{\diff \Delta \hat{\sigma}_{ab}^{(1)}(\hat{s},\hat{v},\hat{w}, \mu_R, \mu_F)}{\diff \hat{v} \diff \hat{w}}
\bigg],
\label{e.deltasigmalepton}
\end{align}
where 
$\diff \Delta \hat{\sigma}_{ab}^{(0)}$ and 
$\diff \Delta \hat{\sigma}_{ab}^{(1)}$ are the lowest order and NLO polarized partonic $W$ production cross sections.
Similar expressions can be obtained for the unpolarized cross section in the denominator of \eref{e.ALW}. 
Integrating over a specified $p_T$ range, \eref{e.deltasigmalepton} can be used to relate the experimental values of the asymmetry $A_L^W$ given in terms of $\eta$ to the polarized and unpolarized PDFs.  
The detailed expressions for the partonic cross sections are taken from Ringer and Vogelsang~\cite{Ringer:2015oaa}.

\subsection{Jet production in polarized hadron collisions}
\label{ss.jetproduction}

The longitudinal double-spin asymmetry for jet production in polarized proton-proton collisions, $\vec{p}\, \vec{p} \to \rm{jet} + X$, is defined as
\begin{align}
A_{LL}^{\rm jet} \equiv \frac{\diff \sigma^{++} - \diff \sigma^{+-}}
{\diff \sigma^{++} + \diff \sigma^{+-}}
\equiv \frac{\diff \Delta \sigma}{\diff \sigma}\, ,
\label{e.ALLjet}
\end{align}
where $\diff \sigma^{+ \pm}$ is the cross section for scattering a proton with helicity $+$ from a proton with helicity $\pm$.
In analogy with the $W$-lepton production cross section in Eq.~(\ref{e.deltasigmalepton}), one can write the polarized differential cross section at high $p_T$ in terms of the jet kinematics as~\cite{Jager:2004jh}
\begin{align}
\frac{\diff^2 \Delta \sigma}{\diff p_T \diff \eta} 
= \frac{2 p_T}{s}
&\sum_{a,b}\int_{vw}^v \frac{\diff \hat{v}}{\hat{v}(1-\hat{v})} \int_{vw/\hat{v}}^1 \frac{\diff \hat{w}}{\hat{w}}~
\Delta f_a(x_a,\mu_F)~\Delta f_b(x_b,\mu_F) \notag\\
& \qquad
\cross 
\bigg[
\frac{\diff \Delta \hat{\sigma}_{ab}^{(0)}(\hat{s},\hat{v})}{\diff \hat{v}}\delta(1-\hat{w})
+
\frac{\alpha_s(\mu_R)}{\pi}
\frac{\diff \Delta \hat{\sigma}_{ab}^{(1)}(\hat{s},\hat{v},\hat{w}, \mu_R, \mu_F; R)}{\diff \hat{v} \diff \hat{w}}
\bigg],
\label{e.deltasigmajet}
\end{align}
where $p_T$ and $\eta$ now refer to the transverse momentum and rapidity of the produced jet, and $s$ is the invariant mass squared of the colliding protons.
The hadronic variables $v$ and $w$, and partonic variables $\hat{s}$, $\hat{v}$, and $\hat{w}$ are as defined in Eqs.~(\ref{e.STU}) and (\ref{e.stu}), with the lepton momentum replaced by the jet momentum, and $R$ refers to the radius of the jet.
The renormalization and factorization scales are again set to $\mu_R = \mu_F = p_T$.

An analogous expression can be obtained for the unpolarized cross section in the denominator of the asymmetry $A_{LL}^{\rm jet}$ in Eq.~(\ref{e.ALLjet}).
The expressions for the perturbatively calculated hard-scattering partonic cross sections are taken from J\"{a}ger {\it et al}~\cite{Jager:2004jh}.

\subsection{Lattice QCD data}
\label{ss.lattice}

In addition to the experimental cross sections and single- and double-spin asymmetries included in the global QCD analysis, we also consider the recent high-precision lattice QCD calculations of matrix elements of nonlocal operators that are related to pseudo-Ioffe time distributions (pseudo-ITDs)~\cite{Radyushkin:2017cyf}.
Pseudo-ITDs are Lorentz invariant amplitudes that can be matched to the PDFs in the $\overline{\rm{MS}}$ scheme when the invariant separation between the field operators, $z^2$, is sufficiently small~\cite{Balitsky:2021cwr}.
Specifically, the gluon and quark singlet Ioffe time helicity distributions are defined as 
\begin{subequations}
\begin{eqnarray}
\mathcal{I}_{\Delta g}(\nu,\mu^2) 
&=& \int_0^1 \dd{x} x \sin(x \nu)\, \Delta g(x,\mu^2),
\\
\mathcal{I}_{\Delta \Sigma}(\nu,\mu^2) 
&=& \int_0^1 \dd{x} x \sin(x \nu)\, \Delta \Sigma(x,\mu^2),
\end{eqnarray}
\end{subequations}
where $\nu = p \cdot z$ is the Ioffe time, and $\mu$ is the renormalization scale.
The gluon pseudo-ITD is a particularly useful observable, as it has direct sensitivity to $\Delta g$ at leading order in $\alpha_s$.

Following Karpie {\it et al.}~\cite{Karpie:2023nyg}, we include lattice QCD data generated on 1901 gauge configurations of an ensemble with (2+1)-dynamical clover Wilson fermions with stout-link smearing and tree-level tadpole-improved gauge action with a lattice volume $32^3 \times 64$.
The pion mass for these configurations is $m_\pi = 358(3)$~MeV, with lattice spacing $a = 0.096(1)$~fm determined using the $w_0$ scale~\cite{BMW:2012hcm}. 
Uncertainties introduced by extrapolations to the physical pion mass are expected to be small compared with systematic uncertainties from other sources~\cite{HadStruc:2022yaw}.
In our analysis the lattice QCD data are treated on the same footing as experimental data, as in recent JAM global analyses that have included lattice constraints~\cite{Lin:2017stx, Bringewatt:2020ixn, JeffersonLabAngularMomentumJAM:2022aix, Gamberg:2022kdb, Barry:2023qqh, Karpie:2023nyg, Cocuzza:2023oam, Cocuzza:2023vqs}.

\section{Methodology}
\label{s.methodology}

In this section we outline the methodology employed in the current analysis for parametrizing and fitting the polarized PDFs to the data.
We discuss the choices for the functional forms in terms of the fitting parameters, and the determination and characterization of the PDFs uncertainties.

\subsection{Nonperturbative inputs}
\label{s.np-input}

As discussed above in Sec.~\ref{s.observables}, our analysis requires reconstructing the twist-2 helicity-dependent parton density and subleading power corrections. 
To achieve this we model these nonperturbative functions using a generic template function, $T$, at the input scale, $\mu$, of the form 
\begin{align}
T(x,\mu^2) 
= \frac{N}{\cal M}\, x^{\alpha}(1-x)^{\beta}(1+\gamma \sqrt{x} + \eta x),
\label{e.template}
\end{align}
with fitted normalization $N$ and shape parameters $\alpha$, $\beta$, $\gamma$, and $\eta$ to characterize the $x$ dependence. 
The template function is chosen to be normalized with respect to the first moment of the function,
\begin{align}
    {\cal M} = {\rm B}[\alpha+1,\beta+1]
              + \gamma {\rm B}[\alpha+\frac32,\beta+1]
              + \eta {\rm B}[\alpha+2,\beta+1],
\end{align}
where $B$ is the Euler beta function, in order to maximally decorrelate the normalization parameter, $N$, from the rest of the shape parameters. 
Using this template function, we fit the quantities
    $\Delta u_v \equiv \Delta u - \Delta \bar{u}$, 
    $\Delta d_v \equiv \Delta d - \Delta \bar{d}$,
    $\Delta S$, 
    $\delta \bar{u}$, 
and 
    $\delta \bar{d}$. 
With these building blocks, the helicity PDFs are then constructed as 
\begin{subequations}
\label{e.param}
\begin{align}
\Delta u       &= \Delta u_v + \Delta \bar{u},
\\
\Delta d       &= \Delta d_v + \Delta \bar{d},
\\         
\Delta \bar{u} &= \Delta S + \delta \bar{u},
\\
\Delta \bar{d} &= \Delta S + \delta \bar{d},
\\
\Delta s       &= \Delta \bar{s} = \Delta S \;.
\label{e.param.Ds}
\end{align}
\end{subequations}
The parametrization follows that used in the recent JAM unpolarized PDF analysis in Ref.~\cite{Anderson:2024evk}.
In this model {\it ansatz}, a symmetric sea component $\Delta S$ is shared among the sea quarks, with flavor-dependent distortions $\delta \bar{u}$ and $\delta \bar{d}$ away from the symmetric limit for the $\bar u$ and $\bar d$ antiquark PDFs, respectively.
At present, the existing data do not provide sensitivity to discriminate between $\Delta s$ and $\Delta \bar{s}$, or suggest the need for introducing any additional flavor distortion.
We also include the quark triplet axial charge $g_A = 1.269(3)$ (see Eq.~(\ref{e.axialcharges.gA})) as a constraint on the fit, treating it as an experimental data point.

The input scale for the helicity PDFs is set equal to the charm quark mass, $\mu = m_c = 1.28$~GeV~\cite{Workman:2022ynf}, and the scale dependence is generated by solving the DGLAP evolution equations in Mellin space.
Following Ref. \cite{Sato:2016wqj}, we use the zero-mass variable flavor scheme at next-to-leading logarithmic accuracy, and parametrize the heavy quark distributions discontinuously at their mass thresholds.
For the power suppressed corrections, we use the same template for each of the $c_i^{\rm HT}$ functions, with all five parameters active and fitted independently for protons and neutrons for both $i=1$ and 2.

\subsection{Uncertainty quantification}
\label{s.uq}

To reconstruct the parton densities and power suppressed effects, we employ the strategy developed in previous JAM analyses. 
A discussion of the most up-to-date developments in the methodology can be found in Ref.~\cite{Anderson:2024evk}.
Specifically, we employ a data resampling technique in which multiple realizations of the data are generated by adding Gaussian noise to the quoted central values and uncertainties. 
For each realization, we perform parameter optimization by minimizing the $\chi^2$ function. 
The resulting collection of optimized parameters forms the posterior distribution for the shape parameters of the nonperturbative quantities, from which statistical estimators, such as Bayesian credible intervals for physical observables, can be computed. 
For all observables computed from the ensemble of parameters, we use the mean values as the central results, and the edges of the confidence intervals are determined using a nonparametric estimate of the inverse cumulative distribution function~\cite{Hyndman}.

To assess the quality of the fit results, we compute the reduced $\chi^2$ and $Z$-score values, defined by
\begin{equation}
\chired \equiv \frac{1}{N_{\rm dat}} 
\sum_{e,i}
\bigg( 
    \frac{d_{e,i}-T_{e,i}}
    {\alpha_{e,i}}
\bigg)^2, 
\qquad {\rm and} \qquad
Z = \sqrt{2}\, {\rm erf}^{-1} (2p-1),
\end{equation}
respectively, where the $p$-value is computed according to the $\chi$ distribution with $N_{\rm dat}$ as the number of degrees of freedom, $d_{e,i}$ is the data point $i$ from experimental dataset $e$, $\alpha_{e,i}$ is the corresponding uncorrelated error, and $T_{e,i}$ is the corresponding mean theory calculated from the replicas (with adjustments from correlated uncertainties taken into account).
As with the strategy in previous JAM studies~\cite{Zhou:2022wzm, Hunt-Smith:2024khs}, in this analysis we do not impose PDF positivity constraints, which have been debated recently in the literature~\cite{Collins:2021vke, deFlorian:2024utd}, and which are only valid at lowest order in $\alpha_s$.
Physical quantities such as cross sections and asymmetries remains bounded within the expected limits only in the kinematic region where the data are available.
In the extrapolation region we do not impose bounds in order to avoid biases on the extracted nonperturbative quantities.

\section{Global QCD analysis}
\label{s.globalQCDanalysis}

In this section we present the results of our global QCD analysis.
We begin by reviewing in Sec.~\ref{s.data} the datasets that are fitted, and the stability of fit, with the resulting $\chi^2$ values and $Z$-scores for the various high-energy scattering observables.  
The extracted PDFs are discussed in Sec.~\ref{s.ppdfs}, and results for structure functions including higher twist corrections are given in Sec.~\ref{s.HTnuc}.

\subsection{Quality and stability of fit}
\label{s.data}

In the current analysis we include measurements of the longitudinal DIS double-spin asymmetries $A_{\parallel}$ and $A_1$ from fixed target experiments on proton, deuterium, and $\hel$ nuclei from 
EMC~\cite{EuropeanMuon:1989yki}, 
SMC~\cite{SpinMuon:1998eqa, SpinMuon:1999udj}, 
COMPASS~\cite{COMPASS:2006mhr, COMPASS:2010wkz, COMPASS:2015rvb}, 
SLAC~\cite{Baum:1983ha, E142:1996thl, E154:1997xfa, E143:1998hbs, E155:1999pwm, E155:2000qdr}, 
HERMES~\cite{HERMES:1997hjr, HERMES:2006jyl},
and Jefferson Lab  \cite{JeffersonLabHallA:2004tea,  JeffersonLabHallA:2016neg, JeffersonLabHallA:2014mam, CLAS:2014qtg, CLAS:2017qga, CLAS:2015otq}.
We also include measurements of the transverse DIS asymmetries \Ape\ 
and $A_2$ from
SLAC~\cite{E142:1996thl, E143:1998hbs, E154:1997xfa, E155:1999eug, E155:2002iec}, 
HERMES~\cite{HERMES:2011xgd},
and Jefferson Lab~\cite{JeffersonLabHallA:2004tea, JeffersonLabHallA:2016neg}.
For all DIS asymmetries we apply cuts on the four-momentum transfer squared of $Q^2 > m_c^2$ and $W^2 > 4$~GeV$^2$.

For polarized SIDIS we include measurements of the longitudinal semi-inclusive $A_1^h$ asymmetry on proton and deuterium targets from HERMES~\cite{HERMES:2004zsh} and COMPASS~\cite{COMPASS:2009kiy, COMPASS:2010hwr} for pions, kaons, and unidentified charged hadrons.
Following Anderson {\it et al.}~\cite{Anderson:2024evk}, the cuts $Q^2 > m_c^2$ and $W^2_{\mbox{\tiny \rm SIDIS}} > 20$~GeV$^2$ are made to ensure the applicability of the leading power formalism, and enhance the separation of the current and target fragmentation regions.
In addition, a cut on the fragmentation variable $0.2 < z < 0.8$ is applied to avoid large-$z$ threshold corrections to exclude data from the target fragmentation region~\cite{Moffat:2021dji}.
Beyond lepton-hadron reactions, hadron-hadron scattering data include jet asymmetries $A_{LL}^{\rm jet}$ in polarized $pp$ collisions from the STAR~\cite{STAR:2006opb, STAR:2007rjc, STAR:2012hth, STAR:2014wox, STAR:2019yqm, STAR:2021mfd} and PHENIX~\cite{PHENIX:2010aru} collaborations at RHIC, and single spin asymmetries $A_L^W$ from STAR~\cite{STAR:2018fty} and $A_L^{W/Z}$ from PHENIX~\cite{PHENIX:2015ade, PHENIX:2018wuz}.

\begin{figure}[t]
\centering
\includegraphics[width=0.725\textwidth]{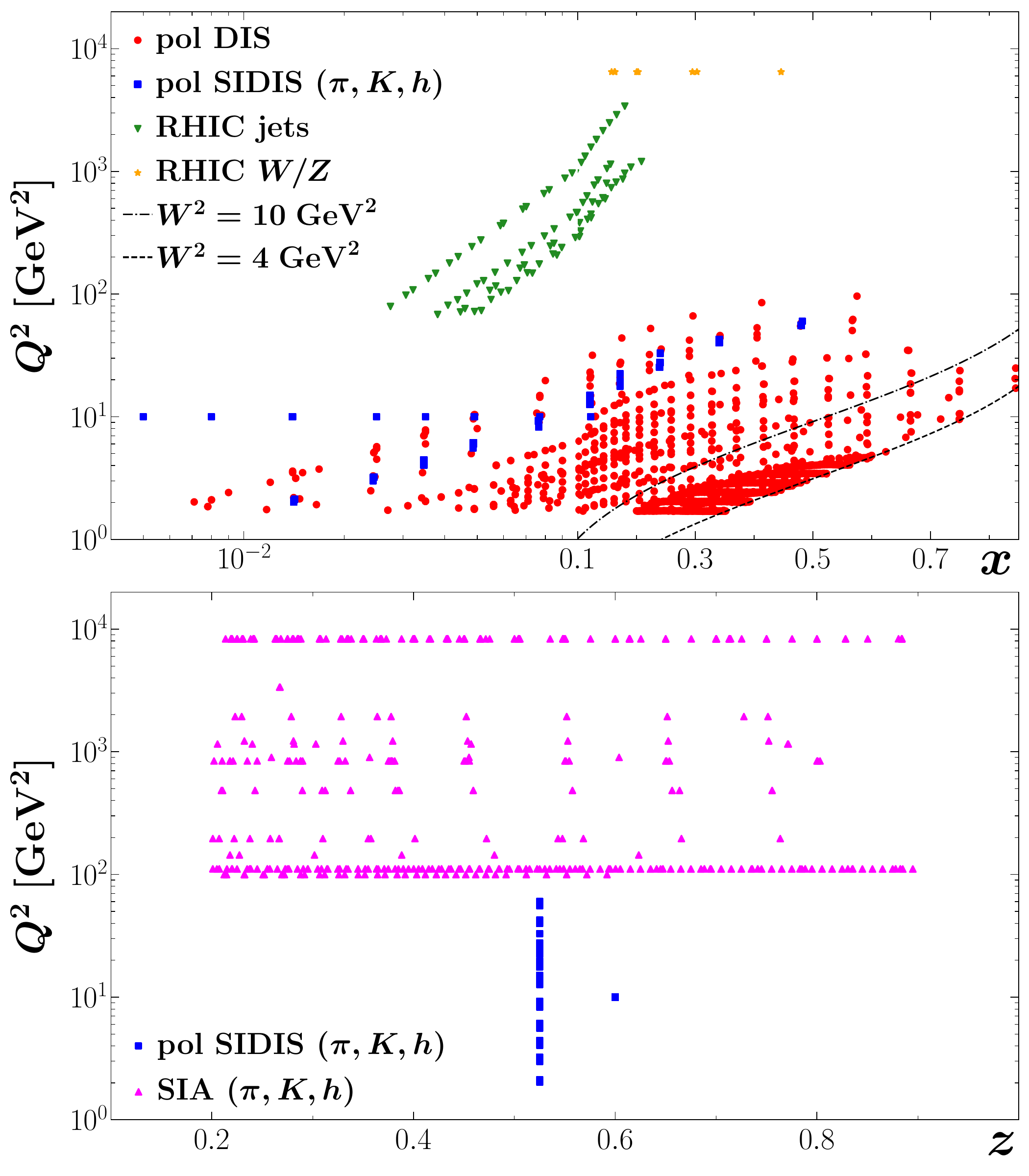}
\caption{Kinematics of polarized datasets included in this analysis with scale $Q^2$ displayed versus the relevant scaling variable.
{(Top panel)} For polarized DIS and SIDIS the four-momentum transfer squared $Q^2$ is shown versus the Bjorken scaling variable; for jet production the transverse momentum squared is shown versus the Feynman variable; and for vector boson production the mass squared of the boson is versus $x_F$. For DIS, cuts on $W^2 > 4$~GeV$^2$ (dashed line) and $W^2 > 10$~GeV$^2$ (dot-dashed line) are also shown for reference.
{(Bottom panel)} The scale $Q^2$ is shown versus the fragmentation variable $z = P \cdot p_h\!~/~\!P \cdot q$ for polarized SIDIS, and versus $z = 2 \, p_h \cdot q/Q^2$ for inclusive annihilation of $e^+ e^-$ pairs into $\pi$, $K$ and unidentified hadrons $h$.}
\label{fig:kinematics}
\end{figure}

Datasets that depend only on unpolarized PDFs and FFs are for the most part the same as those in the earlier JAM analysis of unpolarized PDFs, helicity dependent PDFs, and single-hadron FFs in Ref.~\cite{Cocuzza:2022jye}, but with some notable differences.  
The $W^2$ cut for unpolarized DIS was increased from 3.0~GeV$^2$ to 3.5~GeV$^2$, while a cut of $W^2_{\rm SIDIS} > 20$~GeV$^2$ was applied to unpolarized SIDIS data to match that found in the recent analysis by \mbox{Anderson {\it et al.}~\cite{Anderson:2024evk}}. 
Inclusive $W$+charm production data from $pp$ collisions were also included, as in Ref.~\cite{Anderson:2024evk}.
The kinematics for all the polarized datasets included in this analysis are displayed in~Fig.~\ref{fig:kinematics}, with the scale $Q^2$ (four-momentum transfer squared for DIS and SIDIS; transverse momentum squared for jet production; vector boson mass squared for $W/Z$ production; $e^+ e^-$ invariant mass squared for SIA) plotted versus the relevant scaling variable for each type of dataset (Bjorken $x$ or Feynman $x$, or the fragmentation variable $z$).

Also illustrated are the $W^2 = 4$~GeV$^2$ and 10~GeV$^2$ contours indicating the regions of kinematics that would be excluded for these cuts on the DIS data.
The analysis with a $W^2 > 4$~GeV$^2$ cut on DIS data, for example, would include $\approx 1000$ additional data points than one with a stronger, $W^2 > 10$~GeV$^2$, cut.
To determine the optimal cut, we performed several independent analyses of the global data (including lattice QCD data) using different DIS data cuts, ranging from the most stringent $W^2 > 10$~GeV$^2$ cut down to the most relaxed cut of $W^2 > 3.5$~GeV$^2$.

\begin{table}[b]
\caption{Quality of the global fit (including lattice QCD data) versus $W_{\rm min}^2$ for the inclusive polarized DIS data. The $\chi^2_{\rm red}$ and $Z$-score values and number of data points $N_{\rm dat}$ for each cut refer only to the inclusive polarized DIS data. Results for the preferred value $W_{\rm min}^2 = 4$~GeV$^2$ are highlighted in boldface.}
\begin{tabular}{ c | c | c | c | c | c }
\hhline{======}
~$W_{\rm min}^2$ (GeV$^2$)~ &
3.5 &
{\bf 4} &
5 &
6 &
10
\\
\hline
$N_{\rm dat}^{\mbox{\tiny \rm }}$ &
~2002~ &
~{\bf 1735}~ &
~1287~ &
~1008~ &
~689~
\\
$\chi^2_{\rm red}$ &
1.11 &
{\bf 1.02} &
0.97 &
1.00 &
1.01
\\
$Z$-score &
~$+3.23$~~~ &
~{\bf +0.61}~~~ &
~$-0.83$~~~ &
~$+0.03$~~~ &
~$+0.22$~~~ \\
\hhline{======}
\end{tabular}
\label{tab:W2}
\end{table}

The resulting reduced global $\chi^2$ values are listed in Table~\ref{tab:W2} for various cut values.
Very good fits are obtained for cuts over the range $W_{\rm min}^2=10$~GeV$^2$ down to 4~GeV$^2$, with $\chi^2_{\rm red} \approx 1.0$, and $Z$-scores less than 1.
The worsening of the fit quality as the cut is lowered from $W_{\rm min}^2=4$~GeV$^2$ to 3.5~GeV$^2$ is driven primarily by the large number of points with small uncertainties for the Jefferson Lab Hall~B proton data~\cite{CLAS:2017qga}, many of which are contained in this kinematic region (see also Fig.~\ref{fig:pidis_Apa_proton} and Table~\ref{tab:chi2_Apa} in Appendix~\ref{a.data} below).
While the most stringent cut allows a $\chi^2_{\rm red}$ value for the DIS data of $\approx 1$, this contains only about 1/3 of the data points accessible with a $W^2 > 3.5$~GeV$^2$ cut, which, however, gives a reduced $\chi^2 \approx 1.1$ and a $Z$-score $> 3$.
As a compromise between these two extreme values for the cuts, for our main analysis we opt for a cut of $W^2 > 4$~GeV$^2$ for all polarized DIS data.
Note that unless otherwise indicated, our full fits include both LT and power-suppressed target mass and HT contributions to the structure functions, as discussed in Sec.~\ref{ss.pidis}.

\begin{figure}[t]
\centering
\includegraphics[width=0.70\textwidth]{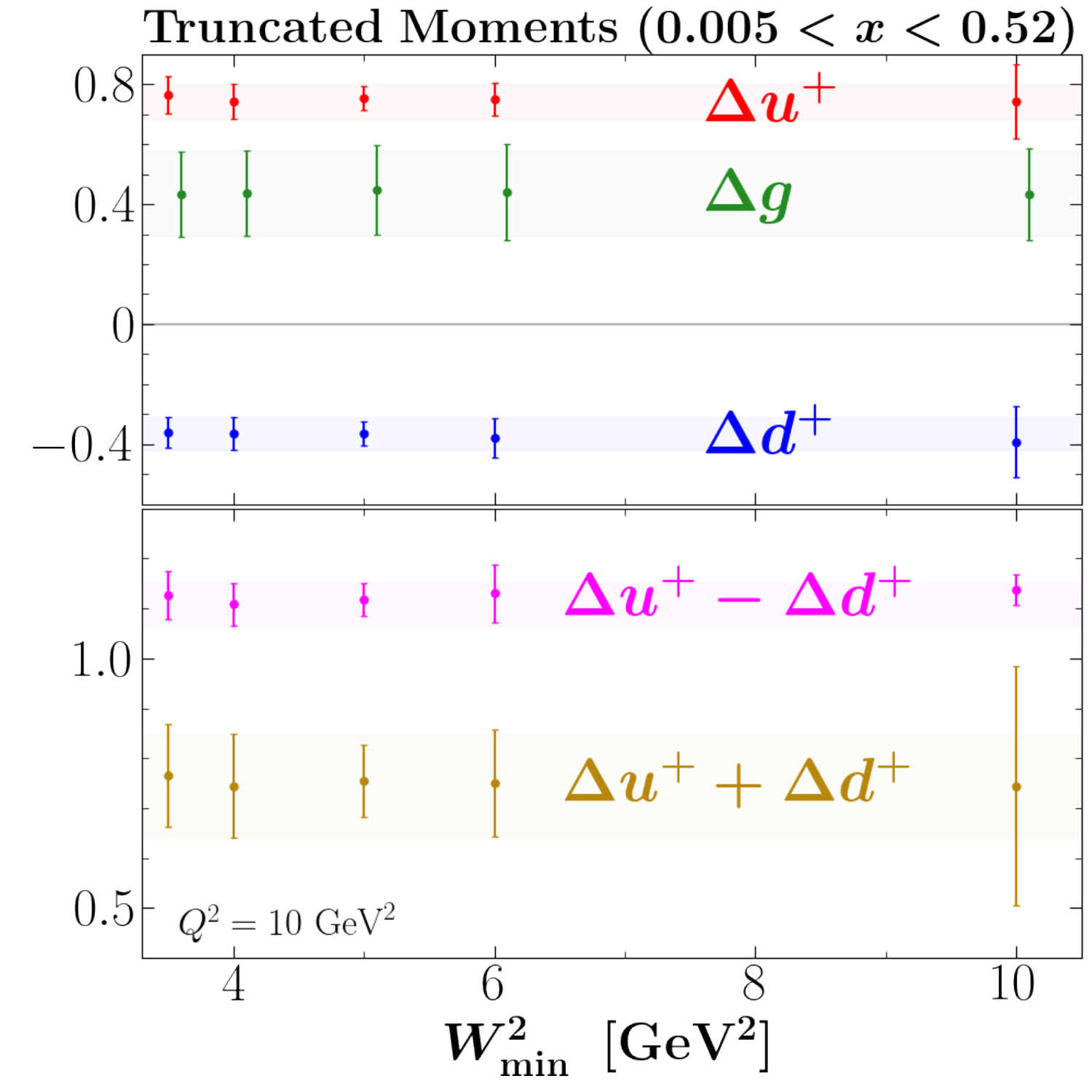}
\vspace*{-0.4cm}
\caption{Stability of various truncated moments, evaluated from $x = 0.005$ to 0.52, with respect to the cut used for the polarized inclusive DIS data, for $W_{\rm min}^2 = 3.5$, 4, 5, 6 and 10~GeV$^2$, at $Q^2 = 10$~GeV$^2$: (top row) lowest moments of PDFs $\Delta u^+$, $\Delta d^+$, and $\Delta g$; (bottom row) isovector and isoscalar combinations of $\Delta u^+$ and $\Delta d^+$. The minimum value of $x$ is chosen to match the lower limit of the experimental data, while the maximum value is set by $x_{\rm max} = Q^2/(W^2 + Q^2 - M^2) = 0.52$ for $W^2 = 10$~GeV$^2$. Note that some of the points have been offset for clarity.}
\label{fig:moments}
\end{figure}

The stability of the fit is further illustrated in Fig.~\ref{fig:moments}, where we show truncated moments of the PDFs, $\Delta f \equiv \int_{x_{\rm min}}^{x_{\rm max}} \dd{x} \Delta f(x,Q^2)$, at $Q^2=10$~GeV$^2$ versus the $W^2$ cut for $W^2_{\rm min} = 3.5$, 4, 5, 6 and 10~GeV$^2$, for each of the flavors $\Delta f = \Delta u^+, \Delta d^+, \Delta g$, as well as the isovector and isoscalar combinations.
The minimum $x$ value down to which the PDFs are kinematically constrained by data is $x_{\rm min} = 0.005$, while the maximum $x$ value is set to $x_{\rm max} = Q^2/(W_{\rm min}^2 + Q^2 - M^2) = 0.52$ for $W_{\rm min}^2 = 10$~GeV$^2$ in order to avoid uncontrolled extrapolations of the PDFs.
This ensures that all the truncated moments, across all values of $W^2_{\rm min}$, are evaluated consistently within the region of $x$ constrained by data.
All of the helicity PDFs in Fig.~\ref{fig:moments} are observed to be stable with respect to variation in the $W_{\rm min}^2$ cut.

\begin{table}[b]
\caption{Summary of $\chi^2_{\rm red}$ and $Z$-score values for reactions included in the analysis for the 3 scenarios: $W^2 > 10$~GeV$^2$ (LT only), $W^2 > 10$~GeV$^2$ (LT only), and the full JAMpol25 fit with $W^2 > 10$~GeV$^2$ (with TMC and HT). The dagger symbol~($\dagger$) in the number of data points $N_{\rm dat}$ column indicates that these datasets have fewer points for the $W^2 > 10$~GeV$^2$ cut (namely, $N_{\rm dat} = 689$ for polarized DIS data, and $N_{\rm dat} = 914$ for all polarized data). \\}
\begin{tabular}{ l | r | c | c | c }
\hhline{=====}
\multicolumn{5}{c}
{~~~~~~~~$\chi^2_{\rm red}$ ($Z$-score)}
\\
\hline
~Data
& ~~$N_{\rm dat}$~~ 
& ~$W^2 > 10$ GeV$^2$~~ 
& ~~$W^2 > 4$ GeV$^2$~~
& ~~$W^2 > 4$ GeV$^2$~~
\\
~
&  
& (LT only) 
& (LT only)
& (with TMC \& HT)
\\
\hline
~Polarized  
& ~1960\textsuperscript{\textdagger}~ 
& 0.96 ($-0.85$) 
& 1.13 ($+3.92$) 
& 0.99 ($-0.30$)
\\
~ --- DIS   
& ~1735\textsuperscript{\textdagger}~ 
& 1.01 ($+0.22$) 
& 1.03 ($+0.99$) 
& 1.02 ($+0.61$)
\\
~ --- SIDIS 
& ~124~~ 
& 0.80 ($-1.68$) 
& 0.82 ($-1.47$) 
& 0.76 ($-2.00$)
\\
~ --- jets  
& ~83~~ 
& 0.81 ($-1.27$) 
& 0.79 ($-1.38$) 
& 0.81 ($-1.28$)
\\
~ --- $W/Z$ boson~~~~ 
& ~18~~ 
& 0.82 ($-0.48$) 
& 0.92 ($-0.14$) 
& 0.74 ($-0.74$)
\\
~Lattice QCD        
& ~48~~ 
& 0.59 ($-2.31$)
& 0.59 ($-2.34$) 
& 0.58 ($-2.40$)
\\
\hline
~\bf Total          
& ~\bf{2008}\textsuperscript{\textdagger}~ 
& \bf{ 0.96 ($-0.86$)}~
& \bf{ 1.13 ($+4.00$)}~ 
& \bf{ 0.99 ($-0.22$)}~
\\
\hhline{=====}
\end{tabular}
\label{tab:chi2_total}
\end{table}

For the preferred value of the $W^2$ cut, $W^2_{\rm min} = 4$~GeV$^2$, which we use in the JAMpol25 analysis, a summary of reduced $\chi^2$ and $Z$-score values for all of the types of reactions and observables fitted in this analysis is given in Table~\ref{tab:chi2_total}, including polarized DIS, SIDIS, jet, and vector boson production data, along with the lattice QCD data on the gluon pseudo ITDs.
In particular, for the longitudinal DIS polarization asymmetries from various global experiments, which form about 60\% of all the polarized data, we find $\chi^2_{\rm red} = 1.01$, while for the transverse DIS polarization asymmetries, which constitute $\approx 1/4$ of the polarized DIS data, we have $\chi^2_{\rm red} = 1.08$.
For the longitudinally polarized SIDIS hadron production asymmetries, the reduced $\chi^2$ value is $\chi^2_{\rm red} = 0.76$, and for the jet and $W/Z$ boson production $\chi^2_{\rm red} = 0.81$ and $\chi^2_{\rm red} = 0.74$, respectively.
Overall, the total $\chi^2_{\rm red} = 0.99$ for 1960 polarized experimental data points.
Tables of reduced $\chi^2$ values for each individual polarized dataset are displayed in Tables~\ref{tab:chi2_Apa}--\ref{tab:chi2_pp} in Appendix~\ref{a.data}.
Comparisons of all the polarized data versus theory are shown in Figs.~\ref{fig:pidis_A1_proton}--\ref{fig:pidis_A2_Ape_helium} in Appendix~\ref{a.data}.
The 48 lattice data points are well fitted with $\chi^2_{\rm red} = 0.58$.
The total reduced $\chi$ ($Z$-score) for this fit is 0.99 ($-0.22$).

Along with the main analysis with $W^2_{\rm min} = 4$~GeV$^2$, which includes target mass and higher twist corrections, we also performed fits with the same cuts but using the LT approximation only.
In this case the $\chi^2_{\rm red}$ increases significantly for the polarized DIS data, but remains similar elsewhere.
This increase comes primarily from the Jefferson Lab eg1b proton and eg1-dvcs deuteron data on \Apa\ (see Table~\ref{tab:chi2_Apa} in Appendix~\ref{a.data} below), which are not as well fitted without including TMCs.
We also compare the results with an analysis in which a more stringent cut of $W^2_{\rm min} = 10$~GeV$^2$ is used.
This naturally leads to a reduction in the number of data points fitted, from $\approx 1700$ for polarized DIS in the full fit to $\approx 700$ in the $W^2_{\rm min} = 10$~GeV$^2$ fit.
The $\chi^2_{\rm red}$ values and $Z$-scores are all acceptable, although the restriction to the higher $W^2$ reduces the constraints in the large-$x$ region, as will be discussed in the following section.

\subsection{Helicity-dependent parton distributions}
\label{s.ppdfs}

\begin{figure}[t]
\centering
\includegraphics[width=0.95\textwidth]{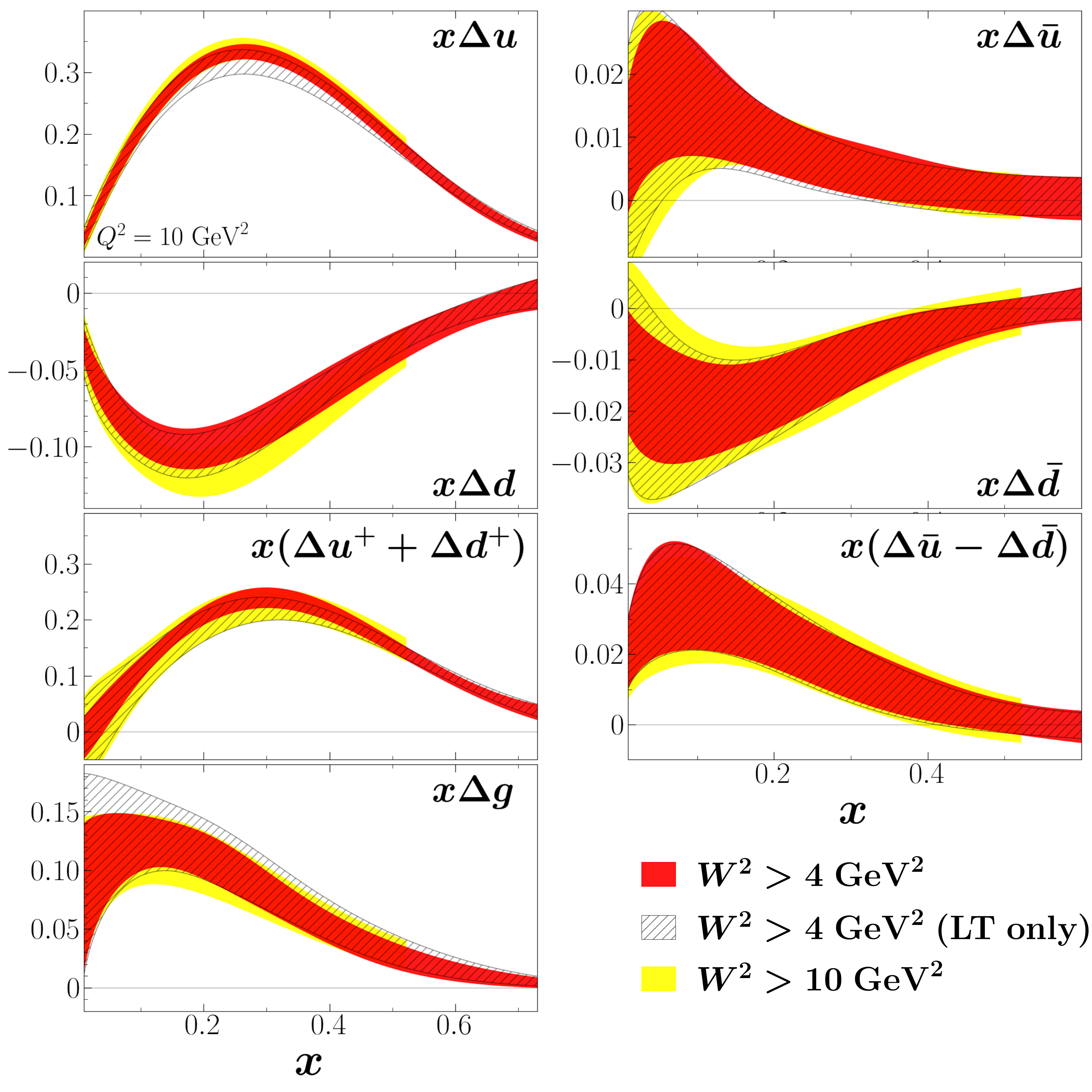}
\caption{Polarized PDFs $x \Delta f$ versus $x$ for various parton flavors $\Delta f = \Delta u$, $\Delta d$, $\Delta \bar u$, $\Delta \bar d$, along with the isosinglet $\Delta u^+ + \Delta d^+$ and nonsinglet $\Delta\bar u - \Delta\bar d$ combinations, and $\Delta g$, at $Q^2 = 10$~GeV$^2$. The JAMpol25 results with the $W^2 > 4$~GeV$^2$ cut on the inclusive polarized DIS data (red bands) are compared with those with the more restrictive $W^2 > 10$~GeV$^2$ cut (yellow bands), and those at LT (black hatched bands). Kinematically the $W^2 > 10$~GeV$^2$ cut constrains the PDFs up to $x = 0.52$ (rightmost edge of the yellow bands), while the $W^2 > 4$~GeV$^2$ cut allows PDFs to be constrained up to $x = 0.845$.} 
\label{fig:ppdfs}
\end{figure}

In this section we present the main results for the helicity dependent PDFs inferred from our global QCD analysis.
The $x$ dependence of the PDFs $x \Delta f$ is displayed in Fig.~\ref{fig:ppdfs} at a scale $Q^2 = 10$~GeV$^2$ for each of the quark flavors $\Delta f = \Delta u$, $\Delta d$, $\Delta \bar u$, $\Delta \bar d$ and $\Delta g$, along with the isosinglet $\Delta u^+ + \Delta d^+$ and nonsinglet $\Delta\bar u - \Delta\bar d$ combinations.
The results of our full JAMpol25 fit, with a $W^2 > 4$~GeV$^2$ cut, are compared with those using the more restrictive $W^2 > 10$~GeV$^2$ cut.
One of the consequences of the increased $W$ range with the less restrictive cut is that the PDFs are constrained to larger values of $x$ --- namely, up to $x_{\rm max} = 0.76$ at $Q^2 = 10$~GeV$^2$, in contrast to $x_{\rm max} = 0.52$ for the $W^2 > 10$~GeV$^2$ cut.
Note that this does not imply that PDFs with the weaker cut should differ from the ones with the stronger cut only in the region $0.52 \lesssim x \lesssim 0.76$; as evident in Fig.~\ref{fig:kinematics}, much of the inclusive DIS data included in the new JAMpol25 analysis lie in the range down to $x \approx 0.2$ at lower $Q^2$ values, so that a reduction in the PDF uncertainties at lower $x$ may also be expected for the new analysis relative to the $W^2 > 10$~GeV$^2$ fit.

Generally, the central values of the helicity PDFs are similar for the two sets of cuts.
The PDF uncertainties, however, are clearly reduced for the JAMpol25 fit, especially for the valence quark distributions. 
For $\Delta u$ and $\Delta d$, this reduction can be as large as 60\% and 40\%, respectively, around $x = 0.4$. 
Consequently, the errors on the isoscalar combination $\Delta u^+ + \Delta d^+$ are reduced by around 60\% as well. 
The uncertainties on $\Delta g$ are also reduced by around 40\%.
Clearly then, the inclusion of the low-$W^2$ polarized DIS data provides a significant improvement in our knowledge of the quark helicity PDFs at both low and high values of $x$.
The inclusion of the TMC and HT corrections also has little impact on the extracted helicity PDFs, with the uncertainty bands similar for both the full JAMpol25 and the LT only results, though slightly larger for some of the PDFs at small $x$ for the latter.

As in the earlier JAM analysis~\cite{Cocuzza:2022jye}, we obtain clearly nonzero $\bar u$ and $\bar d$ antiquark polarizations, with the former being positive and the latter negative, such that the asymmetry $\Delta \bar u - \Delta \bar d$ is sizable and positive over the range $0.1 \lesssim x \lesssim 0.4$.
The uncertainties on the $\Delta \bar u$ and $\Delta \bar d$ PDFs are also slightly reduced for the JAMpol25 analysis, especially for the $\Delta \bar d$ distribution, although we do not expect the lower-$W$ data to have much constraint on the polarized sea quark distributions at high $x$.

In the strange quark sector, a nonzero strange quark polarization may be expected in nature, along with differences between $\Delta s$ and $\Delta \bar s$~\cite{Signal:1987gz, Ji:1995rd, Melnitchouk:1996fj, Melnitchouk:1999mv, Wang:2020hkn, He:2022fne}, and in principle SIDIS kaon production data could be sensitive to a nonzero $\Delta s$ or $\Delta \bar s$. 
In practice, however, the existing datasets described in Sec.~\ref{s.data} do not provide sufficient constraints, without introducing additional theoretical assumptions, such as SU(3) flavor symmetry. 
Employing the {\it ansatz} in Eq.~(\ref{e.param.Ds}), the strange quark polarization is set equal to the flavor symmetric sea quark component, $\Delta S$.
Since the $\bar u$ and $\bar d$ polarizations have opposite signs and similar magnitude, this effectively means that the central values of the strange quark polarization will be close to zero, although with a very large uncertainty that renders $\Delta s$ essentially unconstrained.
Future, higher precision SIDIS kaon production data, at Jefferson Lab and at the future Electron-Ion Collider, may help to resolve a nonzero strange polarization.

In a recent JAM analysis~\cite{Hunt-Smith:2024khs}, it was found that the combination of polarized jet production data from RHIC, low-$W^2$ polarized DIS data from Jefferson Lab, and the lattice QCD measurements of pseudo-ITDs effectively ruled out the negative $\Delta g$ solution, without the need to impose PDF positivity constraints.
In the current analysis we therefore remove the negative solutions from all fits, regardless of the $W^2$ cut. 
The result is that $\Delta g$ is positive throughout the range of $x$ shown, and  the inclusion of the low-$W^2$ data leads to a significant reduction in its uncertainties.

Integrating the PDFs in Fig.~\ref{fig:ppdfs} over $x$, we can obtain moments of the distributions, which can in principle be compared with matrix elements of specific operators that can be computed from lattice QCD.
However, since our PDFs are determined over a finite range of $x$, the most direct and experimentally constrained quantities that we can compute are truncated moments, defined as integrals over the restricted range
        $x_{\rm min} \leq x \leq x_{\rm max}$.
As noted above, the lower $x$ limit down to which our PDFs are kinematically constrained by data is $x_{\rm min} = 0.005$, while the upper limit is determined by the value of $W^2_{\rm min}$ used for the inclusive DIS data, which for $Q^2=10$~GeV$^2$ corresponds to $x_{\rm max}=0.52$ and 0.76 for the $W^2 > 10$~GeV$^2$ and 4~GeV$^2$ cuts, respectively.
We do not compute full moments, as these are subject to unknown uncertainties when extrapolating into unmeasured regions at $x < x_{\rm min}$ and $x > x_{\rm max}$, especially for flavor singlet quantities such as $\Delta g$.

\begin{table}[b]
\caption{Truncated moments of PDFs at $Q^2=10$~GeV$^2$.  The different columns correspond to different values of $x_{\rm min}$, $x_{\rm max}$, and $W^2_{\rm min}$ used in the fit, with the right-most column corresponding to the full JAMpol25 fit.}
\begin{tabular}{ c | l | l | l } 
\hhline{====}
& \multicolumn{3}{c}{$[x_{\rm min},\, x_{\rm max};\, W^2_{\rm min} ({\rm GeV}^2)]$}
\\ 
Truncated Moment
& ~~[0.005, 0.52; 10]~~ 
& ~~[0.005, 0.52; 4]~~ 
& ~~[{\bf 0.005}, {\bf 0.76}; {\bf 4}]~~ 
\\
\hline
$\Delta u^+$ &
~~~~~\phantom{+}0.75(12) & 
~~~~~\phantom{+}0.75(6) & 
~~~~~\phantom{+}${\bf 0.78(6)}$ 
\\
$\Delta d^+$ & 
~~~~~$-0.39(12)$ &
~~~~~$-0.36(5)$ & 
~~~~~${\bf -0.37(5)}$ 
\\ 
$\Delta g$ &
~~~~~\phantom{+}0.44(15) &
~~~~~\phantom{+}0.44(14) &
~~~~~\phantom{+}${\bf 0.44(14)}$ 
\\ 
\hline
$\Delta u^+ + \Delta d^+$ & 
~~~~~\phantom{+}0.41(37) &
~~~~~\phantom{+}0.42(17) &
~~~~~\phantom{+}${\bf 0.42(17)}$ 
\\
$\Delta u^+ - \Delta d^+$ &
~~~~~\phantom{+}1.14(3) &
~~~~~\phantom{+}1.11(4) &
~~~~~\phantom{+}${\bf 1.15(4)}$ 
\\
\hhline{====}
\end{tabular}
\label{tab:moments}
\end{table}

The lowest truncated moments of the PDFs $\Delta u^+$, $\Delta d^+$, and $\Delta g$, along with the isoscalar $\Delta u^+ + \Delta d^+$ and isovector $\Delta u^+ - \Delta d^+$ combinations, are listed in Table~\ref{tab:moments} for different values of $x_{\rm min}$ and $x_{\rm max}$ at $Q^2=10$~GeV$^2$.
Comparing the results of the $W^2_{\rm min} = 10$ GeV$^2$ fit, with $x_{\rm min} = 0.005$ and $x_{\rm max} = 0.52$ matching the kinematics of the data, with the JAMpol25 fit with $W^2_{\rm min} = 4$~GeV$^2$, we see that the central values are similar for all the helicity PDFs, but with smaller uncertainties for the lower-$W^2$ cut, as may be expected. 
Considering the entire constrained interval $[0.005, 0.76]$, the lowest truncated moments remain very similar for the helicity PDFs, indicating that the contributions from the region $x \in [0.52, 0.76]$ are relatively small.
On the other hand, the higher twist contributions to the $g_1$ and $g_2$ structure functions themselves may not necessarily be identical, as we discuss in the next section.

\begin{figure}[t]
\centering
\includegraphics[width=1.0\textwidth]{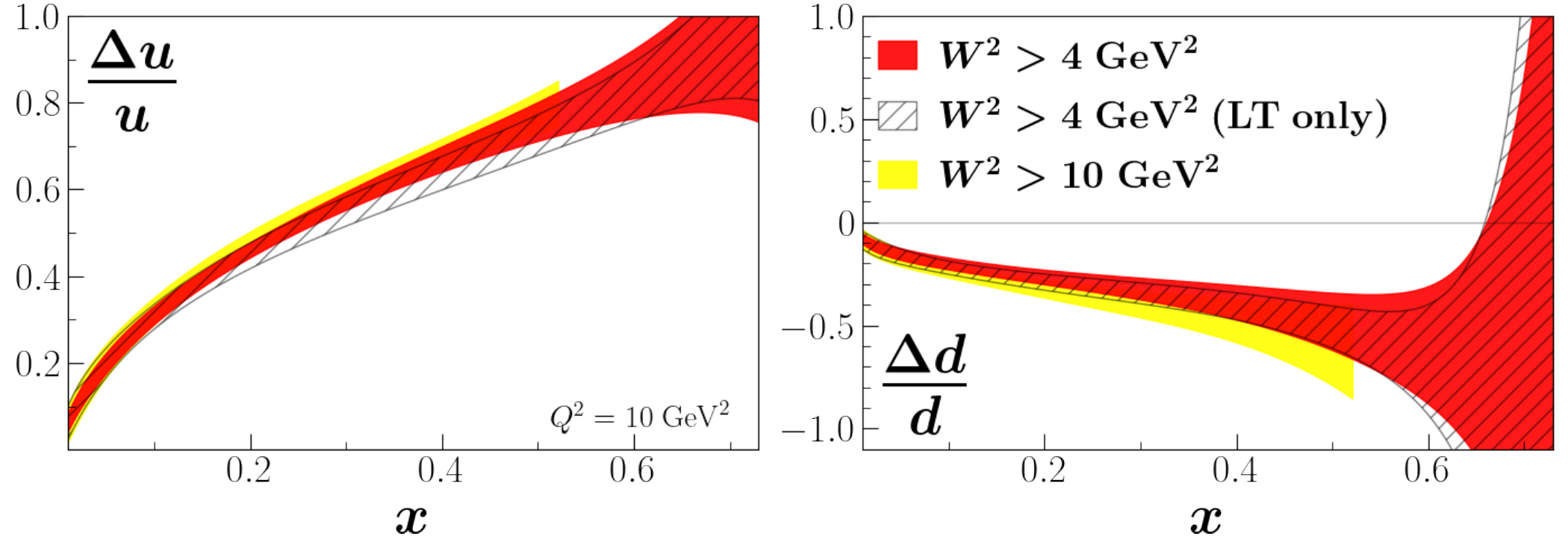}
\caption{Polarized to unpolarized PDF ratios $\Delta u/u$ and $\Delta d/d$ versus $x$ at $Q^2 = 10$~GeV$^2$, comparing the results with cuts $W^2 > 4$~GeV$^2$ (red bands) and $W^2 > 10$~GeV$^2$ (yellow bands) on the inclusive polarized DIS data, and the fit with $W^2 > 4$~GeV$^2$ at LT (black hatched bands).}
\label{fig:polarization}
\end{figure}

\begin{figure}[t]
\centering
\includegraphics[width=0.98\textwidth]{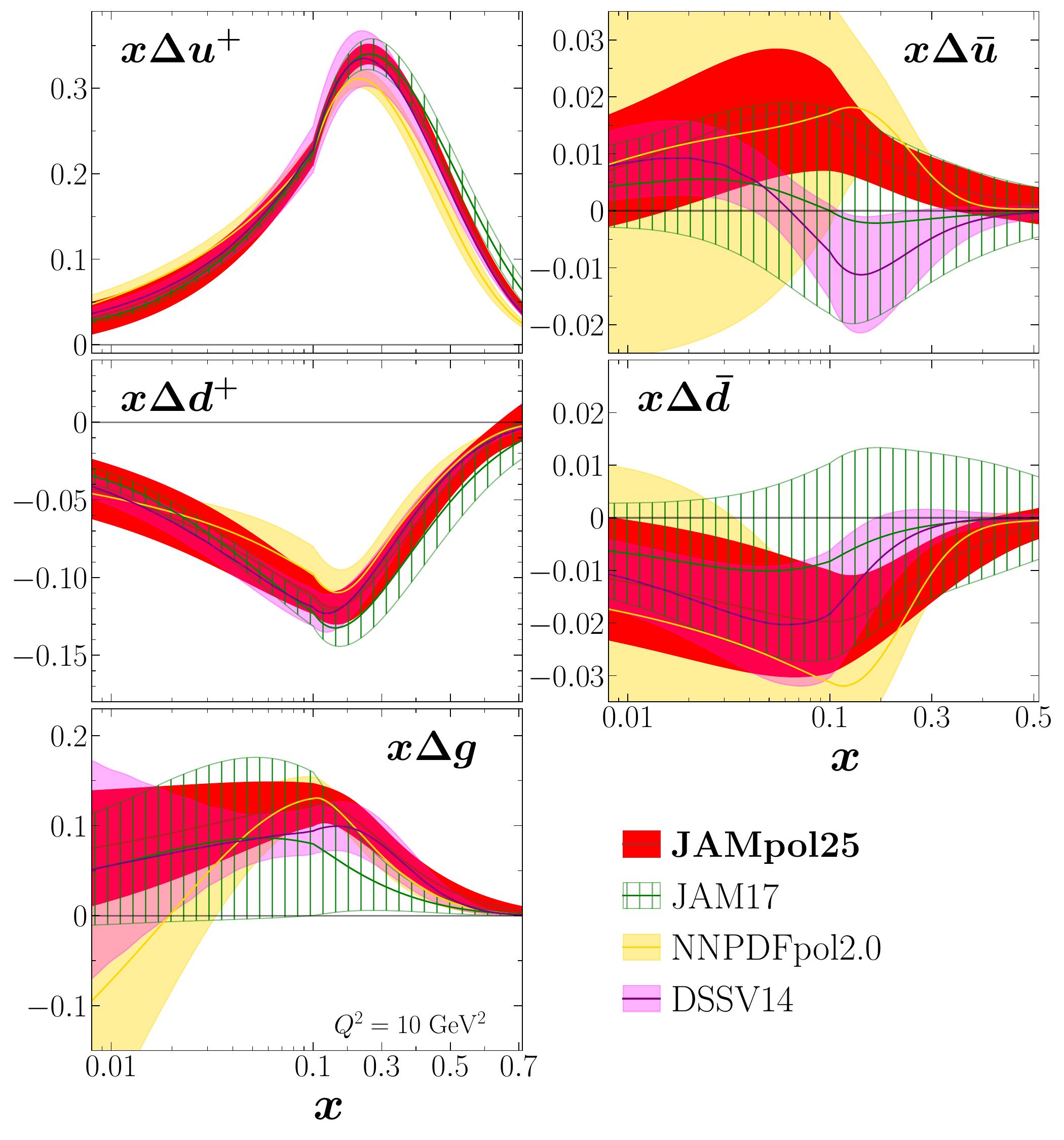}
\caption{Helicity-depdendent PDFs $x\Delta f$ (for $\Delta f = \Delta u^+$, $\Delta d^+$, $\Delta\bar u$, $\Delta\bar d$, $\Delta g$) from the current JAMpol25 analysis (red bands) compared with the previous JAM17 helicity PDF fit~\cite{Ethier:2017zbq} (green hatched bands) and parametrizations from NNPDFpol2.0~\cite{Cruz-Martinez:2025ahf} (yellow bands) and DSSV14~\cite{deFlorian:2014yva} (magenta bands) at a common scale $Q^2 = 10$~GeV$^2$.}
\label{fig:ppdfs_compare}
\end{figure}

With the extended range of $x$ accessible through the lowering of the $W^2$ cut, we can also study the behavior of the ratio of polarized to unpolarized quark PDFs in the deep valence region, $x \to 1$, as illustrated in Fig.~\ref{fig:polarization}.
Various nonperturbative and perturbative models exist which make specific predictions for the asymptotic $x \to 1$ behavior of $\Delta u/u$ and $\Delta d/d$~\cite{Melnitchouk:1995fc, Melnitchouk:1996zg,Close:1988br, Close:2003wz, Close:1979bt, Cloet:2005pp}.
For example, assuming dominance of the 3-quark Fock state component of the proton wave function, models with helicity conservation for the scattered parton predict the dominance of the helicity-aligned configurations in the large-$x$ region, which leads to the expectations that $\Delta u/u \to 1$ and $\Delta d/d \to 1$ in the asymptotic $x \to 1$ limit.
Other nonperturbative models, such as that based on spin-flavor SU(6) symmetry, the equal probabilities of the different components of the wave function lead to the nonrelativistic expectations $\Delta u/u \to~2/3$ and $\Delta d/d \to -1/3$ as $x \to 1$.
Notice that the predictions for the $d$ quark PDF ratios between the models actually differ in sign at high $x$~\cite{Jimenez-Delgado:2014xza}.
Empirically, $\Delta d/d$ remains negative for most of the range of $x$ covered by experiment, although the existing data do not rule out either behavior for $x \gtrsim 0.8$.

In Fig.~\ref{fig:ppdfs_compare} we compare the helicity PDFs from the JAMpol25 analysis with results from some other global analysis groups~\cite{Nocera:2014gqa, deFlorian:2014yva}, as well as from the previous JAM17 fit~\cite{Ethier:2017zbq} which simultaneously fitted polarized DIS, SIDIS and SIA data with a more restrictive $W$ cut of $W^2 > 10$~GeV$^2$.
For $\Delta u^+$ and $\Delta d^+$, there is a significant reduction in uncertainties at higher $x$ values, $x \gtrsim 0.1$, from the inclusion of the low $W^2$ DIS data.
For the light sea quark distributions, we find $\Delta \bar{u}$ to be clearly positive and $\Delta \bar{d}$ clearly negative, in contrast to some previous analyses that were less conclusive about the signs.
We also see significantly smaller errors on $\Delta g$ compared to other groups, largely due to the inclusion of the lattice QCD data.
In some regions the PDF uncertainties from the JAMpol25 analysis exceed those of JAM17, despite the inclusion of more experimental data, which reflects the use of a more flexible parameterization for the helicity PDFs in the current analysis.

\subsection{Structure functions and sum rules}
\label{s.HTnuc}

\begin{figure}[t]
\centering
\includegraphics[width=0.6\textwidth]{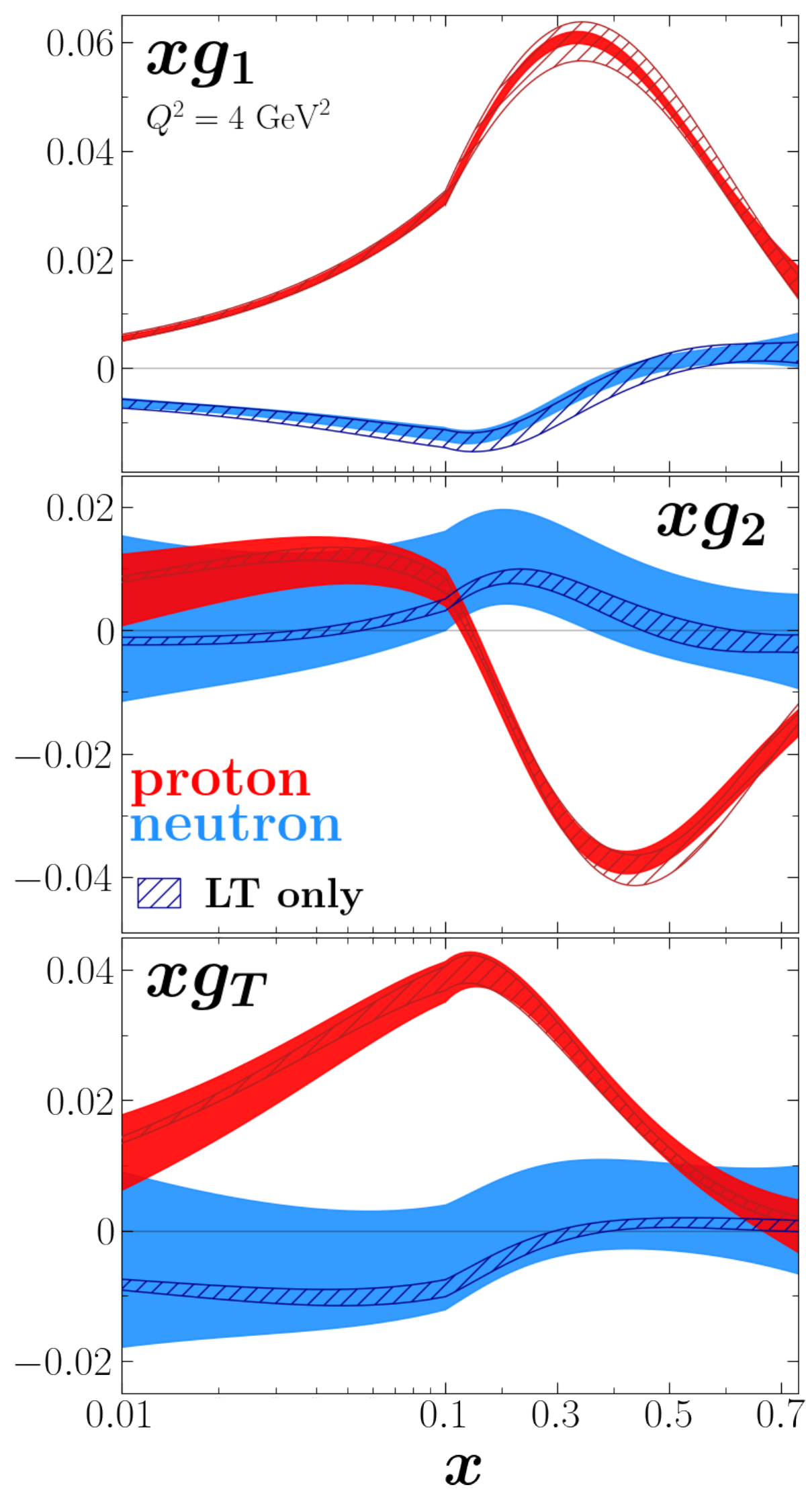}
\caption{Polarized structure functions $xg_1$ (top panel), $xg_2$ (middle panel), and $xg_T$ (bottom panel) evaluated at $Q^2 = 4$~GeV$^2$, for the proton (red) and neutron (blue), with the full JAMpol25 results (solid bands) compared with those at LT (hatched bands).}
\label{fig:pstfs}
\end{figure}

The polarized proton and neutron $g_1$ and $g_2$ structure functions, along with their sum, the transverse structure function $g_T$, are shown in Fig.~\ref{fig:pstfs} at a constant $Q^2 = 4$~GeV$^2$ for the full JAMpol25 analysis, and compared with the results with contributions from LT only,
consistent with the results found in Fig.~\ref{fig:ppdfs}.
Both the proton and neutron $g_1$ structure functions are rather stable with the addition of HT effects, with slightly larger uncertainties for the LT analysis.
This indicates that there are indeed sufficient data available to constrain the HT contributions for the proton and neutron $g_1$ and the proton $g_2$ structure functions.
For the neutron $g_2$ structure function, on the other hand, there is a significant increase in the uncertainties with the inclusion of the HT corrections. 
This is mainly because the neutron $g_2$ is primarily constrained by transverse DIS polarization asymmetries measured on deuteron and $\hel$ targets, for which the data are sparse and have sizable uncertainties (see Table~\ref{tab:chi2_Ape} and Figs.~\ref{fig:pidis_Ape_deuteron} and \ref{fig:pidis_A2_Ape_helium}).
These effects carry over to the transverse $g_T$ structure function, where the proton results remain similar with and without HT corrections.
For the neutron, however, without including the HT corrections one may naively conclude that $g_T$ is negative below $x \approx 0.2$, whereas the addition of HT effects shows that the uncertainties are underestimated, and that the neutron $g_T$ is still consistent with zero across the entire range of $x$ considered.

\begin{figure}[t]
\centering
\includegraphics[width=0.65\textwidth]{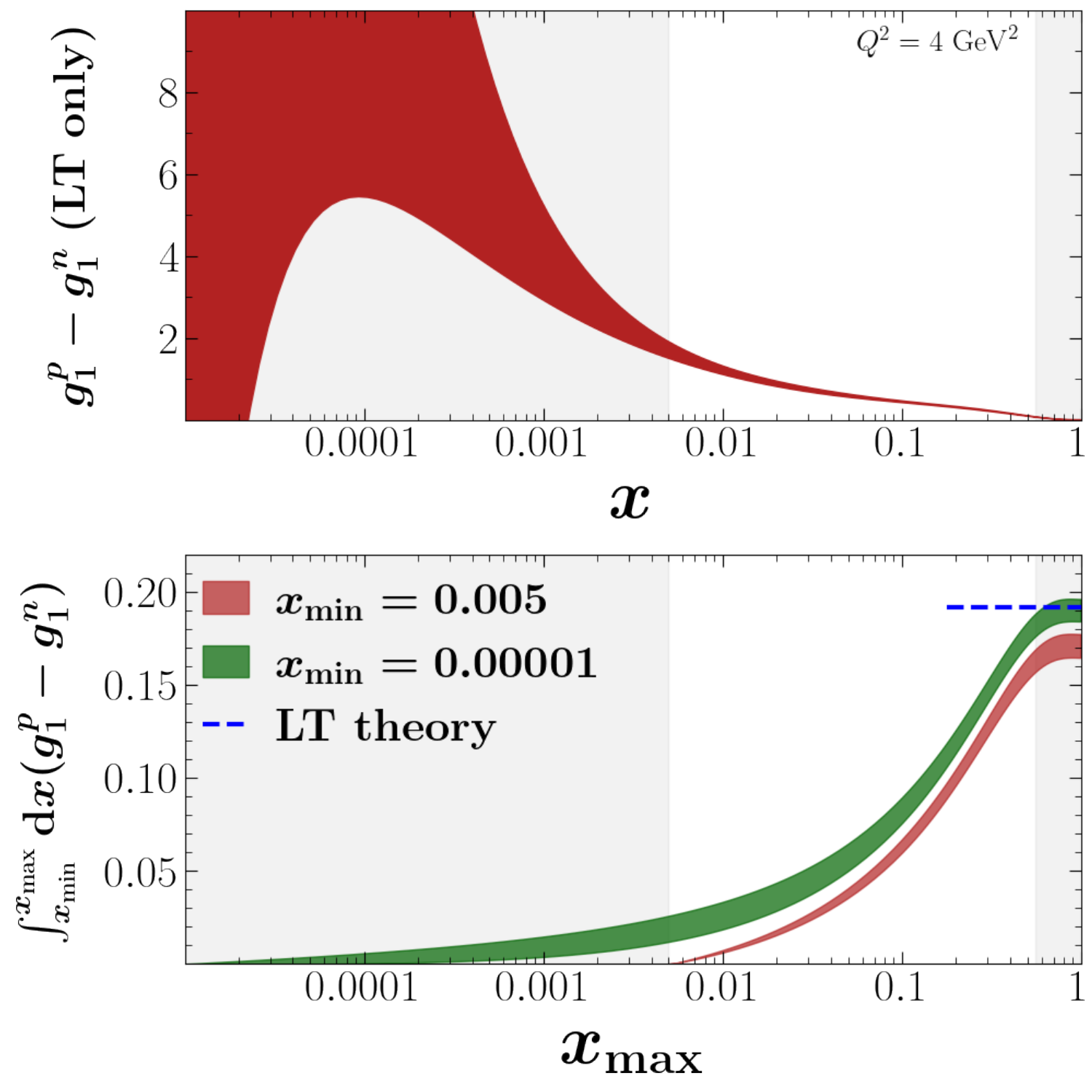}
\caption{
(Top) The isovector polarized structure function, $g_1^p - g_1^n$, for the LT only analysis, corresponding to the integrand of the Bjorken sum rule, as a function of $x$ for $Q^2 = 4$~GeV$^2$. (Bottom) Truncated Bjorken sum rule as a function of $x_{\rm max}$, starting from $x_{\rm min}=0.005$ (red) and $x_{\rm min} = 0.00001$ (green).  The result for the full moment (integrated from $x=0$ to 1) from the NLO LT calculation is also shown as the blue dashed line~\cite{Lampe:1998eu}. The regions of extrapolation at low and high $x$ are indicated by the vertical gray shaded bands.}
\label{fig:bjorken_sumrule}
\end{figure}

Taking the difference between the $g_1$ structure functions of the proton and neutron, the nonsinglet isovector combination $g_1^p - g_1^n$ for the LT only fit, corresponding to the integrand of the Bjorken sum rule~\cite{Bjorken:1966jh}, is shown in Fig.~\ref{fig:bjorken_sumrule} as a function of $x$ at fixed value of $Q^2=4$~GeV$^2$.
The integrand is observed to be positive across all $x$, although the uncertainties expand significantly for $x \lesssim 0.001$ as one goes into the unmeasured extrapolation region.
Also shown is the cumulative truncated Bjorken integral,
        $\int_{x_{\rm min}}^{x_{\rm max}} \dd{x} (g_1^p - g_1^n)$,
as a function of $x_{\rm max}$, integrated from the lowest value of $x$ for which there are experimental constraints, $x_{\rm min} = 0.005$, as well as the lower value $x_{\rm min} = 0.00001$.
For $x_{\rm min} \to 0$ and $x_{\rm max} \to 1$, the integral for the LT contribution should approach the full Bjorken sum, given at NLO by 
        $\frac16 g_A\, (1 - \alpha_s/\pi)$ \cite{Lampe:1998eu} 
(see Eq.~(\ref{e.Gamma1LT})). 
While the $x_{\rm min} = 0.005$ result slightly undershoots the LT theory calculation, one sees that the $x_{\rm min} = 0.00001$ result is compatible with it. 
This indicates that there is still a non-negligible contribution to the moment from the low-$x$ region of extrapolation, and that our results within the non-extrapolated region ($x_{\rm min} = 0.005$) are consistent with the theoretical expectation.

\begin{figure}[t]
\centering
\includegraphics[width=1.0\textwidth]{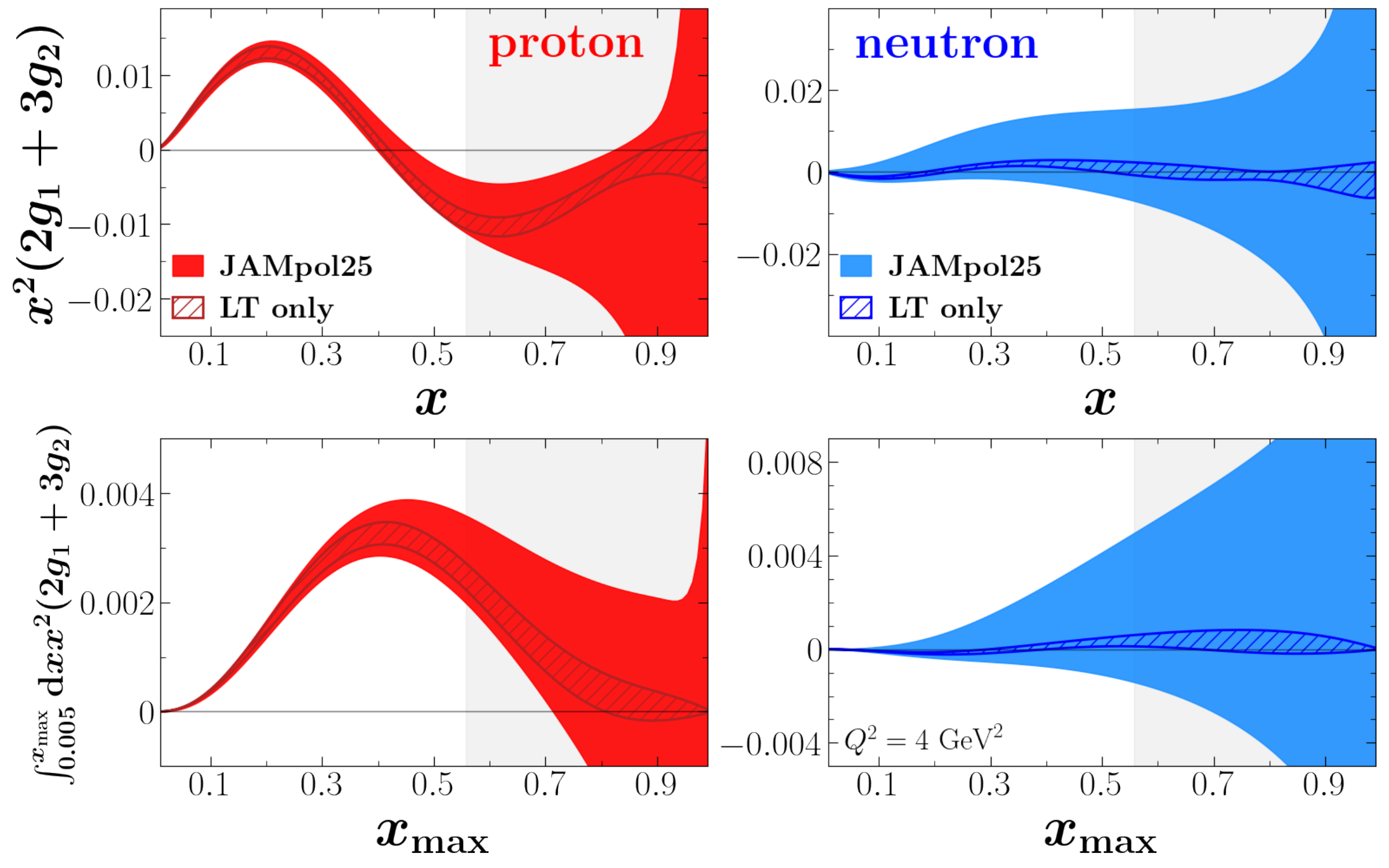}
\caption{(Top row) Integrand of the $d_2$ moment for the proton (left) and neutron (right) at $Q^2 = 4$~GeV$^2$ for the full JAMpol25 fit (solid bands) and the LT approximation (no HT correction or TMC, hatched bands). (Bottom row) Truncated integral as a function of $x_{\rm max}$ starting from $x = 0.005$ for the proton (left) and neutron (right). The region of extrapolation into the low-$W$ region is indicated by the gray shaded area.}
\label{fig:d2_integrand}
\end{figure}

In analogy with the Bjorken integral, in Fig.~\ref{fig:d2_integrand} we show the integrand of the $d_2$ matrix element, defined in Eq.~(\ref{e.d2}), for the proton and neutron as a function of $x$ at $Q^2=4$~GeV$^2$, for both the full JAMpol25 fit and the LT only result.
Also shown is the truncated integral as a function of $x_{\rm max}$ on a linear scale.
The lower limit is fixed to $x = 0.005$ in order to avoid extrapolation into the unmeasured region, though it should be noted that the contribution below $x = 0.005$ to the integral is expected to be negligible because of the $x^2$ weighting.
This causes the LT only result to (approximately) converge to zero as $x_{\rm max} \to 1$, while the full result (with HT corrections and TMCs) has increasingly large uncertainties.
For the neutron the integrand is always consistent with zero and thus also is the truncated moment, regardless of the inclusion of HT corrections and TMCs.
On the other hand, for the proton the integral is nonzero throughout the entire range of $x_{\rm max}$ when HT corrections and TMCs are not included.
With these included, the result remains positive up until $x \approx 0.8$, where the extrapolation of the HT corrections causes the uncertainties to increase dramatically.

\clearpage
\section{Conclusions}
\label{s.conclusions}

We have presented a new comprehensive global QCD analysis of spin-dependent parton distribution functions within the JAM Bayesian iterative Monte Carlo framework, combining all available data on polarized inclusive and semi-inclusive DIS, and inclusive $W$ and jet production in polarized $pp$ collisions, along with new lattice QCD data on gluonic pseudo Ioffe-time distributions.
To consistently utilize the polarized SIDIS data, we also simultaneously fit the parton to hadron fragmentation functions, which are constrained by SIDIS data as well as single-inclusive $e^+ e^-$ annihilation data into pions, kaons, and charged hadrons~\cite{Sato:2016wqj, Moffat:2021dji}.

With a DIS cut of $W^2 > 4$~GeV$^2$, we include for the first time the recent Jefferson Lab data from the SANE experiment in Hall~C~\cite{SANE:2018pwx}, the interpretability of which requires a careful treatment of subleading power corrections at low $Q^2$ and high $x$.
This demanded the derivation of target mass corrections for spin structure functions within the collinear factorization framework (see Appendix~\ref{a.TMC-CF}), which is necessary for our theoretical framework to be applicable in the low-$W$ and low-$Q^2$ domain.
Including the TMCs alongisde the fitted phenomenological higher twist corrections to order $1/Q^2$, we studied the dependence of the polarized PDFs on the $W^2$ cut chosen for the DIS data, and verified the stability of the PDFs for $W^2 > 4$~GeV$^2$ and $Q^2 > m_c^2$.

The low-$W^2$ data, concentrated in the region $0.2 < x < 0.6$ but extending to $x$ as large as 0.845, have a large impact on the uncertainties of the helicity PDFs, with reductions as high as 60\% for the $\Delta u$ and 40\% for the $\Delta d$ PDFs.
Our analysis allows us to determine the $\Delta u/u$ and $\Delta d/d$ ratios with significantly smaller uncertainties up to $x \approx 0.6-0.7$.
The errors on the $\Delta d$ PDF at higher $x$ values become very large, preventing definitive conclusions being made about the behavior of $\Delta d/d$ as $x \to 1$.
We confirm positive $\bar u$ and negative $\bar d$ polarizations for $x \lesssim  0.4$, driven mainly by the $W$ boson asymmetry data, with reduced uncertainties at intermediate $x$ values, especially for the $\Delta \bar d$ distribution.
Without imposing theoretical constraints, such as SU(3) flavor symmetry, we find that the existing data are not able to resolve a clear signal of a nonzero strange polarization, $\Delta s$.

Comparing our full results with those that omit HT corrections, we do not observe overwhelming phenomenological evidence for the need of the HT terms to describe the data down to $W^2 = 4$~GeV$^2$, with only a modest improvement in the global $\chi^2_{\rm red}$ for polarized data, from $\chi^2_{\rm red} = 1.01$ ($Z$-score +0.63) to 0.98 ($-0.61$) with the inclusion of the HTs.
The inclusion of HT corrections does, however, provide insight into the uncertainties on the structure functions.
We find that while for the proton spin structure functions the difference between the full and no HT results is relatively small, omitting the HT terms leads to significant underestimation of the uncertainties on the neutron $g_2$ and $g_T$ structure functions.
In particular, without including the HT corrections one may naively extract a neutron $g_T$ that is negative at $x \lesssim 0.2$, while in the full analysis with HT effects the neutron $g_T$ is consistent with zero across the entire range of $x$ considered.

Furthermore, the small negative (positive) HT correction to the proton (neutron) $g_1$ structure function at high $x$ values leads to a small negative HT contribution to the Bjorken integral.
For the truncated $d_2$ moment, we find that both the proton and neutron integrals are consistent with zero for the range covered by the current experimental data.
We note, however, the potentially important role played by nucleon resonances at high values of $x$, which at $Q^2=4$~GeV$^2$ means the extrapolation beyond $x \approx 0.55$.
Future higher energy data from Jefferson Lab at 12~GeV and the EIC should reduce this extrapolation region and allow the $d_2$ moment to be extracted over a larger range of $x$.

Following Ref.~\cite{Hunt-Smith:2024khs}, we have also included lattice QCD data on the gluonic pseudo Ioffe-time distributions, which, together with the jet production in polarized $pp$ collisions and high-$x$ DIS data, were found to be important in constraining the polarized gluon PDF to be positive, without the need for assumptions about PDF positivity.
Note, however, that at present the gluon Ioffe time distribution data do not provide constraints on PDFs in the large-$x$ region beyond resolving the sign of $\Delta g$.
A future extension of the current analysis to include dijet production data in polarized $pp$ collisions could potentially remove the need for lattice QCD data in unambiguously determining the sign of $\Delta g$.

Our results have demonstrated that polarized data, in particular the polarized DIS data at large $x$ and low $W$, can be successfully described within an NLO framework, indicating the applicability of QCD factorization, and consistent with the observations in Refs.~\cite{Borsa:2024mss, Cruz-Martinez:2025ahf} that NNLO corrections are rather small.
Of course, this situation may change in the future as the precision of experimental data improves with the ongoing JLab 12~GeV program and the future EIC, as well as with the inclusion of new observables that are more sensitive to NNLO effects. 
We leave for future work the extension of this analysis to include NNLO effects.

Upcoming polarized DIS and SIDIS data from the 12~GeV program at Jefferson Lab should shed further light on the intermediate- and high-$x$ regions, and in particular higher precision measurements of transversely polarized asymmetries on $^3$He would be beneficial to more reliably determine the neutron $d_2$ matrix element.
Data from a future EIC data would, on the other hand, provide constraints on the $x$ dependence of structure functions at lower $x$ values, and reduce the large unertainties on the singlet quark and gluon polarizations arising from extrapolations through the unmeasured region.

\begin{acknowledgments}

We thank P.~C.~Barry, A.~Metz, and Y.~Zhou for helpful discussions, and F.~Ringer and W.~Vogelsang for the code used for calculating the $W$-lepton cross sections.
This work was supported by the U.S. Department of Energy Contract No.~DE-AC05-06OR23177, under which Jefferson Science Associates, LLC operates Jefferson Lab, and the National Science Foundation under grant number PHY-1516088.
The work of N.S. was supported by the DOE, Office of Science, Office of Nuclear Physics in the Early Career Program. 
Support was also provided by the Australian Research Council through the Centre of Excellence for Dark Matter Particle Physics, CE200100008, and Discovery Project DP230101791.

\end{acknowledgments}

\appendix
\section{Target mass corrections in collinear factorization}
\label{a.TMC-CF}

Following the derivation by Moffat {\it et al.}~\cite{Moffat:2019qll} of the unpolarized $F_1$ and $F_2$ structure functions at finite values of $M^2/Q^2$, in this appendix we outline the steps in the corresponding derivation of the spin-dependent structure functions $g_1$ and $g_2$ in the presence of hadron masses.
As usual, we define the Nachtmann scaling variable as
\begin{equation}
\xn = \frac{2x}{1+\rho},
\qquad \rho^2 = 1 + \frac{4x^2 M^2}{Q^2}\, .
\end{equation}
Neglecting lepton masses, the hadronic tensor for polarized DIS in the one-photon exchange approximation is given by
\begin{eqnarray}
W_{\mu\nu} 
&=& i \epsilon_{\mu \nu \alpha \beta} \frac{q^{\alpha}}{P \cdot q} 
\bigg[ S^{\beta}\, g_1(x,Q^2) 
    + \Big( S^{\beta} - \frac{S \cdot q}{P \cdot q} P^{\beta} \Big)\, g_2(x,Q^2) 
\bigg],
\label{e.hadrontensorCF}
\end{eqnarray}
where the structure functions $g_1$ and $g_2$ contain all $M^2/Q^2$ target mass corrections.

In the Breit frame the four-momenta of the target nucleon $P = (P^+, P^-, \bm{P}_\perp)$ and virtual photon $q = (q^+, q^-, \bm{q}_\perp)$ can be written in light-front coordinates as
\begin{eqnarray}
P &=& \bigg( \frac{Q}{\sqrt{2} \xn}, \frac{\xn M^2}{\sqrt{2} Q}, \bm{0}_{\perp} \bigg),
\qquad
q = \bigg( \!-\frac{Q}{\sqrt{2}}, \frac{Q}{\sqrt{2}}, \bm{0}_{\perp} \bigg),
\label{e.breitPq}
\end{eqnarray}
respectively.
The covariant spin vector $S$ is given by
\begin{eqnarray}
S = M\, \bigg( \lambda \frac{P^+}{M}, -\lambda \frac{M}{2 P^+}, \bm{S}_{\perp} \bigg),
\label{e.covariantS}
\end{eqnarray}
and satisfies the relations $P \cdot S$ = 0 and $S^2 = -M^2$.
In the massless target limit, $M \to 0$, the ``$-$'' components of $P$ and $S$ vanish, and we can define ``massless'' four-vectors $\widetilde{P}$ and $\widetilde{S}$ by
\begin{eqnarray}
\widetilde{P} = \bigg( \frac{Q}{\sqrt{2} \xn}, 0, \bm{0}_{\perp} \bigg),
~~~~~~
\widetilde{S} = M \bigg( \lambda \frac{P^+}{M}, 0, \bm{S}_{\perp} \bigg),
\label{e.MTA_PS}
\end{eqnarray}
keeping the transverse component of the spin vector.

With this notation one can write the hadronic tensor as
\begin{eqnarray}
W_{\mu \nu}
= i \epsilon_{\mu \nu \alpha \beta} \frac{q^{\alpha}}{\widetilde{P} \cdot q}
\bigg[ \widetilde{S}^{\beta}\, \widetilde{g}_1(x,Q^2) 
    + \Big( \widetilde{S}^{\beta} - \frac{\widetilde{S} \cdot q}{\widetilde{P}
        \cdot q} \widetilde{P}^{\beta}  
      \Big)\, \widetilde{g}_2(x,Q^2)
\bigg],
\label{e.hadrontensortilde}
\end{eqnarray}
where the structure functions $\widetilde{g}_{1,2}$ do not contain the $M^2/Q^2$ corrections, in contrast to those in \eref{e.hadrontensorCF}.  
The hadronic tensors in \eref{e.hadrontensorCF} and \eref{e.hadrontensortilde} must be equal, as they are simply the same object expressed in terms of two choices of basis tensors.

Using the relations
\begin{subequations}
\begin{eqnarray}
\widetilde{P}^{\mu} 
&=& \frac{1}{Q^2 + \xn^2 M^2} \big( Q^2 P^{\mu} - \xn M^2 q^{\mu} \big),
\\
\widetilde{S}^{\mu} 
&=& S^{\mu} + \lambda \frac{\xn M^2}{Q^2 + \xn^2 M^2} \big( \xn P^{\mu} + q^{\mu} \big),
\label{e.PSrelations}
\end{eqnarray}
\end{subequations}
one can then write
\begin{subequations}
\begin{eqnarray}
i \epsilon_{\mu \nu \alpha \beta} \frac{q^{\alpha}}{\widetilde{P} \cdot q} \widetilde{S}^{\beta} &=& i \epsilon_{\mu \nu \alpha \beta} \frac{q^{\alpha}}{P \cdot q} 
\bigg[ \frac{(Q^2 - \xn^2 M^2)(Q^2 + 3 \xn^2 M^2)}{(Q^2 + \xn^2 M^2)^2} 
S^{\beta}
\notag\\
& & \hspace*{2cm} -\, \frac{\xn^2 M^2}{Q^2} \frac{(Q^2 - \xn^2 M^2)^2}{(Q^2 + \xn^2 M^2)^2} 
\Big( 
S^{\beta} - \frac{S \cdot q}{P \cdot q} P^{\beta}   
\Big) 
\bigg],
\label{e.CFrelation1}
\end{eqnarray}
and
\begin{eqnarray}
i \epsilon_{\mu \nu \alpha \beta} \frac{q^{\alpha}}{\widetilde{P} \cdot q} 
\big(
\widetilde{S}^{\beta} 
- \frac{\widetilde{S} \cdot q}{\widetilde{P} \cdot q} \widetilde{P}^{\beta}
\big)
&=& 
i \epsilon_{\mu \nu \alpha \beta}
\frac{q^{\alpha}}{P \cdot q} 
\bigg[ 4\xn^2 M^2 \frac{Q^2 - \xn^2 M^2}{(Q^2 + \xn^2 M^2)^2} 
S^{\beta}
\notag\\
& & \hspace*{1cm} +\, \frac{(Q^2 - \xn^2 M^2)^3}{Q^2(Q^2 + \xn^2 M^2)^2} 
\Big( 
S^{\beta} - \frac{S \cdot q}{P \cdot q} P^{\beta}   
\Big) 
\bigg].
\label{e.CFrelation2}
\end{eqnarray}
\end{subequations}
Substituting Eqs.~(\ref{e.CFrelation1}) and (\ref{e.CFrelation2}) into \eref{e.hadrontensortilde} and comparing the tensor structures to \eref{e.hadrontensorCF} leads to the relations
\begin{subequations}
\begin{eqnarray}
g_1(x,Q^2) 
&=& \frac{(Q^2 - \xn^2 M^2)(Q^2 + 3 \xn^2 M^2)}{(Q^2 + \xn^2 M^2)^2}\, 
    \widetilde{g}_1(x,Q^2) 
 +  4 \xn^2 M^2 \frac{(Q^2 - \xn^2 M^2)}{(Q^2 + \xn^2 M^2)^2}\, 
    \widetilde{g}_2(x,Q^2),
\notag\\
& &
\\
g_2(x,Q^2) 
&=& - \frac{\xn^2}{Q^2} \frac{(Q^2 - \xn^2 M^2)^2}{(Q^2 + \xn^2 M^2)^2}\,
    \widetilde{g}_1(x,Q^2)
 +  \frac{(Q^2 - \xn^2 M^2)^3}{Q^2(Q^2 + \xn^2 M^2)^2}\, 
    \widetilde{g}_2(x,Q^2),
\end{eqnarray}
\end{subequations}
which can be rewritten in terms of $\rho$ as
\begin{subequations}
\begin{eqnarray}
g_1(x,Q^2)  
&=& \frac{2\rho -1}{\rho^2}\, \widetilde{g}_1(x,Q^2) 
 +  \frac{2(\rho -1)}{\rho^2}\, \widetilde{g}_2(x,Q^2),
\label{e.g1CFrelation1}
\\
g_2(x,Q^2)  
&=& \frac{1-\rho}{(1+\rho)\rho^2}\, \widetilde{g}_1(x,Q^2) 
 +  \frac{2}{(1+\rho)\rho^2}\, \widetilde{g}_2(x,Q^2).
\label{e.g2CFrelation2}
\end{eqnarray}
\end{subequations}
Equations~(\ref{e.g1CFrelation1}) and (\ref{e.g2CFrelation2}) then represent the general, model-independent relation between the massless limit structure functions and those evaluated at finite values of $M^2/Q^2$, without any kinematical or dynamical approximations.

Within the standard collinear factorization framework~\cite{Collins:2011zzd}, in the LT approximation the structure functions can be written in factorized form in terms of PDFs and hard coefficient functions~(\ref{e.g1LT}), but with the lower limit of integration in the convolution integral~(\ref{e.convolution}) equal to the Nachtmann scaling variable $\xn$.
Namely, without making any massless target approximations, in collinear factorization one has
\begin{subequations}
\begin{eqnarray}
\widetilde g_1(x,Q^2) \to g_1^{\mbox{\tiny \rm LT}}(\xn,Q^2),
\\
\widetilde g_2(x,Q^2) \to g_2^{\mbox{\tiny \rm LT}}(\xn,Q^2).
\end{eqnarray}
\end{subequations}
Using the Wandzura-Wilczek relation is to write $g_2^{\mbox{\tiny \rm LT}}$ in terms of $g_1^{\mbox{\tiny \rm LT}}$~\cite{Wandzura:1977qf} then leads to
\begin{subequations}
\begin{eqnarray}
g_1^{\mbox{\tiny \rm TMC}} (x,Q^2)
&=& \frac{2\rho - 1}{\rho^2}\,
    g_1^{\mbox{\tiny \rm LT}}(\xn,Q^2) 
  + \frac{2(\rho -1)}{\rho^2}
    \Big( -g_1^{\mbox{\tiny \rm LT}}(\xn,Q^2)
        + \int_{\xn}^1 \frac{\diff z}{z} g_1^{\mbox{\tiny \rm LT}}(z,Q^2)
    \Big),
\notag\\
& &
\\
\!\!\!g_2^{\mbox{\tiny \rm TMC}}(x,Q^2)
&=& \frac{1-\rho}{(1+\rho)\rho^2}\, 
    g_1^{\mbox{\tiny \rm LT}}(\xn,Q^2) 
 +  \frac{2}{(1+\rho)\rho^2}\, 
    \Big( -g_1^{\mbox{\tiny \rm LT}}(\xn,Q^2)
        + \int_{\xn}^1 \frac{\diff z}{z} 
    g_1^{\mbox{\tiny \rm LT}}(z,Q^2)
    \Big).
\notag\\
& &
\end{eqnarray}
\end{subequations}
After simplifying, we obtain
\begin{subequations}
\begin{eqnarray}
g_1^{\mbox{\tiny \rm TMC}}(x,Q^2) 
&=& \frac{1}{\rho^2} 
    g_1^{\mbox{\tiny \rm LT}}(\xn,Q^2) 
 +  \frac{2(\rho-1)}{\rho^2} \int_{\xn}^1 \frac{\diff z}{z} 
    g_1^{\mbox{\tiny \rm LT}}(z,Q^2), 
\\
g_2^{\mbox{\tiny \rm TMC}}(x,Q^2) 
&=& -\frac{1}{\rho^2} 
    g_1^{\mbox{\tiny \rm LT}}(\xn,Q^2) 
 +  \frac{2}{(1+\rho) \rho^2} \int_{\xn}^1 \frac{\diff z}{z} 
    g_1^{\mbox{\tiny \rm LT}}(z,Q^2),
\end{eqnarray}
\end{subequations}
as in Eqs.~(\ref{e.CFtmc}).

\section{Data to theory comparisons}
\label{a.data}

In this appendix we present the full set of comparisons of the polarized data versus theory, and the individual reduced $\chi^2$ and $Z$-score values for the spin-dependent observables used in this analysis.
Tables~\ref{tab:chi2_Apa} and \ref{tab:chi2_Ape} give the $\chi^2_{\rm red}$ values and $Z$-scores for longitudinally polarized ($A_1$ and \Apa) and transversely polarized ($A_2$, \Ape, and \Atpe), respectively, inclusive DIS asymmetries  for proton, deuteron, and $^3$He targets.
The number of data points ($N_{\rm dat}$) corresponds to a cut of $W^2 > 4$~GeV$^2$ on the DIS data.
The results for the full JAMpol25 analysis, which includes TMC and HT effects, are listed along with those from a fit using the LT approximation only and those using a $W^2 > 3.5$ GeV$^2$ cut. 
In the latter case, there are significant increases in the $\chired$ for the JLab eg1-dvcs and SANE proton $A_{\parallel}$ data, increasing the total $A_{\parallel}$ $\chired$ from 1.01 to 1.11. 
There is little change in the $A_{\perp}$ data between the two $W^2$ cuts.

The $\chi^2_{\rm red}$ and $Z$-score values for the longitudinally polarized SIDIS asymmetries, $A_1^h$, for the production of hadrons $h$ $(=\pi^\pm, K^\pm, h^\pm)$ from $p$, $d$, and $^3$He targets are listed in Table~\ref{tab:chi2_SIDIS}.
Note that for the SLAC 155x data the measured transverse asymmetry \Atpe\ has a small contribution from longitudinal polarization, since the target polarizations were not completely perpendicular to the beam line in that experiment (see Ref.~\cite{E155:2002iec} for details).
Finally, Table~\ref{tab:chi2_pp} lists the $\chi^2_{\rm red}$ values and $Z$-scores for inclusive jet and weak boson production in longitudinally polarized $pp$ collisions.
Here the jet data are binned in the jet rapidity, $y_{\rm jet}$, and transverse momentum, $p_T$, while the $W/Z$ production data are binned in $p_T$ and the electron rapidity,~$\eta_e$.

\begin{figure*}[t]
\centering
\includegraphics[width=0.95\textwidth]{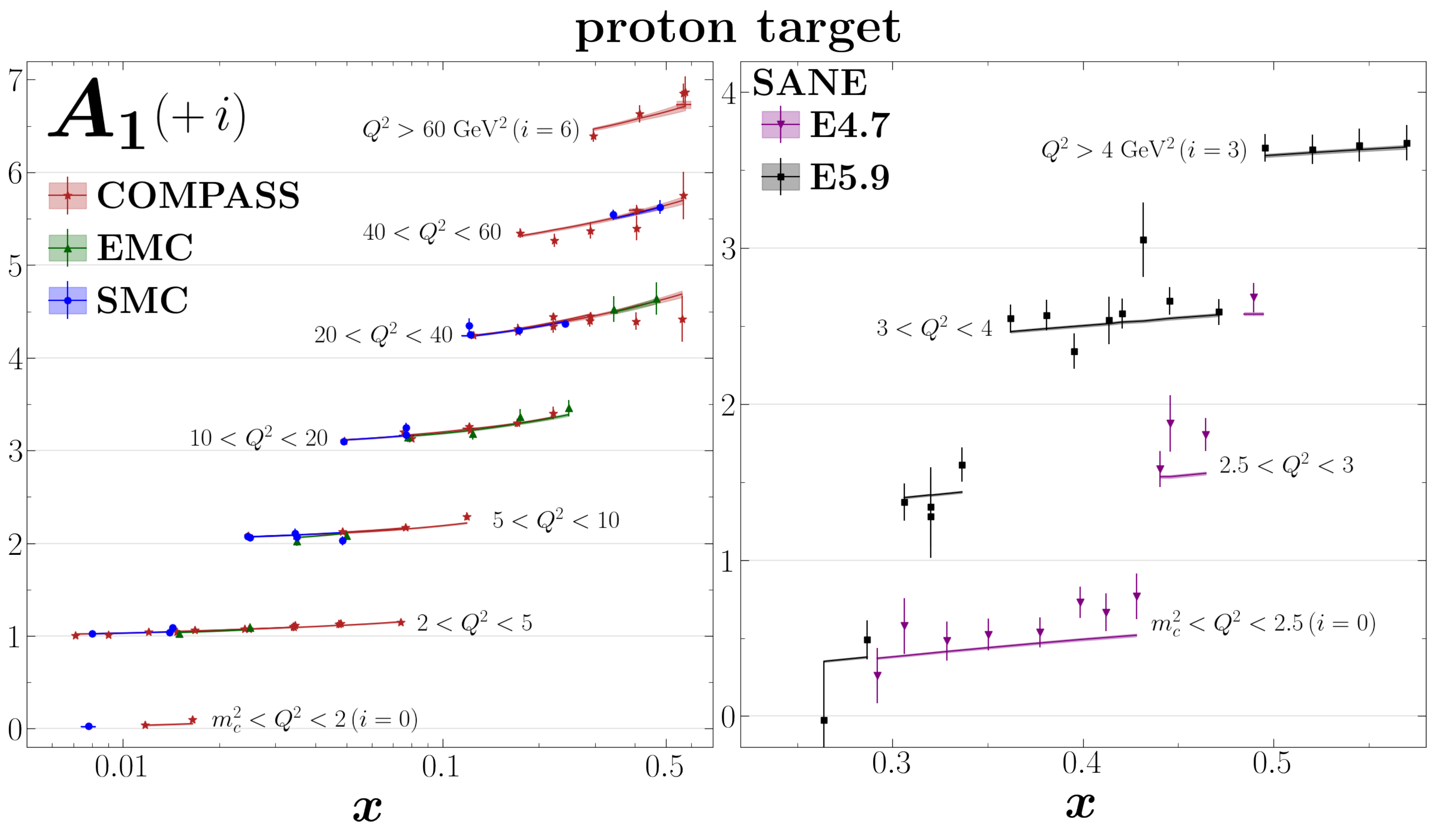}
\caption{Proton $A_1$ polarization asymmetry from COMPASS~\cite{COMPASS:2010wkz, COMPASS:2015rvb} (red), EMC~\cite{EuropeanMuon:1989yki} (green), SMC~\cite{SpinMuon:1998eqa, SpinMuon:1999udj} (blue), and the SANE experiment at Jefferson Lab~\cite{SANE:2018pwx} at energies of 4.7~GeV (purple) and 5.9~GeV (black) versus Bjorken~$x$, compared with the JAM fit (colored lines and 1$\sigma$ bands). Data in different $Q^2$ bins are offset by $i$ for clarity.
}
\label{fig:pidis_A1_proton}
\end{figure*}

\begin{figure*}[h]
\centering
\includegraphics[width=0.90\textwidth]{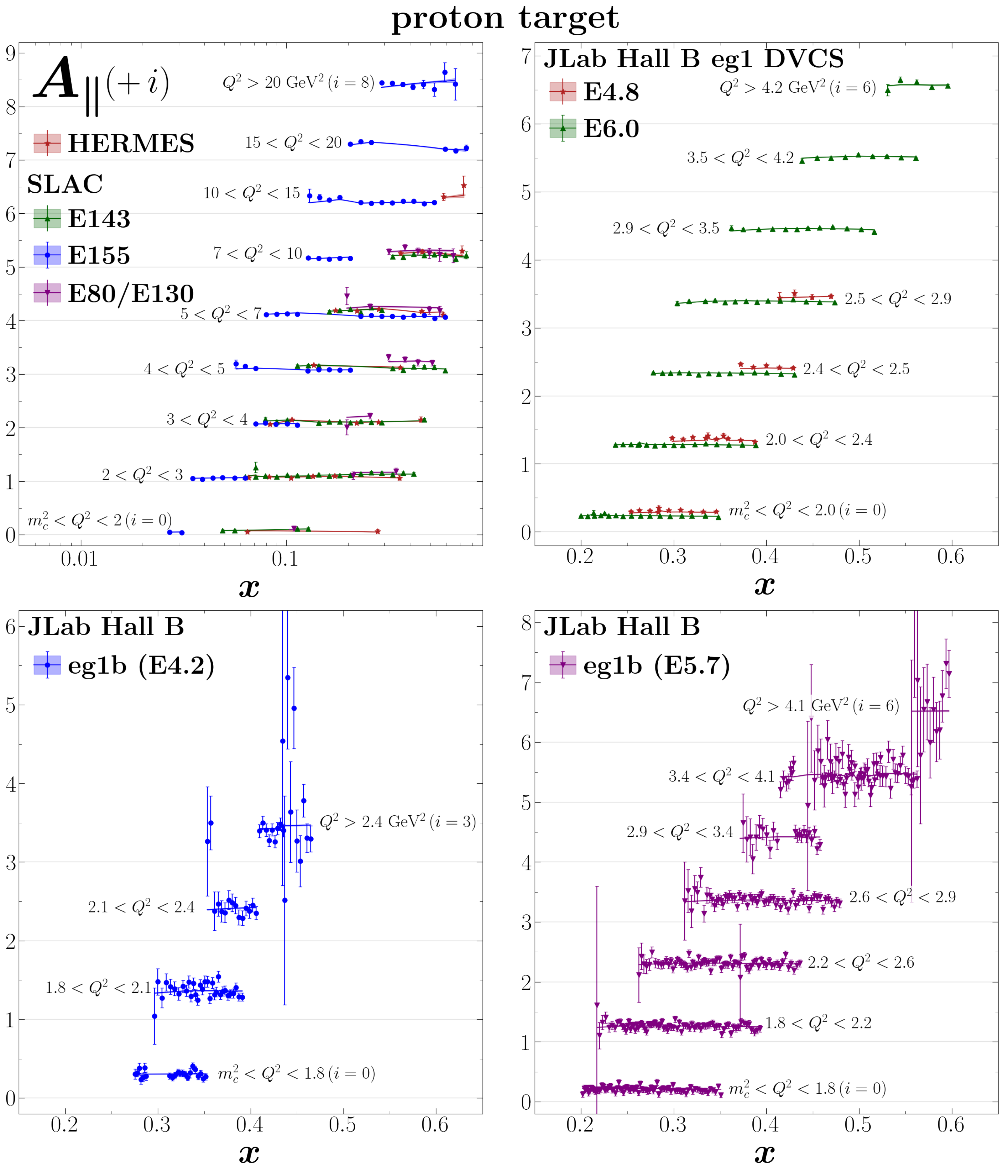}
\caption{Polarized DIS data for the proton observable $A_{\parallel}$ from HERMES \cite{HERMES:2006jyl}, SLAC \cite{Baum:1983ha, E143:1998hbs, E155:2000qdr}, and Jefferson Lab \cite{CLAS:2014qtg, CLAS:2017qga} plotted as a function of Bjorken $x$ against the JAM result (colored lines and 1$\sigma$ bands). The top left panel shows data from HERMES (red) and SLAC at energies 143~GeV (green), 155~GeV (blue), and 80/130~GeV (purple).  The top right panel shows data from the DVCS experiment at Jefferson Lab Hall~B at energies 4.8~GeV (red) and 6.0 GeV (green).  The bottom panels show the eg1b experiment at Jefferson Lab at energies 4.2~GeV (left, blue) and 5.7~GeV (right, purple).  Data in different $Q^2$ bins are shifted by $i$ for clarity.
}
\label{fig:pidis_Apa_proton}
\end{figure*}

\begin{figure*}[t]
\centering
\includegraphics[width=0.45\textwidth]{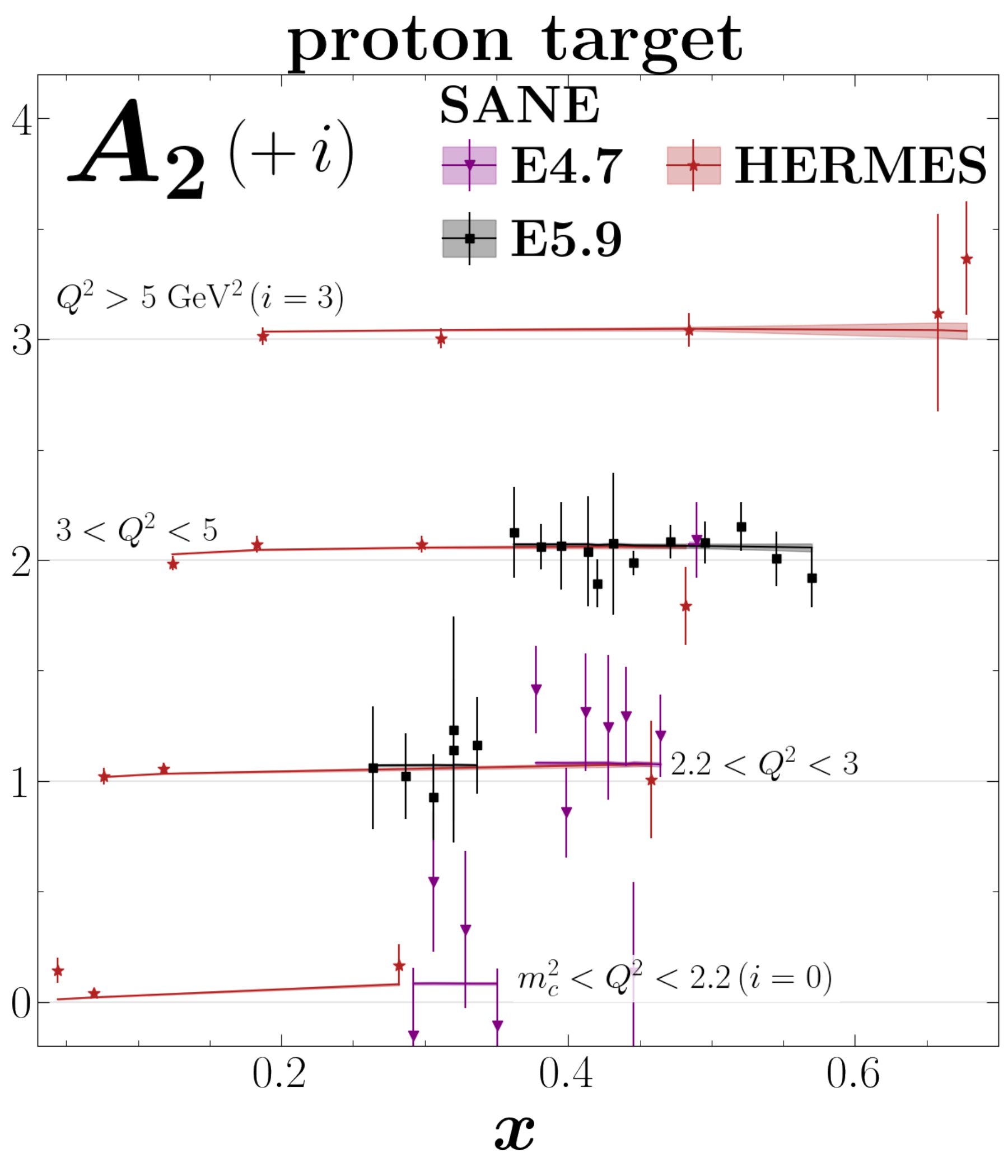}
\caption
{Polarized DIS data for the proton $A_2$ asymmetry from HERMES \cite{HERMES:2011xgd} (red) and SANE \cite{SANE:2018pwx} at energies 4.7 GeV (purple) and 5.9 GeV (black) plotted as a function of Bjorken $x$ against the JAM result (colored lines and 1$\sigma$ bands). Data in different $Q^2$ bins are increased by $i$ for clarity.
}
\label{fig:pidis_A2_proton}
\end{figure*}

\begin{figure*}[b]
\centering
\includegraphics[width=0.8\textwidth]{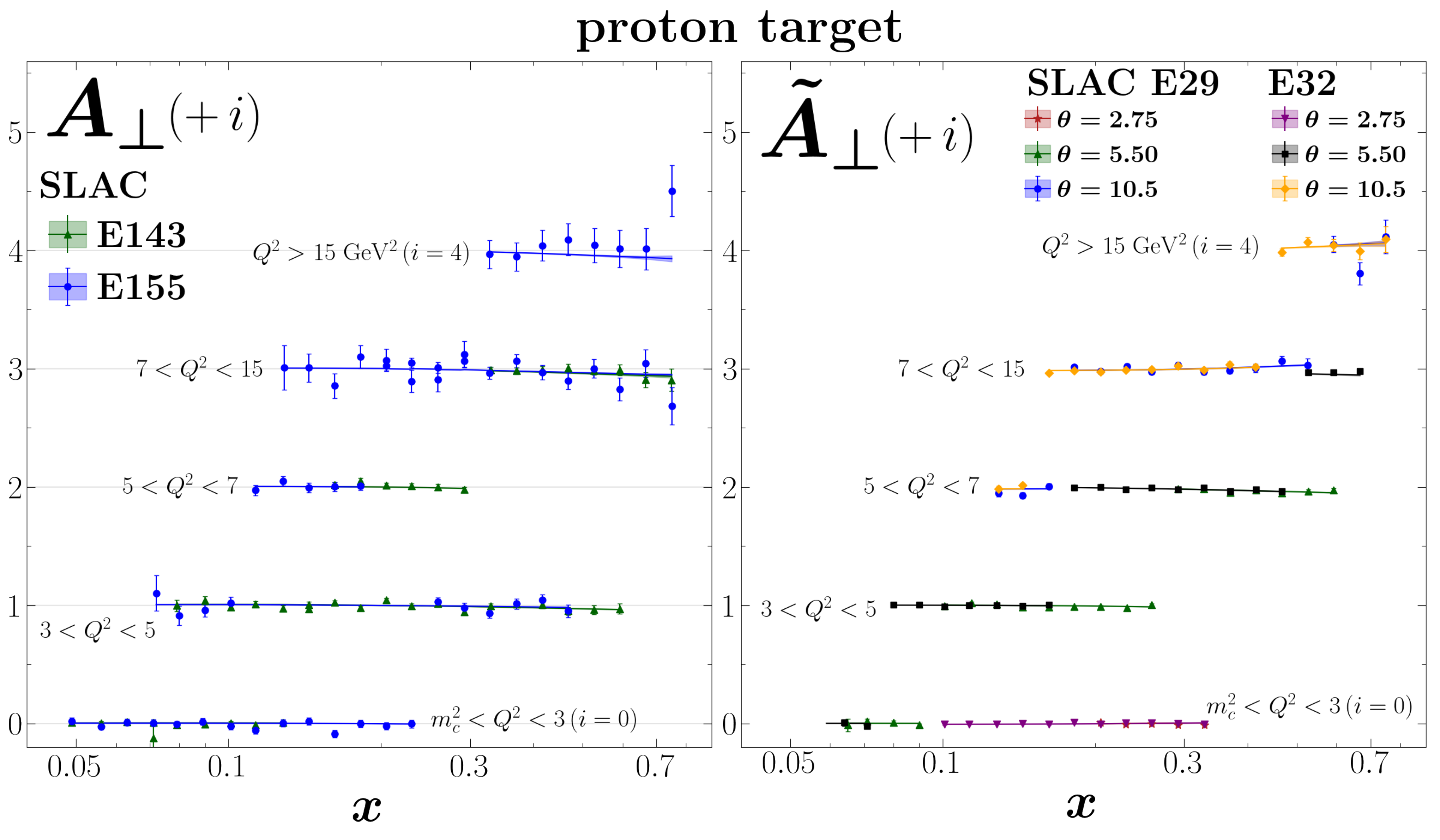}
\caption
{Polarized DIS data for the proton observables $A_\perp$ (left panel) and $\widetilde{A}_\perp$ (right panel) (see Ref.~\cite{E155:2002iec} for more details on this observable) from SLAC \cite{E143:1998hbs, E155:1999eug, E155:2002iec} plotted as a function of Bjorken $x$ against the JAM result (colored lines and 1$\sigma$ bands). On the left panel, results are shown at energies 143~GeV (green) and 155~GeV (blue).  On the right panel, results are shown at energies 29~GeV for $\theta = 2.75, 5.50, 10.5$ (red, green, and blue, respectively) and at 32~GeV for $\theta = 2.75, 5.50, 10.5$ (purple, black, and orange, respectively).  Data in different $Q^2$ bins are increased by $i$ for clarity.
}
\label{fig:pidis_Ape_proton}
\end{figure*}

\begin{figure*}[t]
\centering
\includegraphics[width=0.5\textwidth]{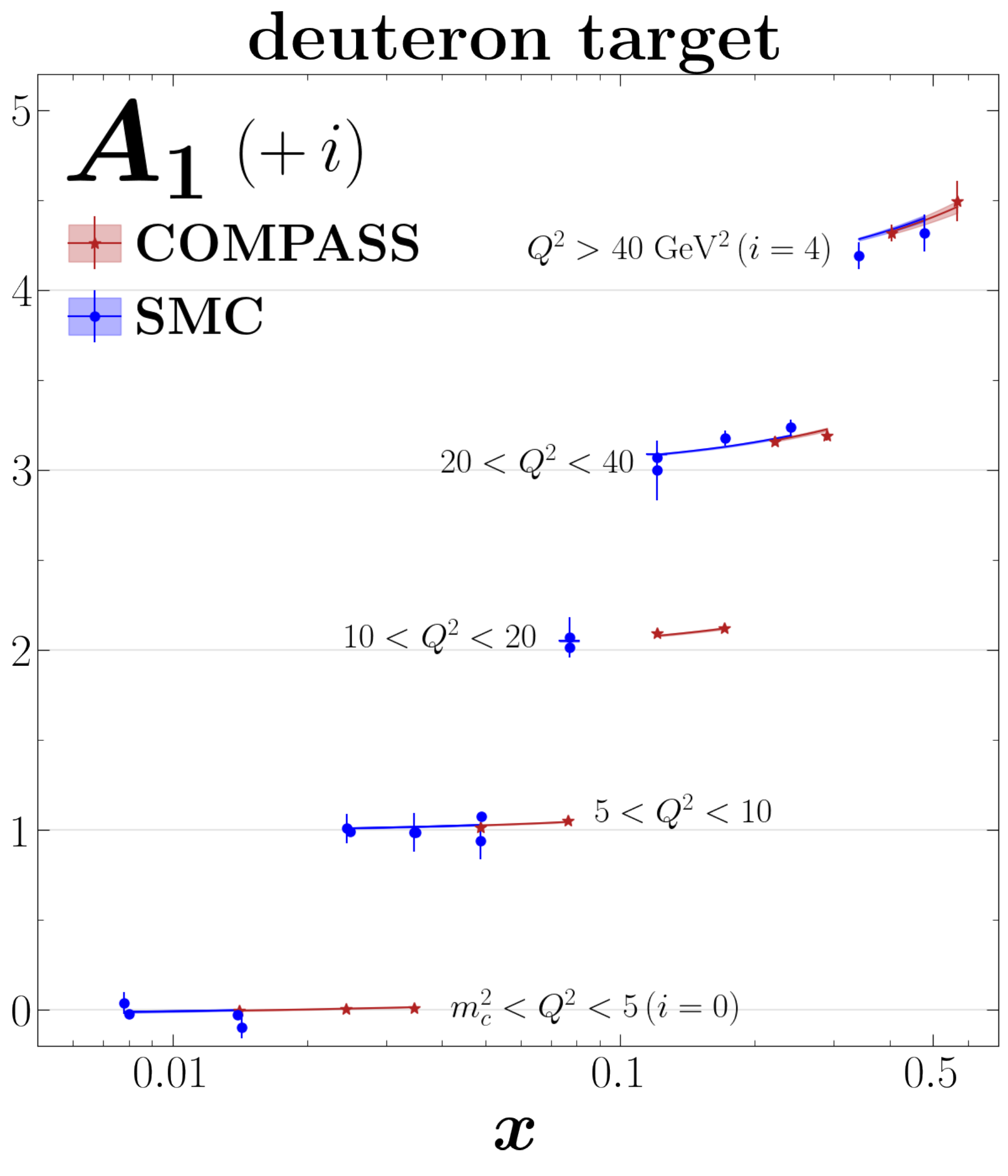}
\caption
{Polarized DIS data for the deuteron $A_1$ asymmetry from COMPASS \cite{COMPASS:2006mhr} (red) and SMC \cite{SpinMuon:1998eqa, SpinMuon:1999udj} (blue) plotted as a function of Bjorken $x$ against the JAM result (colored lines and 1$\sigma$ bands). Data in different $Q^2$ bins are increased by $i$ for clarity.
}
\label{fig:pidis_A1_deuteron}
\end{figure*}

\begin{figure*}[b]
\centering
\includegraphics[width=0.9\textwidth]{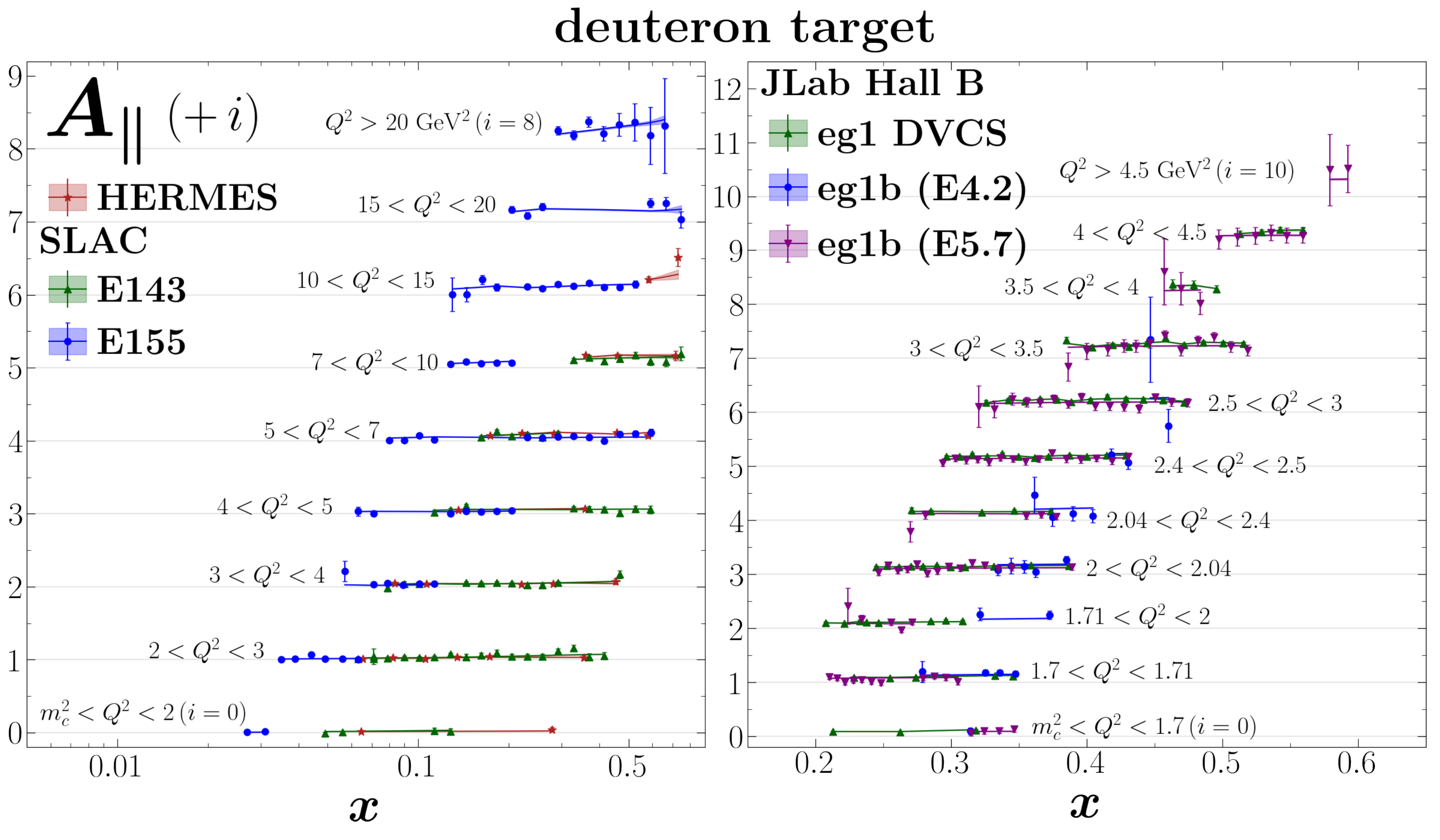}
\caption
{Polarized DIS data for the deuteron observable $A_{\parallel}$ from from HERMES \cite{HERMES:2006jyl}, SLAC \cite{E143:1998hbs, E155:1999pwm}, and Jefferson Lab \cite{CLAS:2014qtg, CLAS:2015otq} plotted as a function of $x$ against the JAM result (colored lines and 1$\sigma$ bands). The left panel shows the results from HERMES (red) and SLAC at energies 143 GeV (green) and 155 GeV (blue).  The right panel shows results for the JLab Hall B DVCS experiment (green) and eg1b experiment at energies 4.2 GeV (blue) and 5.7 GeV (purple). Data in different $Q^2$ bins are increased by $i$ for clarity.
}
\label{fig:pidis_Apa_deuteron}
\end{figure*}

\begin{figure*}[h]
\centering
\includegraphics[width=0.9\textwidth]{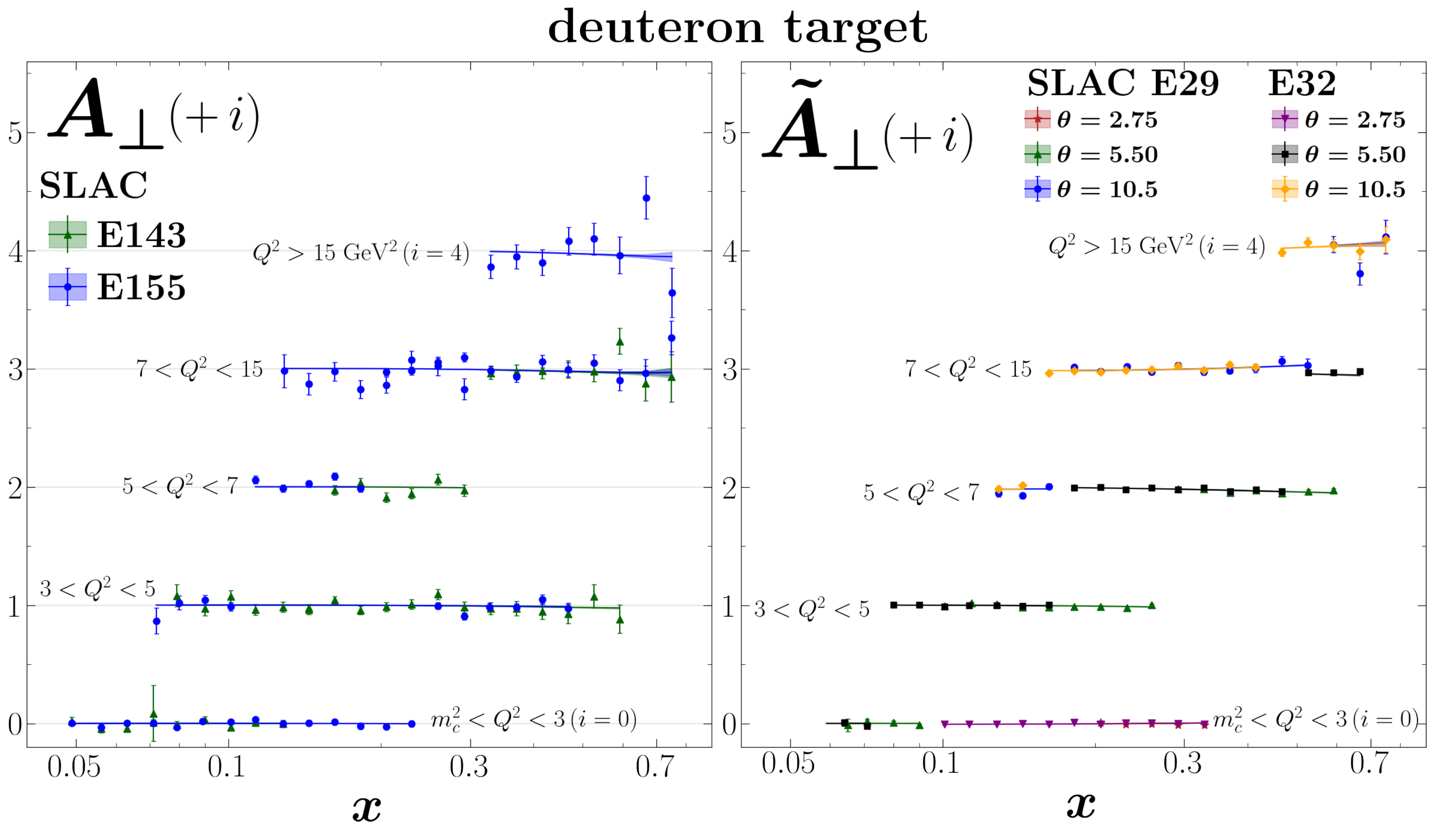}
\caption
{Polarized DIS data for the deuteron observables $A_\perp$ (left panel) and $\widetilde{A}_\perp$ (right panel) from SLAC~\cite{E143:1998hbs, E155:1999eug, E155:2002iec} plotted as a function of Bjorken $x$ against the JAM result (colored lines and 1$\sigma$ bands). On the left panel, results are shown at energies 143~GeV (green) and 155~GeV (blue).  On the right panel, results are shown at energies 29~GeV for $\theta = 2.75^\circ, 5.50^\circ, 10.5^\circ$ (red, green, and blue, respectively) and at 32~GeV for $\theta = 2.75^\circ, 5.50^\circ, 10.5^\circ$ (purple, black, and orange, respectively).  Data in different $Q^2$ bins are increased by $i$ for clarity.
}
\label{fig:pidis_Ape_deuteron}
\end{figure*}

\begin{figure*}[h]
\centering
\includegraphics[width=0.9\textwidth]{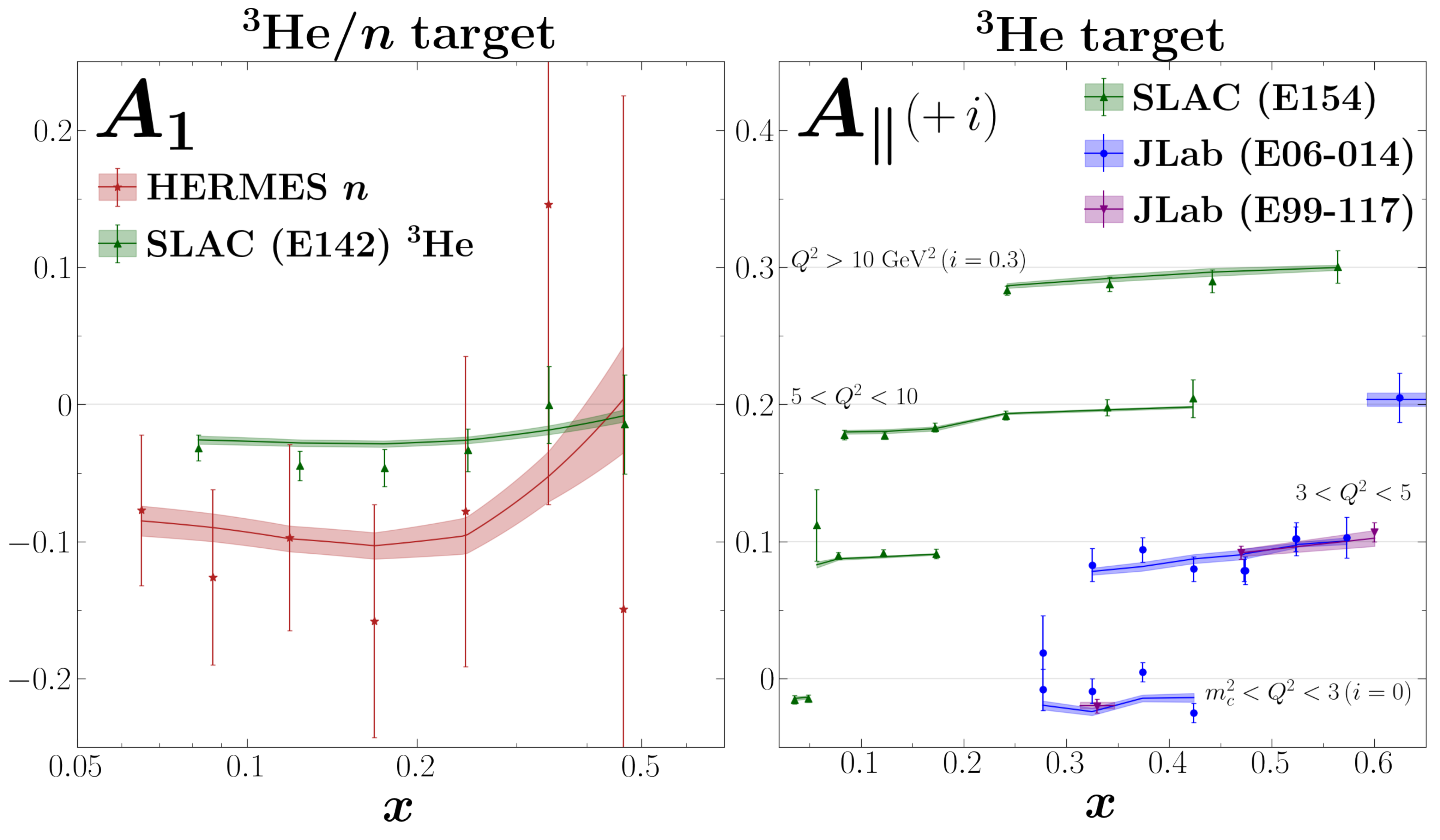}
\caption
{Polarized DIS data for the helium and neutron observable $A_1$ (left panel) and helium $A_{\parallel}$ (right panel) from HERMES~\cite{HERMES:1997hjr}, SLAC~\cite{E142:1996thl, E154:1997xfa}, and Jefferson Lab \cite{JeffersonLabHallA:2004tea, JeffersonLabHallA:2016neg} plotted as a function of Bjorken $x$ against the JAM result (colored lines 1$\sigma$ bands). The left panel shows results for HERMES on a neutron target (red) and SLAC on a helium target (green).  The right panel shows results from SLAC (green) and Jefferson Lab E06-014 (blue) and E99-117 (purple). Data in different $Q^2$ bins are increased by $i$ for clarity.
}
\label{fig:pidis_A1_Apa_helium}
\end{figure*}

\begin{figure*}[h]
\centering
\includegraphics[width=0.9\textwidth]{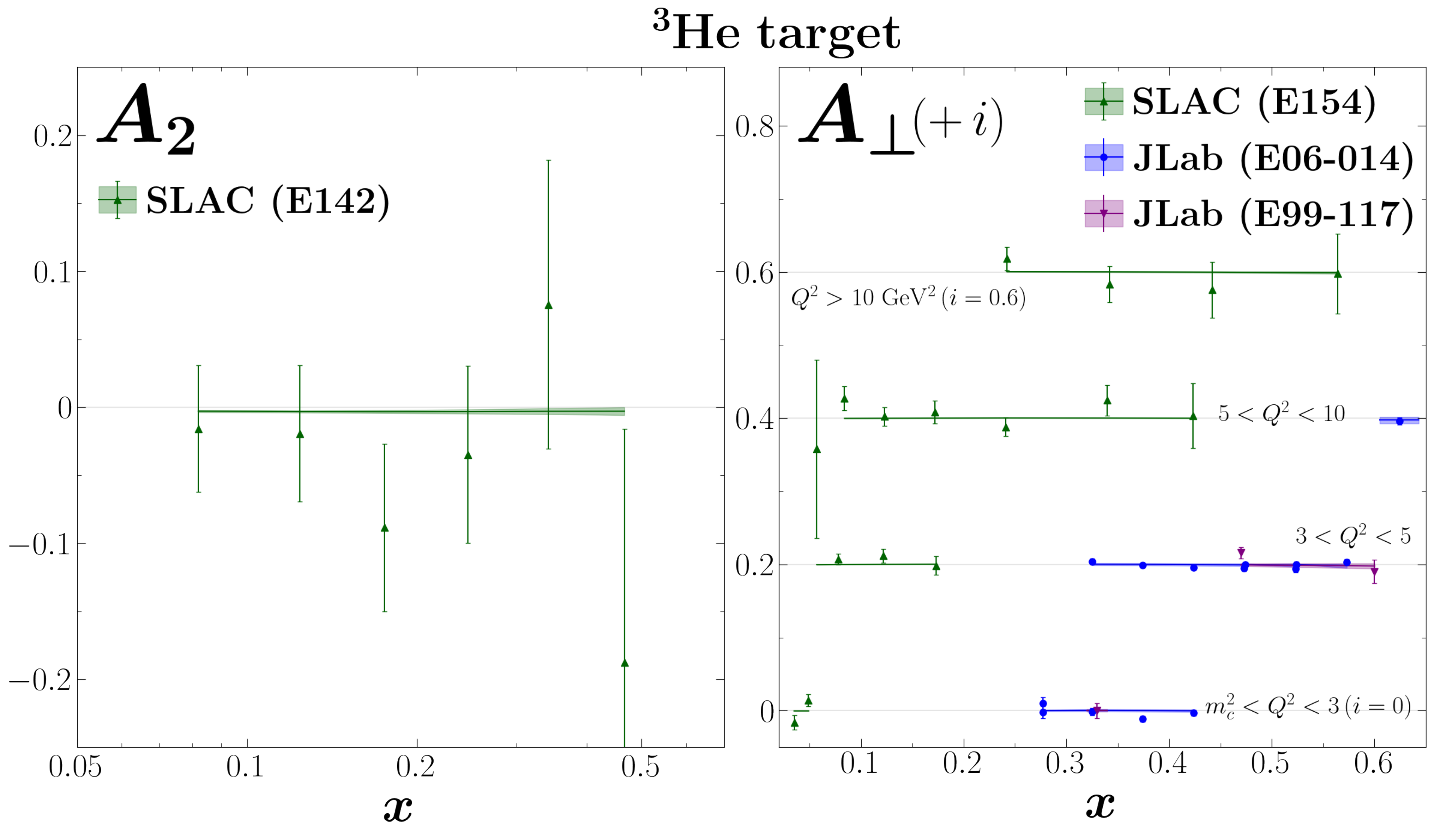}
\caption
{Polarized DIS data for the helium observables $A_2$ (left panel) and $A_{\perp}$ (right panel) from SLAC~\cite{E142:1996thl, E154:1997xfa} and Jefferson Lab~\cite{JeffersonLabHallA:2004tea, JeffersonLabHallA:2016neg} plotted as a function of Bjorken $x$ against the JAM result (colored lines and 1$\sigma$ bands). The left panel shows results from SLAC (green). The right panel shows results from the SLAC E154 (green) and Jefferson Lab E06-014 (blue) and E99-117 (purple) experiments. Data in different $Q^2$ bins are increased by $i$, in increments of $i=0.2$, for clarity.}
\label{fig:pidis_A2_Ape_helium}
\end{figure*}

Comparisons between the data from the individual experiments and the fitted values are shown in Figs.~\ref{fig:pidis_A1_proton}--\ref{fig:pidis_A2_Ape_helium}.
The most basic polarization observable, the inclusive DIS asymmetry $A_1$, is shown in Fig.~\ref{fig:pidis_A1_proton} for the proton data from EMC~\cite{EuropeanMuon:1989yki}, SMC~\cite{SpinMuon:1998eqa, SpinMuon:1999udj}, and COMPASS~\cite{COMPASS:2010wkz, COMPASS:2015rvb} experiments, along with the new lower-$Q^2$ data from the SANE experiment at Jefferson Lab~\cite{SANE:2018pwx}, which probed deeper into the high-$x$ region. 
Good agreement is obtained for all the datasets, although the theory predictions for the SANE data tend to be slightly lower than the observed points, particularly for the $E=4.7$~GeV data.
Data on the parallel asymmetry $A_\parallel$ for the proton are shown in Fig.~\ref{fig:pidis_Apa_proton} from HERMES~\cite{HERMES:2006jyl}, SLAC~\cite{Baum:1983ha, E143:1998hbs, E155:2000qdr}, and Jefferson Lab~\cite{CLAS:2014qtg, CLAS:2017qga} experiments.
These data are generally all described well.

The transverse $A_2$ asymmetry for the proton is shown in Fig.~\ref{fig:pidis_A2_proton} from the HERMES~\cite{HERMES:2011xgd} experiment and the more recent SANE data from Hall~C at Jefferson Lab~\cite{SANE:2018pwx}.
Unlike for the longitudinal $A_1$ asymmetry in Fig.~\ref{fig:pidis_A1_proton}, our fit goes through the SANE data and shows no indication of underestimating the data with any consistency.
The comparison with the transverse proton DIS asymmetries $A_\perp$ and $\widetilde{A}_\perp$ from SLAC~\cite{E143:1998hbs, E155:1999eug, E155:2002iec} are shown in Fig.~\ref{fig:pidis_Ape_proton}.
While the vast majority of points are fitted well, one datum in the SLAC E155 $A_\perp$ dataset at the very highest $x$ and $Q^2$ values is anomalously higher than our fit.

The polarized DIS data for the deuteron $A_1$ asymmetry from COMPASS~\cite{COMPASS:2006mhr} and SMC~\cite{SpinMuon:1998eqa, SpinMuon:1999udj} are shown in Fig.~\ref{fig:pidis_A1_deuteron}.
The COMPASS data are fitted better than the SMC data for this observable, with some SMC points at the same $x$ value are slightly separated from the theory prediction.
For the polarized deuteron longitudinal asymmetry $A_\parallel$, Fig.~\ref{fig:pidis_Apa_deuteron} shows the comparison with data from HERMES~\cite{HERMES:2006jyl}, SLAC~\cite{E143:1998hbs, E155:1999pwm}, and CLAS at Jefferson Lab~\cite{CLAS:2014qtg, CLAS:2015otq}.
The data for the transversely polarized deuteron DIS asymmetries $A_\perp$ and $\widetilde{A}_\perp$ are shown in Fig.~\ref{fig:pidis_Ape_deuteron} from the SLAC E143~\cite{E143:1998hbs} and E155~\cite{E155:1999eug, E155:2002iec} experiments.

For polarized $^3$He (and neutron) targets, Fig.~\ref{fig:pidis_A1_Apa_helium} shows DIS data for the $A_1$ and $A_\parallel$ asymmetries from HERMES~\cite{HERMES:1997hjr}, SLAC~\cite{E142:1996thl, E154:1997xfa}, and Jefferson Lab Hall~A~\cite{JeffersonLabHallA:2004tea, JeffersonLabHallA:2016neg}.
For the HERMES data the $^3$He asymmetry was converted, assuming a model $^3$He wave function, to an asymmetry on the free neutron~\cite{HERMES:1997hjr}.
The fit generally gives a good description of the $\hel$ data.
Good description are also found for the $^3$He nuclear $A_2$ and $A_\perp$ asymmetries in Fig.~\ref{fig:pidis_A2_Ape_helium} for  SLAC~\cite{E142:1996thl, E154:1997xfa} and Jefferson Lab~\cite{JeffersonLabHallA:2004tea, JeffersonLabHallA:2016neg} data.

\begin{figure*}[h]
\centering
\hspace*{-0.2cm}
\includegraphics[width=0.95\textwidth]{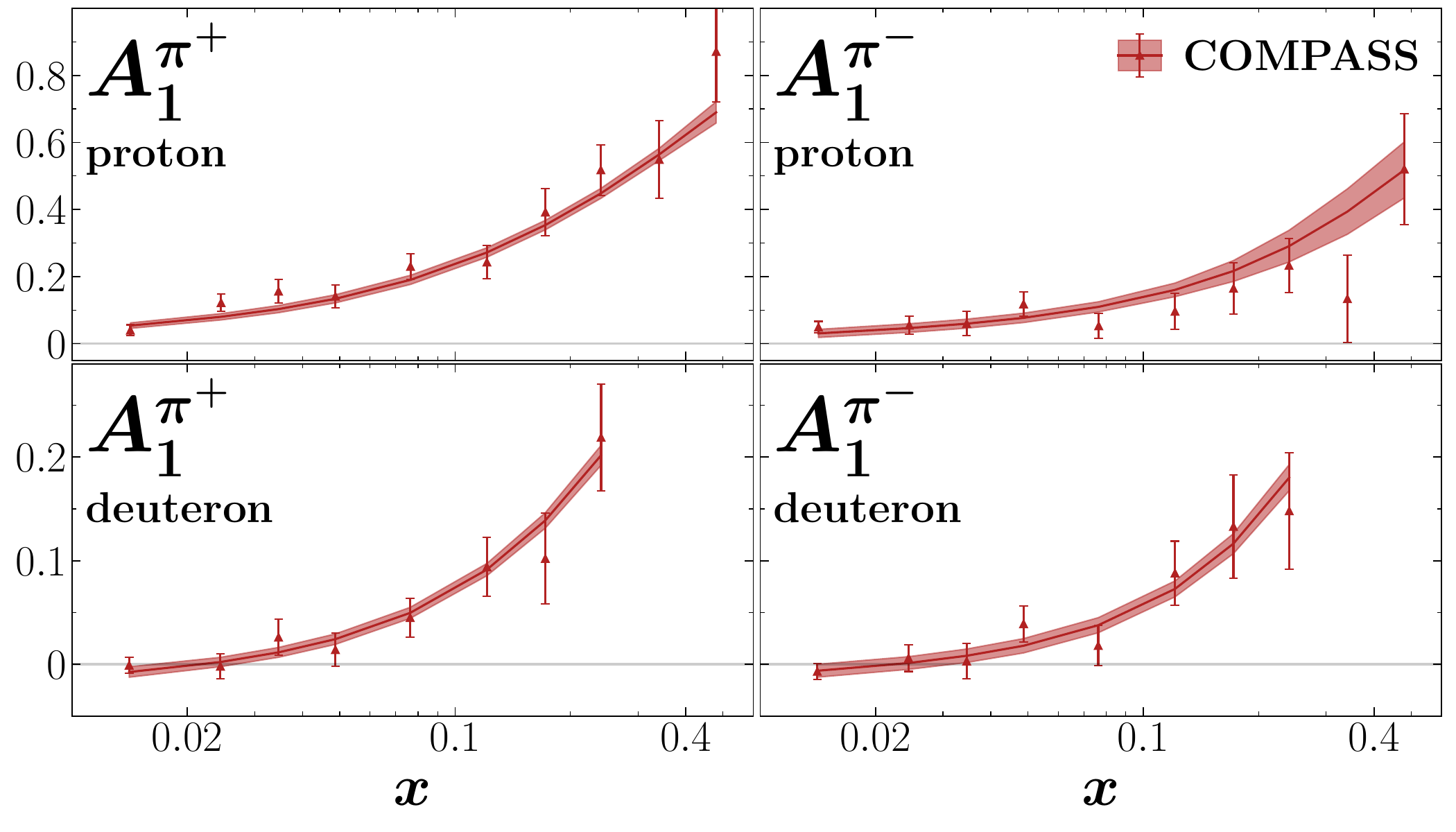}
\vspace*{-0.4cm}
\caption{Polarized SIDIS asymmetries $A_1^{\pi^{\pm}}$ from COMPASS \cite{COMPASS:2009kiy, COMPASS:2010hwr} (red) plotted as a function of Bjorken $x$ and compared to the JAM fit (colored lines and 1$\sigma$ bands).
}
\label{fig:psidis_pion}
\end{figure*}

\begin{figure*}[h]
\centering
\hspace*{-0.2cm}
\includegraphics[width=0.95\textwidth]{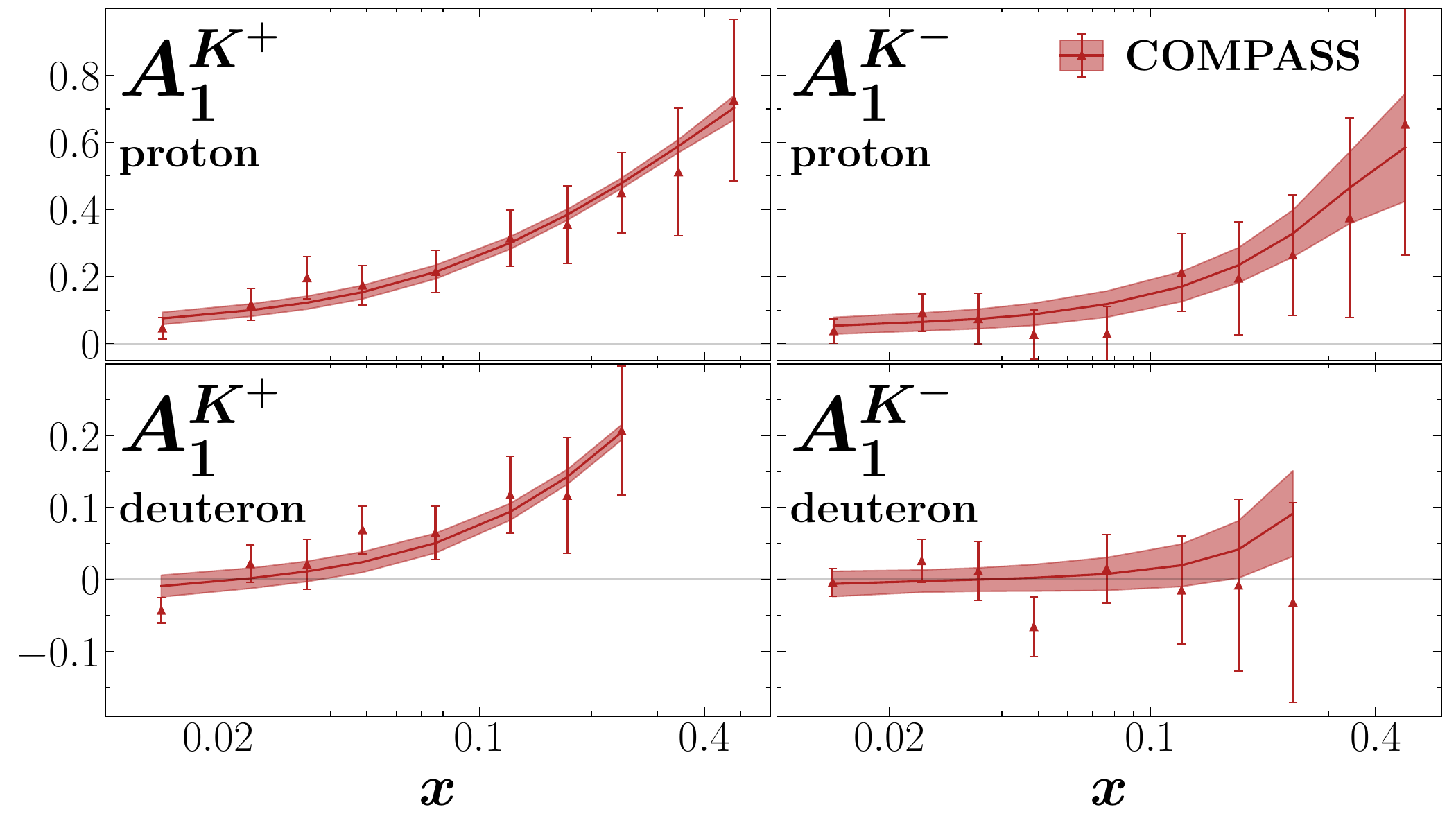}
\vspace*{-0.4cm}
\caption{Polarized SIDIS asymmetries $A_1^{K^\pm}$ from COMPASS \cite{COMPASS:2009kiy, COMPASS:2010hwr} (red) plotted as a function of Bjorken $x$ and compared to the JAM fit (colored lines and 1$\sigma$ bands).
}
\label{fig:psidis_kaon}
\end{figure*}

\begin{figure*}[h]
\centering
\hspace*{-0.2cm}
\includegraphics[width=0.95\textwidth]{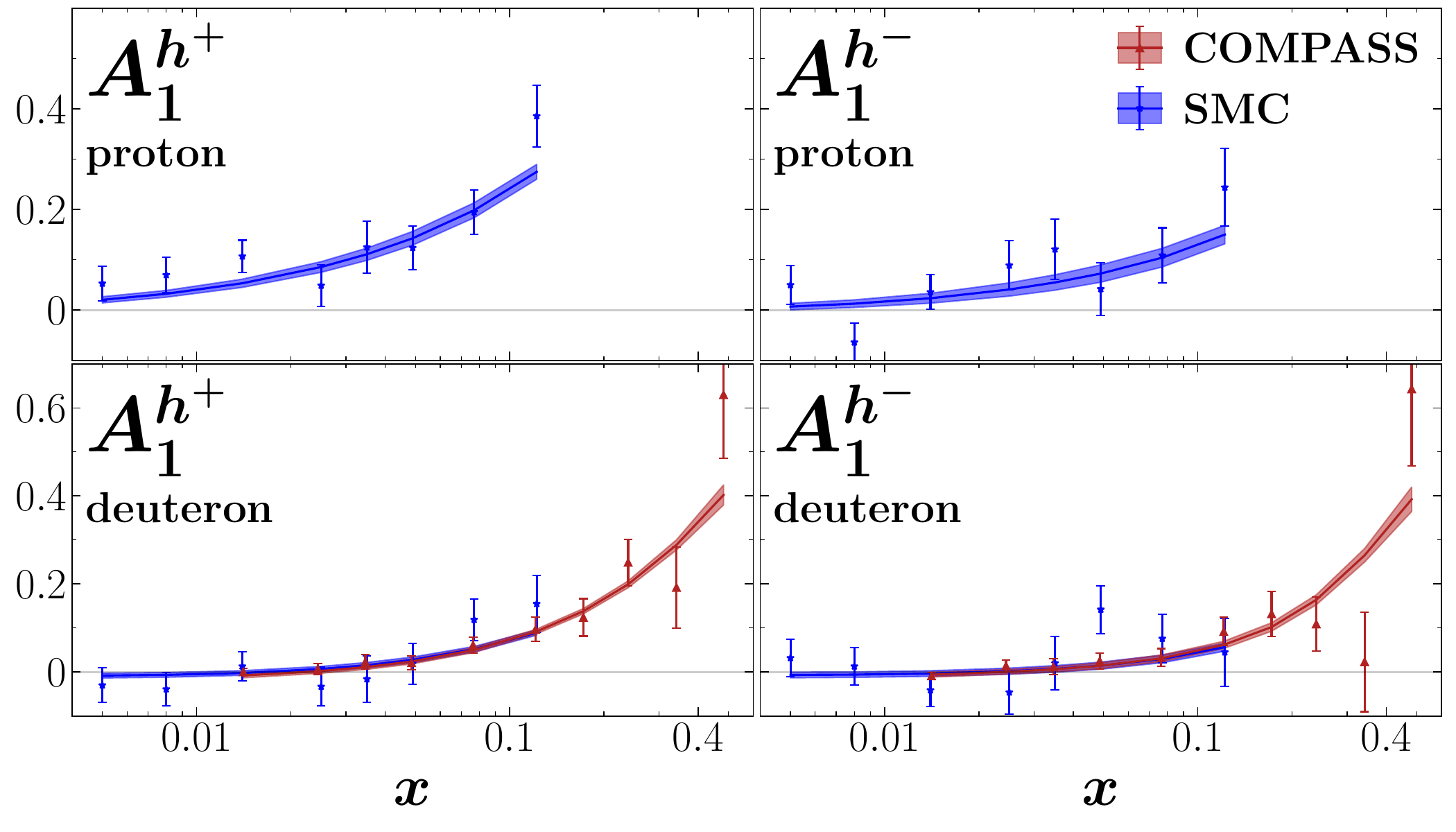}
\caption{Polarized SIDIS asymmetries $A_1^{h^{\pm}}$ from COMPASS \cite{COMPASS:2009kiy} (red) and SMC \cite{SpinMuon:1997yns} (blue) as a function of Bjorken $x$ and compared to the JAM fit (colored lines and 1$\sigma$ bands).}
\label{fig:psidis_hadron}
\end{figure*}

Turning now to the polarized SIDIS asymmetries $A_1^h$, in Fig.~\ref{fig:psidis_pion} we show the data versus theory comparison for the production of charged pions, $h = \pi^\pm$, for $p$ and $d$ data from COMPASS~\cite{COMPASS:2009kiy, COMPASS:2010hwr}.
The agreement with data is relatively good across all panels.
The polarized SIDIS kaon production asymmetries, $A_1^{K^\pm}$, are shown in Fig.~\ref{fig:psidis_kaon} for COMPASS~\cite{COMPASS:2009kiy, COMPASS:2010hwr} data on proton and deuteron targets. 
Because of the large uncertainties bands, these datasets are rather easy to fit, which often results in $\chired$ well below $1$.
For the production of unidentified hadrons $h^\pm$, Fig.~\ref{fig:psidis_hadron} shows polarized SIDIS asymmetries $A_1^{h^\pm}$ from COMPASS~\cite{COMPASS:2009kiy} and SMC~\cite{SpinMuon:1997yns}.
Again the agreement between the data and the fitted values is quite good, with both reflecting the increasing numerical values of the asymmetries in the high-$x$ region.

\begin{figure*}[h]
\centering
\hspace*{-0.2cm}
\includegraphics[width=1.00\textwidth]{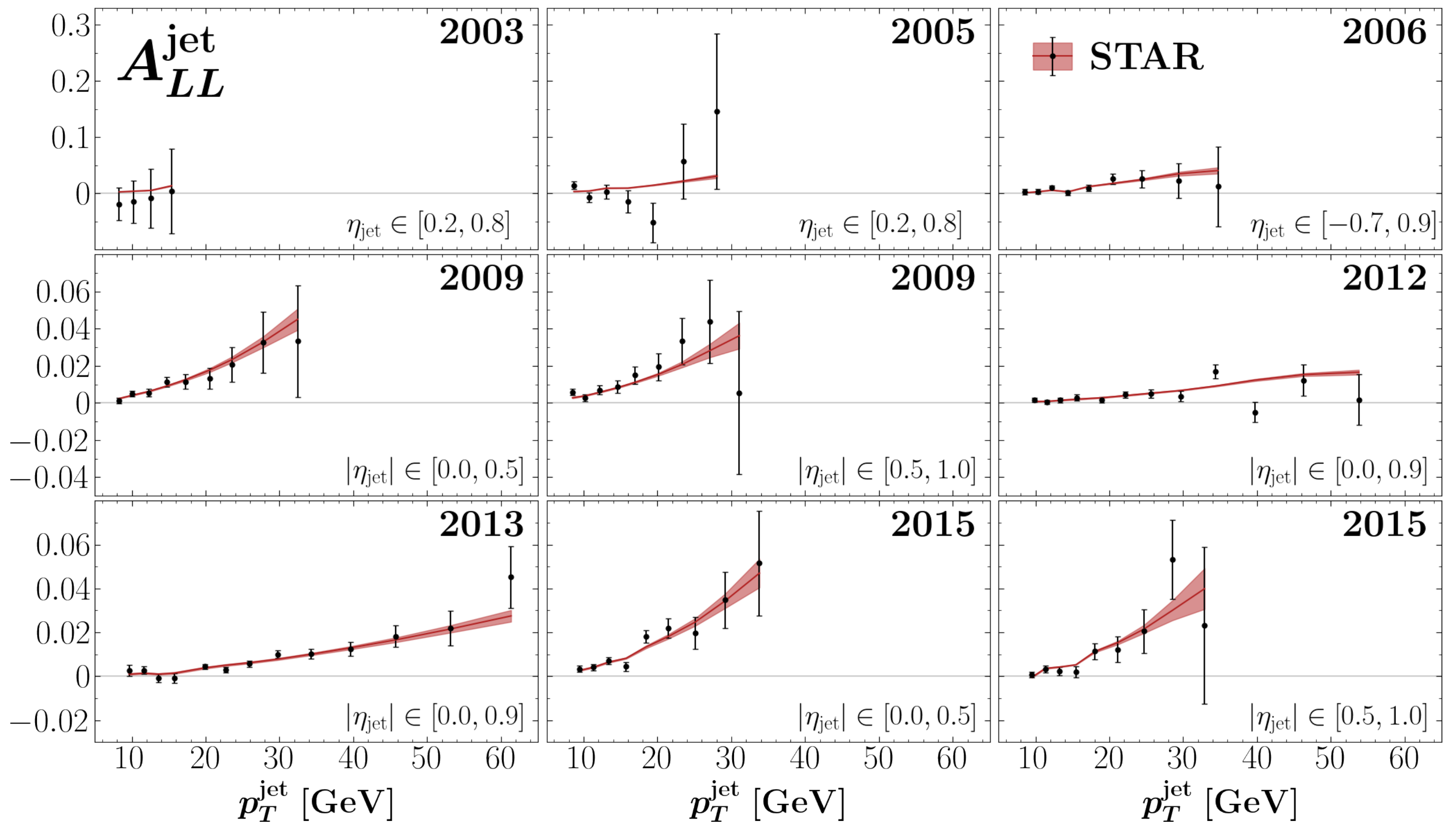}
\caption{Polarized double longitudinal jet asymmetries $A_{LL}^{\rm jet}$ from STAR \cite{STAR:2006opb,STAR:2007rjc,STAR:2012hth,STAR:2014wox,STAR:2019yqm,STAR:2021mfd,STAR:2021mqa} (black circles) plotted as a function of $p_T^{\rm jet}$ and compared to the JAM fit (red lines and 1$\sigma$ bands).  Each subplot shows the year when the data was taken and the pseudorapidity bins $\eta_{\rm jet}$ or $|\eta_{\rm jet}|$.}
\label{fig:pjets}
\end{figure*}

For polarized $pp$ scattering data, Fig.~\ref{fig:pjets} shows the longitudinally polarized jet asymmetries $A_{LL}^{\rm jet}$ at $\sqrt{s} = 200$ and 510~GeV from the STAR Collaboration~\cite{STAR:2006opb, STAR:2007rjc, STAR:2012hth, STAR:2014wox, STAR:2019yqm, STAR:2021mfd, STAR:2021mqa} at RHIC.
These data are also fitted quite well, with only a single datum in the mid-range of the 2005 dataset falling outside of the theoretical uncertainty bands.
The $\chi^2_{\rm red}$ values are generally below unity, with only the STAR data from Refs.~\cite{STAR:2019yqm, STAR:2007rjc} giving a $\chi^2_{\rm red}$ exceeding~1.

\begin{figure}[t]
\includegraphics[width=0.95\textwidth]{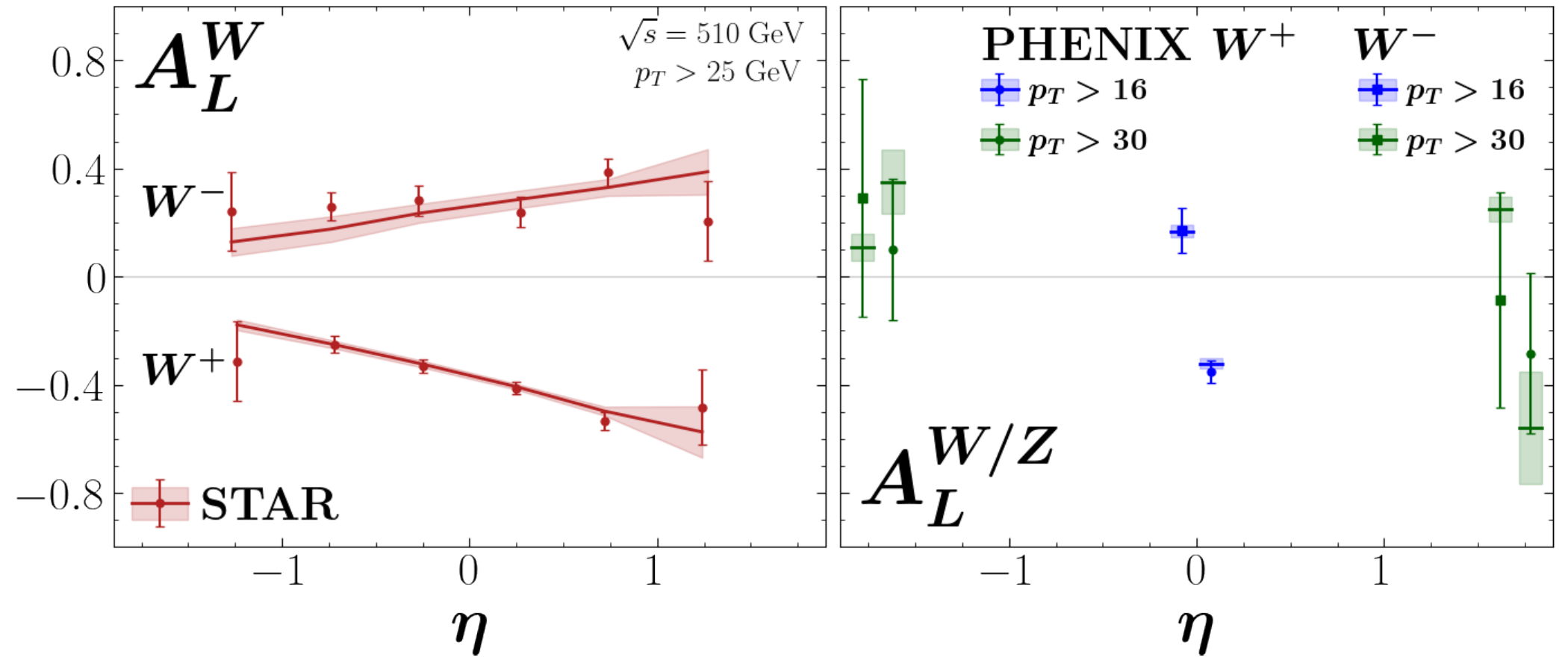}
\vspace*{-0.5cm}
\caption{Single spin asymmetries $A_L^W$ and $A_L^{W/Z}$ versus lepton pseudorapidity $\eta$.
{\it Left panel}: Asymmetries from STAR~\cite{STAR:2018fty} (red) at $\sqrt{s} = 510$~GeV and integrated over $p_T > 25$~GeV, compared with the full JAM fit (red solid lines and $1\sigma$ uncertainty bands).
{\it Right panel}: Asymmetries from PHENIX~\cite{PHENIX:2015ade, PHENIX:2018wuz} at $\sqrt{s} = 510$~GeV and integrated over $p_T > 16$~GeV (blue points) or $p_T > 30$~GeV (purple points), compared with the full JAM fit (red points). The $W^+/Z$ asymmetries are shown with circles, while the $W^-/Z$ asymmetries are shown with squares.
}
\label{fig:polWlep}
\end{figure}

The single spin asymmetries $A_L^W$ and $A_L^{W/Z}$ for vector boson production in longitudinally polarized $pp$ scattering at $\sqrt{s}=510$~GeV are shown in Fig.~\ref{fig:polWlep}.
The data from the STAR~\cite{STAR:2018fty} and PHENIX~\cite{PHENIX:2015ade, PHENIX:2018wuz} Collaborations, integrated over specific $p_T$ ranges, are well described by our fitted results.

\begin{figure}[t]
\includegraphics[width=0.65\textwidth]{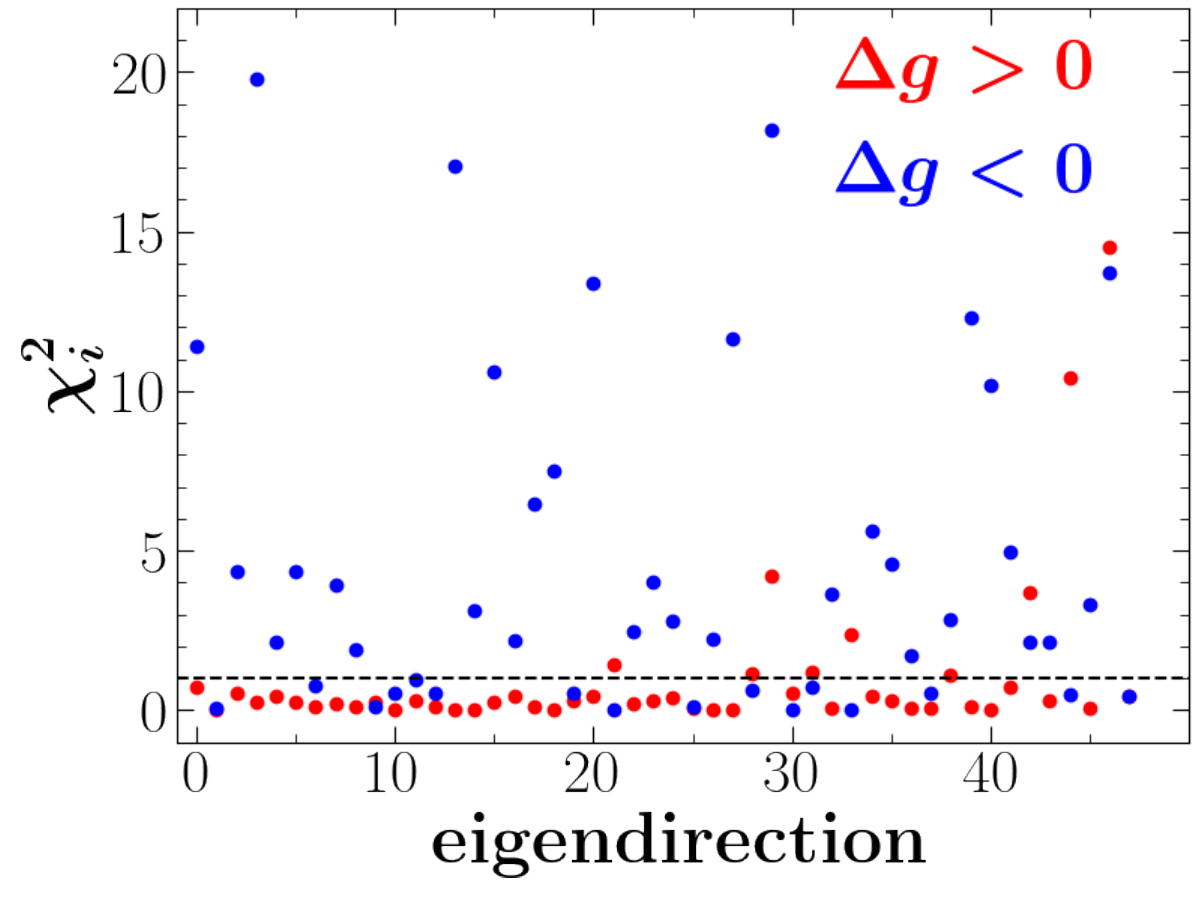}
\vspace*{-0.5cm}
\caption{Comparison of $\chi^2$ contributions for positive (red dots) and negative (blue dots) $\Delta g$ mean theory predictions for lattice QCD data versus eigendirection; $\chi^2_i$ represents the $i$-th lattice data point.
}
\label{fig:lattice} 
\end{figure}

Finally, in Fig.~\ref{fig:lattice} we show the contributions to the total $\chi^2$ from each eigendirection of the covariance matrix.
Here, for completeness we also include the results for the negative $\Delta g$ solutions separately.
Since for the lattice QCD data we only have access to the covariance matrix between the 48 lattice points rather than individual uncertainties on each point, we rotate the residuals in the eigenspace of the covariance matrix~\cite{Karpie:2023nyg}. 
Compared with the baseline + lattice QCD scenario, there is clearly a larger disagreement between the theory predictions and the lattice QCD data for the $\Delta g < 0$ solutions, across a majority of data points.  This justifies removing the negative $\Delta g$ solutions from our analysis.

\begin{table}[p]
\footnotesize
\caption{Reduced $\chi^2$ and $Z$-score values for longitudinal DIS polarization asymmetries $A_1$ and \Apa\ for $p$, $d$, and $^3$He (or ``neutron'') targets. The number of data points ($N_{\rm dat}$) corresponds to cuts of \mbox{$W^2 > 4$~GeV$^2$} and $Q^2 > m_c^2$ on the DIS data. With a cut of $W^2 > 3.5$~GeV$^2$, this increases to 1516 data points. Results are shown for the full JAMpol25 fit including TMC+HT contributions and for the LT only analysis.}
\resizebox{14cm}{!} 
{\begin{tabular}{ l c c r | c | c | c}
\hhline{=======}
& & & & \textbf{JAMpol25} & LT only & ~$W^2 > 3.5$ GeV$^2$~ \\
~Experiment & Target~ & ~Observable~ & ~$N_{\rm dat}$~ & ~~$\chired$ ($Z$-score)~ & ~~$\chired$ ($Z$-score)~ & ~~$\chired$ ($Z$-score)~ \\
\hline
~EMC \cite{EuropeanMuon:1989yki}      & $p$    & $A_1$          & 10~~  &~0.29 ($-2.11$) & 0.35 ($-1.82$)  & 0.29 ($-2.16$) \\
~SMC \cite{SpinMuon:1998eqa}          & $p$    & $A_1$          & 11~~  &~0.40 ($-1.73$) & 0.46 ($-1.44$)  & 0.39 ($-1.78$) \\
~SMC \cite{SpinMuon:1998eqa}          & $d$    & $A_1$          & 11~~  &~1.70 ($+1.51$) & 1.73 ($+1.55$)  & 1.77 ($+1.61$) \\
~SMC \cite{SpinMuon:1999udj}          & $p$    & $A_1$          &  7~~  &~1.44 ($+0.90$) & 1.43 ($+0.89$)  & 1.41 ($+0.86$) \\
~SMC \cite{SpinMuon:1999udj}          & $d$    & $A_1$          &  7~~  &~0.67 ($-0.52$) & 0.65 ($-0.57$)  & 0.67 ($-0.51$) \\
~COMPASS \cite{COMPASS:2010wkz}       & $p$    & $A_1$          & 11~~  &~0.96 ($+0.04$) & 1.21 ($+0.60$)  & 0.90 ($-0.11$) \\
~COMPASS \cite{COMPASS:2006mhr}       & $d$    & $A_1$          & 11~~  &~0.61 ($-0.91$) & 0.55 ($-1.12$)  & 0.55 ($-1.12$) \\
~COMPASS \cite{COMPASS:2015rvb}       & $p$    & $A_1$          & 35~~  &~0.92 ($-0.27$) & 0.85 ($-0.56$)  & 0.94 ($-0.17$) \\
~SLAC E80/E130 \cite{Baum:1983ha}     & $p$    & \Apa           & 19~~  &~0.59 ($-1.40$) & 0.63 ($-1.21$)  & 0.55 ($-1.59$) \\
~SLAC E142 \cite{E142:1996thl}        & $\hel$ & $A_1$          &  6~~  &~0.86 ($-0.05$) & 0.69 ($-0.42$)  & 0.87 ($-0.03$) \\
~SLAC E143 \cite{E143:1998hbs}        & $p$    & \Apa           & 56~~  &~0.87 ($-0.68$) & 0.87 ($-0.65$)  & 0.91 ($-0.42$) \\
~SLAC E143 \cite{E143:1998hbs}        & $d$    & \Apa           & 56~~  &~1.08 ($+0.47$) & 1.10 ($+0.59$)  & 1.07 ($+0.42$) \\
~SLAC E154 \cite{E154:1997xfa}        & $\hel$ & \Apa           & 16~~  &~0.48 ($-1.73$) & 0.44 ($-1.93$)  & 0.45 ($-1.87$) \\
~SLAC E155 \cite{E155:2000qdr}        & $p$    & \Apa           & 65~~  &~1.21 ($+1.16$) & 1.47 ($+2.41$)  & 1.20 ($+1.15$) \\
~SLAC E155 \cite{E155:1999pwm}        & $d$    & \Apa           & 65~~  &~1.01 ($+0.11$) & 1.03 ($+0.22$)  & 0.99 ($+0.01$) \\
~HERMES \cite{HERMES:1997hjr}         & $``n"$ & $A_1$          &  7~~  &~0.25 ($-1.90$) & 0.26 ($-1.87$)  & 0.25 ($-1.91$) \\
~HERMES \cite{HERMES:2006jyl}         & $p$    & \Apa           & 25~~  &~0.55 ($-1.80$) & 0.58 ($-1.64$)  & 0.51 ($-2.16$) \\
~HERMES \cite{HERMES:2006jyl}         & $d$    & \Apa           & 25~~  &~0.95 ($-0.08$) & 1.10 ($+0.44$)  & 0.94 ($-0.17$) \\
~JLab E99-117 \cite{JeffersonLabHallA:2004tea} & $\hel$ & \Apa  &  3~~  &~0.15 ($-1.49$) & 0.22 ($-1.17$)  & 0.12 ($-1.60$) \\
~JLab E06-014 \cite{JeffersonLabHallA:2016neg} & $\hel$ & \Apa  & 14~~  &~1.75 ($+1.75$) & 2.04 ($+2.26$)  & 1.47 ($+1.28$) \\
~JLab eg1-dvcs \cite{CLAS:2014qtg}    & $p$    & \Apa           & 117~~ &~1.66 ($+4.29$) & 1.66 ($+4.27$)  & 2.29 ($+7.39$) \\
~JLab eg1-dvcs \cite{CLAS:2014qtg}    & $d$    & \Apa           & 73~~  &~0.36 ($-5.21$) & 1.33 ($+1.86$)  & 0.37 ($-5.74$) \\
~JLab eg1b ($E\!=\!4.2$ GeV) \cite{CLAS:2017qga}   & $p$    & \Apa  & 88~~  &~1.22 ($+1.43$) & 1.47 ($+2.75$)  & 1.51 ($+3.72$) \\
~JLab eg1b ($E\!=\!5.7$ GeV) \cite{CLAS:2017qga}   & $p$    & \Apa  & 385~~ &~0.95 ($-0.66$) & 1.30 ($+3.86$)  & 0.96 ($-0.65$) \\
~JLab eg1b ($E\!=\!4.2$ GeV) \cite{CLAS:2015otq}   & $d$    & \Apa  & 20~~  &~0.89 ($-0.26$) & 0.84 ($-0.42$)  & 1.07 ($+0.37$) \\
~JLab eg1b ($E\!=\!5.7$ GeV) \cite{CLAS:2015otq}   & $d$    & \Apa  & 92~~  &~0.84 ($-1.09$) & 0.88 ($-0.76$)  & 0.82 ($-1.42$) \\
~JLab SANE ($E\!=\!4.7$ GeV)\cite{SANE:2018pwx}    & $p$    & $A_1$ & 12~~  &~1.99 ($+2.03$) & 1.54 ($+1.27$)  & 2.79 ($+3.81$) \\
~JLab SANE ($E\!=\!5.9$ GeV) \cite{SANE:2018pwx}   & $p$    & $A_1$ & 18~~  &~0.87 ($-0.29$) & 0.80 ($-0.52$)  & 0.93 ($-0.12$) \\
\hline
~Total & & & 1265~~ & ~{\bf 1.01 (\boldmath$+0.26$)} & 1.20 ($+4.74$) & 1.11 ($+2.94$) \\
\hhline{=======}
\end{tabular} }
\label{tab:chi2_Apa}
\end{table}

\begin{table}[b]
\footnotesize
\caption{Reduced $\chi^2$ and $Z$-score values for transverse DIS polarization asymmetries $A_2$, \Ape, and \Atpe\ for $p$, $d$, and $^3$He targets. The number of data points ($N_{\rm dat}$) corresponds to a cuts of \mbox{$W^2 > 4$~GeV$^2$} and $Q^2 > m_c^2$ on the DIS data. With a cut of $W^2 > 3.5$ GeV$^2$, this increases to 486 data points.  Results are shown for the full JAMpol25 fit including TMC+HT contributions and for the LT only analysis.\\}
\begin{tabular}{l c c r | c | c| c }
\hhline{=======}
& & & & \textbf{JAMpol25} & LT only & ~$W^2 > 3.5$ GeV$^2$ \\
~Experiment & Target~ & ~Obs.~ & ~$N_{\rm dat}$ & ~$\chired$ ($Z$-score)~ & ~$\chired$ ($Z$-score)~  & ~$\chired$ ($Z$-score) \\
\hline
~SLAC E142 \cite{E142:1996thl}  & $\hel$ & $A_2$                 &  6~~  & ~0.67 ($-0.45$) & 0.68 ($-0.43$) & 0.67 ($-0.44$)  \\
~SLAC E143 \cite{E143:1998hbs}  & $p$    & \Ape                  & 43~~  & ~1.05 ($+0.29$) & 1.05 ($+0.31$) & 1.06 ($+0.36$)  \\
~SLAC E143 \cite{E143:1998hbs}  & $d$    & \Ape                  & 43~~  & ~0.94 ($-0.23$) & 0.93 ($-0.24$) & 0.94 ($-0.22$)  \\
~SLAC E154 \cite{E154:1997xfa}  & $\hel$ & \Ape                  & 16~~  & ~1.11 ($+0.42$) & 1.11 ($+0.41$) & 1.11 ($+0.42$)  \\
~SLAC E155 \cite{E155:1999eug}  & $p$    & \Ape                  & 58~~  & ~0.95 ($-0.21$) & 0.96 ($-0.18$) & 0.94 ($-0.25$)  \\
~SLAC E155 \cite{E155:1999eug}  & $d$    & \Ape                  & 58~~  & ~1.55 ($+2.60$) & 1.55 ($+2.60$) & 1.53 ($+2.53$)  \\
~SLAC E155x \cite{E155:2002iec} & $p$    & \Atpe                 & 92~~  & ~1.36 ($+2.22$) & 1.41 ($+2.50$) & 1.39 ($+2.43$)  \\
~SLAC E155x \cite{E155:2002iec} & $d$    & \Atpe                 & 92~~  & ~0.80 ($-1.41$) & 0.79 ($-1.46$) & 0.80 ($-1.45$)  \\
~HERMES \cite{HERMES:2011xgd}   & $p$    & $A_2$                 & 15~~  & ~1.01 ($+0.16$) & 0.99 ($+0.09$) & 1.24 ($+0.74$)  \\
~JLab E99-117 \cite{JeffersonLabHallA:2004tea}  & $\hel$ & \Ape  &  3~~  & ~1.49 ($+0.79$) & 1.15 ($+0.44$) & 1.48 ($+0.78$)  \\
~JLab E06-014 \cite{JeffersonLabHallA:2016neg}  & $\hel$ & \Ape  & 14~~  & ~1.18 ($+0.58$) & 1.93 ($+2.06$) & 1.15 ($+0.51$)  \\
~JLab SANE ($E=4.7$ GeV) \cite{SANE:2018pwx} & $p$ & $A_2$       & 12~~  & ~1.29 ($+0.79$) & 1.30 ($+0.80$) & 1.33 ($+1.00$)  \\
~JLab SANE ($E=5.9$ GeV) \cite{SANE:2018pwx} & $p$ & $A_2$       & 18~~  & ~0.42 ($-2.13$) & 0.41 ($-2.21$) & 0.34 ($-2.87$)  \\
\hline
~Total & & & 470~~ & {\bf ~1.08 (\boldmath$+1.22$)} & 1.11 ($+1.65$) & 1.09 (+1.38) \\
\hhline{=======}
\end{tabular}
\label{tab:chi2_Ape}
\end{table}

\begin{table*}
\footnotesize
\caption{Reduced $\chi^2$ and $Z$-score values for longitudinal SIDIS asymmetries $A_1^h$ for production of hadrons $h$ $(=\pi^\pm, K^\pm, h^\pm)$ from $p$, $d$, and $^3$He targets. The number of data points ($N_{\rm dat}$) corresponds to a cuts of \mbox{$W^2_{\mbox{\tiny \rm SIDIS}} > 20$~GeV$^2$}, $Q^2 > m_c^2$, and $0.2 < z < 0.8$ on the SIDIS data.\\}
\begin{tabular}{l c c r c }
\hhline{=====}
~Experiment & Target~ & ~Hadron~ & ~$N_{\rm dat}$ & ~~$\chired$ ($Z$-score) \\
\hline
~COMPASS \cite{COMPASS:2010hwr}  & $p$ & $\pi^+$   & 10~~  & 0.98 ($+0.10$) \\
~COMPASS \cite{COMPASS:2010hwr}  & $p$ & $\pi^-$   & 10~~  & 1.12 ($+0.41$) \\
~COMPASS \cite{COMPASS:2009kiy}  & $d$ & $\pi^+$   &  8~~  & 0.35 ($-1.60$) \\
~COMPASS \cite{COMPASS:2009kiy}  & $d$ & $\pi^-$   &  8~~  & 0.42 ($-1.32$) \\
\hline
~COMPASS \cite{COMPASS:2010hwr}  & $p$ & $K^+$     & 10~~  & 0.28 ($-2.21$) \\
~COMPASS \cite{COMPASS:2010hwr}  & $p$ & $K^-$     & 10~~  & 0.27 ($-2.24$) \\
~COMPASS \cite{COMPASS:2009kiy}  & $d$ & $K^+$     &  8~~  & 0.83 ($-0.20$) \\
~COMPASS \cite{COMPASS:2009kiy}  & $d$ & $K^-$     &  8~~  & 0.62 ($-0.71$) \\
\hline
~SMC \cite{SpinMuon:1997yns}     & $p$ & $h^+$     & 8~~  & 1.15 ($+0.45$) \\
~SMC \cite{SpinMuon:1997yns}     & $p$ & $h^-$     & 8~~  & 1.17 ($+0.50$) \\
~SMC \cite{SpinMuon:1997yns}     & $d$ & $h^+$     & 8~~  & 0.69 ($-0.54$) \\
~SMC \cite{SpinMuon:1997yns}     & $d$ & $h^-$     & 8~~  & 1.16 ($+0.47$) \\
~COMPASS \cite{COMPASS:2009kiy}  & $d$ & $h^+$     & 10~~  & 0.67 ($-0.69$) \\
~COMPASS \cite{COMPASS:2009kiy}  & $d$ & $h^-$     & 10~~  & 1.02 ($+0.20$) \\
\hline
~Total & & & 124~~ & \boldmath{~$0.76$ ($-2.00$)} \\
\hhline{=====}
\end{tabular}
\label{tab:chi2_SIDIS}
\end{table*}

\begin{table}[b]
\footnotesize
\caption{Reduced $\chi^2$ and $Z$-score values for jet and weak boson production in polarized $pp$ collisions for the JAMpol25 analysis. Inclusive jets data are binned according to $y_{\rm jet}$ and $p_T$, while $W/Z$ production are binned according to $p_T$ and the electron rapidity, $\eta_e$. The number of data points ($N_{\rm dat}$) corresponds to a cut of $p_T > 8$~GeV on the jet data. There are no cuts on the $W/Z$ production data.\\}
\begin{tabular}{l l c c c}
\hhline{=====}
~Reaction & Experiment & Observable & ~~$N_{\rm dat}$~~ & ~~$\chired$ ($Z$-score) \\
\hline
~Inclusive jets~~ & ~STAR \cite{STAR:2006opb}     & \Apajet      &  4~  & ~0.22 ($-1.46$) \\
                  & ~STAR \cite{STAR:2007rjc}     & \Apajet      &  7~  & ~1.49 ($+0.98$) \\
                  & ~STAR \cite{STAR:2012hth}     & \Apajet      &  9~  & ~0.33 ($-1.80$) \\
                  & ~STAR \cite{STAR:2014wox}     & \Apajet      & 18~~ & ~0.56 ($-1.46$) \\
                  & ~STAR \cite{STAR:2019yqm}     & \Apajet      & 12~~ & ~1.55 ($+1.29$) \\
                  & ~STAR \cite{STAR:2021mfd}     & \Apajet      & 18~~ & ~0.77 ($-0.63$) \\
                  & ~STAR \cite{STAR:2022pgd}     & \Apajet      & 13~~ & ~0.71 ($-0.68$) \\
                  & ~PHENIX \cite{PHENIX:2010aru} & \Apajet      &  2~  & ~0.38 ($-0.49$) \\
\hline
~$W/Z$ production~~ & ~STAR \cite{STAR:2018fty}    & $A_L^{W}$   & 12~~ & ~0.84 ($-0.27$) \\
                  & ~PHENIX \cite{PHENIX:2015ade} & ~~$A_L^{W/Z}$ & 2~  & ~0.24 ($-0.79$) \\
                  & ~PHENIX \cite{PHENIX:2018wuz} & ~~$A_L^{W/Z}$ & 4~  & ~0.68 ($-0.27$) \\
\hline
~Total & & & 101~~ & \boldmath{~~$0.80$ ($-1.48$)} \\
\hhline{=====}
\end{tabular}
\label{tab:chi2_pp}
\end{table}

\clearpage

\clearpage
\bibliography{bibliography}

\end{document}